\pdfoutput=1
\documentclass[cernpreprint, atlasdraft=true, texlive=2016, UKenglish, texmf]{atlasdoc}
 
\usepackage[abbrevref=false]{atlaspackage}
\usepackage{atlasbiblatex}
 
\usepackage{atlasphysics}
 
\usepackage{pdflscape}
\usepackage{multirow}
\usepackage{bigstrut}
\usepackage{adjustbox}
\usepackage{PMGRefs-defs} 
 
\graphicspath{{logos/}{figures/}}
 
\usepackage{ANA-HDBS-2018-55-PAPER-defs}

\AtlasTitle{Search for Higgs bosons decaying
into new spin-0 or spin-1 particles in four-lepton final states with the ATLAS detector
with 139~fb$^{-1}$ of $pp$ collision data at $\sqrt{s}=13$~TeV}
 
\AtlasAbstract{
Searches are conducted for new spin-0 or spin-1 bosons
using events   where a Higgs boson with mass $125$~GeV
decays into four leptons ($\ell =$ $e$,
$\mu$). This decay is presumed to occur via an intermediate state which
contains two on-shell,
promptly decaying bosons: $H \rightarrow XX/ZX \rightarrow 4\ell$,
where the new boson $X$ has a mass between 1 and 60~GeV.
The search  uses $pp$ collision data
collected with the ATLAS detector at the LHC with an integrated luminosity of
139~fb$^{-1}$ at a centre-of-mass energy $\sqrt{s}=13$~TeV.
The data are found to be consistent with Standard Model expectations.
Limits are set on fiducial cross sections and on the branching ratio
of the Higgs~boson to decay into $XX/ZX$, improving those from previous
publications by a factor between two and four.  Limits are also set
on mixing parameters relevant in extensions of the Standard
Model containing a dark sector where $X$ is interpreted to be a dark boson.
}
 
\author{The ATLAS Collaboration}
 
\AtlasRefCode{HDBS-2018-55}

\AtlasJournal{JHEP}
\AtlasJournalRef{JHEP 03 (2022) 041}
\AtlasDOI{10.1007/JHEP03(2022)041}

\PreprintIdNumber{CERN-EP-2021-193}
\hypersetup{pdftitle={ATLAS document},pdfauthor={The ATLAS Collaboration}}
 
\begin{document}
 
\maketitle
 
\tableofcontents

\section{Introduction}

Although the Higgs~boson was discovered at the Large Hadron Collider (LHC)
in 2012~\cite{HIGG-2012-27,CMS-HIG-12-028},
there is good reason to believe that the description
of the Higgs sector of the Standard Model (SM) is still incomplete.
Besides the well-known issues of naturalness and baryon asymmetry,
astrophysical observations implying the existence of dark matter
motivate extensions to the Higgs sector of the SM,
particularly those that propose the existence of a `dark' (i.e., hidden)
sector, with its own hidden-sector particles~\cite{Martin2014,Cline2014}.
 
An attractive way to search for new physics in the Higgs sector
is through non-standard (`exotic') decays of the Higgs~boson.
Existing precision measurements of the properties of the Higgs~boson
still allow a branching ratio of up to about 30\% to non-standard
decays (assuming that the couplings of the Higgs~boson
to the $W$~and $Z$~bosons are not
larger than their SM values)~\cite{HIGG-2014-06,CMS-HIG-14-009,HIGG-2015-07}.
Further, since the SM predicts
a very narrow decay width for the Higgs~boson, even a small coupling
to a new light state could result in a significant branching ratio
to that state.  In addition, new hidden-sector particles may preferentially
couple to the Higgs~boson, making it a `portal' to explore this
new physics~\cite{Shrock:1982kd,Strassler:2006im,Schabinger:2005ei,Patt:2006fw}.
Such exotic decays of the Higgs boson
are predicted by many proposed extensions to the SM, including
models with a first-order electroweak phase
transition~\cite{Profumo:2007wc,Blinov:2015sna}, models with neutral
naturalness~\cite{Burdman:2006tz,Craig:2015pha,Curtin:2015fna}, and
models with a hidden
sector~\cite{Fayet:2004bw,Finkbeiner:2007kk,ArkaniHamed:2008qn,Dudas:2012t1,Curtin:2014cca,Curtin:2013fra,Davoudiasl:2013aya,Davoudiasl:2012ag,wells2008find,gopalakrishna2008higgs,Evans:2017kti},
as well as by several models of dark
matter~\cite{Silveira:1985rk,Pospelov:2007mp,Draper:2010ew,Ipek:2014gua,Martin:2014sxa,Clowe},
including some posited to explain observed excesses of astrophysical
positrons~\cite{Adriani:2008zr,ATIC:2008t1,Accardo:2014lma}.
They are also predicted by the
Next-to-Minimal Supersymmetric Standard Model
(NMSSM)~\cite{Dobrescu:2000yn,Ellwanger:2003jt,Dermisek:2005ar,Chang:2008cw,Morrissey:2008gm,Dawson:2018dcd}.
 
This paper reports three related searches, each of which looks
for a SM Higgs boson $H$ decaying
via a new boson into a final state consisting of
four charged leptons  ($\ell\equiv e$, $\mu$).
All use the full LHC Run~2 data set of about \SI{139}{\ifb} that the
ATLAS detector collected from proton--proton collisions at a
centre-of-mass energy of $\sqrt{s} = \SI{13}{\tev}$.
Following the models motivating these analyses,
the new boson could be either a dark-sector vector boson
or a scalar boson, denoted by $X$.
The three searches considered are:
\begin{itemize}
\item High-mass (HM): $H\ra XX\ra 4\ell$
($\SI{15}{\gev} < m_X < \SI{60}{\gev}$).
\item Low-mass (LM): $H\ra XX\ra 4\mu$
($\SI{1}{\gev} < m_X < \SI{15}{\gev}$).
\item Single $Z$~boson (ZX): $H\ra ZX\ra 4\ell$
($\SI{15}{\gev} < m_X < \SI{55}{\gev}$).
\end{itemize}
The LM analysis uses only the $4\mu$ final state because the selection efficiency
for isolated muons is significantly larger than that for isolated electrons
in this mass range (see \cref{sec:HM_Selection}).
These searches are sensitive to any intermediate bosons
within the considered mass ranges that are narrow, on-shell, and
decay promptly.  This paper provides model-independent fiducial
cross-section limits, as well as limits based on the specific models
described in \cref{sec:theory}.

This work extends previous searches performed by ATLAS with \SI{20}{\ifb} of data collected
at $\sqrt{s}=\SI{8}{\tev}$~\cite{EXOT-2013-15} and with \SI{36}{\ifb}
of data collected at $\sqrt{s}=\SI{13}{\tev}$~\cite{EXOT-2016-22}.
In addition to a larger data sample and improved lepton identification,
the signal region selection of the HM~analysis has been re-optimized.
Other similar searches, including searches for pairs of light bosons decaying
into muons, \tauleptons, photons, and/or jets,
as well as searches for a single light boson decaying into a pair of muons,
using both
$\sqrt{s}=\SI{8}{\tev}$ and \SI{13}{\tev} data, have been performed by
ATLAS~\cite{HDBS-2018-47,HIGG-2017-09,HIGG-2017-05,HIGG-2016-03,HIGG-2014-02},
CMS~\cite{CMS-HIG-13-010, CMS-HIG-16-015,CMS-HIG-18-024,CMS-EXO-20-014},
and LHCb~\cite{LHCB-PAPER-2017-038}.
Further searches for a SM Higgs~boson decaying into undetected particles
are reported in Refs.~\cite{HIGG-2018-54,CMS-HIG-17-023}.
 
This section is followed by a summary of the theoretical models used in the
interpretation of the results (\cref{sec:theory}).
Next, the detector is described  (\cref{sec:detector}),
followed by discussions of features that are common to all three
analyses, including
the samples of data and simulated events (\cref{sec:DataAndMCSamples}),
the reconstruction
of lepton candidates and of their combinations
(\cref{sec:LeptonReconstruction,sec:KinematicVariables}),
the event selections (\cref{sec:EventSelection}),
and the common systematic uncertainties (\cref{sec:systematics}).
Next, aspects specific to each analysis are
described
(\cref{sec:HM_analysis,sec:LM_analysis,sec:ZX_analysis}).
Finally, the ways the analyses are combined to extract limits, and the
interpretations of the results in terms of the theoretical models, are
presented in \cref{sec:Limits}, and a summary is given in \cref{sec:Conclusion}.

\section{Benchmark models}
\label{sec:theory}

\subsection{Dark bosons}
 
Many theories of dark matter posit a hidden
sector~\cite{Fayet:2004bw,Finkbeiner:2007kk,ArkaniHamed:2008qn,Dudas:2012t1,Curtin:2014cca,Curtin:2013fra,Davoudiasl:2013aya,Davoudiasl:2012ag,wells2008find,gopalakrishna2008higgs,Alexander:2016aln},
which does not interact with SM particles except via a mediator
or portal interaction (besides gravity).  A concrete realization
of such a mediator involves adding a field with a U(1)$_\text{d}$ dark gauge
symmetry~\cite{Curtin:2014cca,Curtin:2013fra,Davoudiasl:2013aya,Davoudiasl:2012ag,wells2008find,gopalakrishna2008higgs}
which mixes kinetically with the SM U(1)$_\text{Y}$ hypercharge gauge field with some
strength $\epsilon$~\cite{Galison:1983pa,Holdom:1985ag,Dienes:1996zr}.
The gauge boson of this symmetry is the
$\Zd$ vector boson, also called a `dark photon'.
 
The coupling strength of the $\Zd$ boson to SM particles, and hence its
lifetime (assuming no significant decays to non-SM particles), is determined by the mixing parameter $\epsilon$.
The decays of the $\Zd$ boson, on the other hand, are determined
by the gauge couplings, and the decay branching ratios
are largely independent of $\epsilon$
for $\epsilon \ll 1$.  Over the $\Zd$ mass range
$\SI{1}{\gev} < m_{\Zd} < \SI{60}{\gev}$, the branching ratio for
decays into electron or muon pairs can be 10\%--15\%~\cite{Curtin:2014cca}.
Over the same mass range, the decay is prompt
for $\epsilon \gtrsim 10^{-5}$~\cite{Curtin:2014cca}.
For smaller values of $\epsilon$, the decay vertex would be significantly
displaced from the interaction point,
while for $\epsilon \lesssim 10^{-8}$ the lifetime of the $\Zd$ boson
becomes long enough for it to likely escape the detector.
Also, the decay width of the $\Zd$~boson is very small ($\ll \SI{1}{\GeV}$)
for $\epsilon \ll 1$ and $m_{\Zd} < \SI{60}{\GeV}$.
ATLAS and CMS have
searched for these long-lifetime signatures in collisions at energies of both
\SI{8}{\tev}~\cite{EXOT-2013-22,EXOT-2014-09,SUSY-2014-02,CMS-EXO-12-037}
and \SI{13}{\tev}~\cite{EXOT-2017-03,SUSY-2017-04,EXOT-2017-28,EXOT-2017-32,CMS-HIG-18-003}.
 
If the U(1)$_\text{d}$ symmetry is broken by an additional dark Higgs~boson $s$,
then there could be mixing with strength $\kappa$ between the SM Higgs~boson
and the dark Higgs~boson~\cite{Curtin:2014cca,Curtin:2013fra,Davoudiasl:2013aya,Davoudiasl:2012ag,wells2008find,gopalakrishna2008higgs}.
The observed Higgs~boson would be one of the mass eigenstates and could
also decay into dark-sector particles, including dark Higgs~bosons
that subsequentially decay into SM~fermions. The dark Higgs~boson would
inherit the Yukawa couplings from the SM Higgs~boson and decay
preferentially into high-mass fermion pairs.
 
A further possibility is mass mixing between the $\Zd$~boson and
the SM $Z$~boson~\cite{Davoudiasl:2013aya,Davoudiasl:2012ag}.
If the mass term for this mixing is written as $\epsilon_Z m_Z^2 Z\Zd$,
with $\epsilon_Z = \delta m_{\Zd} / m_Z$, then $\delta$ is the model
parameter describing the mixing.
 
\begin{figure}[!htbp]
\begin{center}
\subfloat[]{\includegraphics[width=0.25\textwidth]{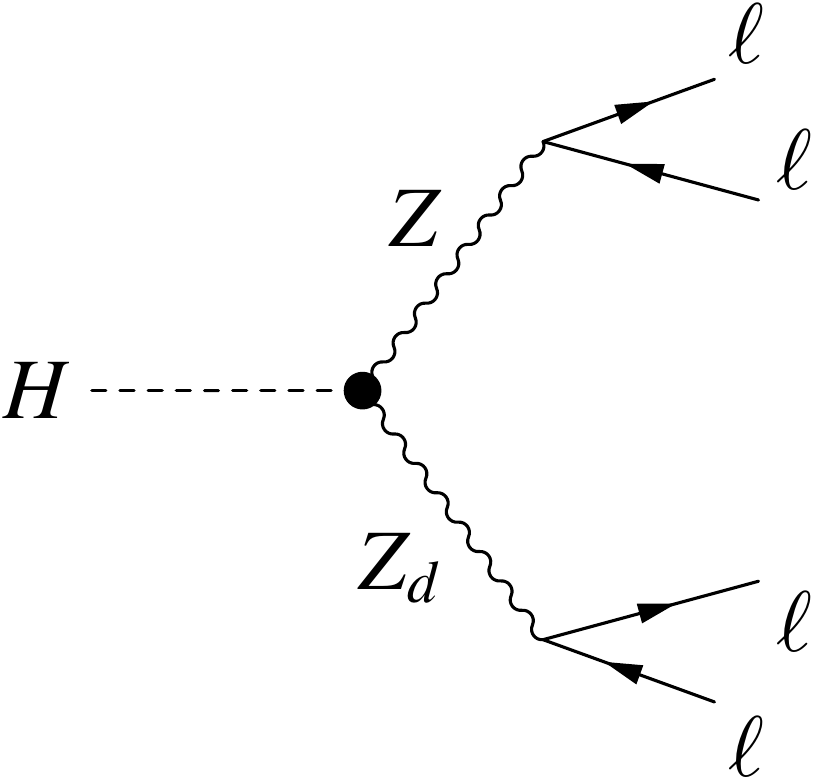}\label{fig:zx-diag}}
\hskip 0.1\textwidth
\subfloat[]{\includegraphics[width=0.25\textwidth]{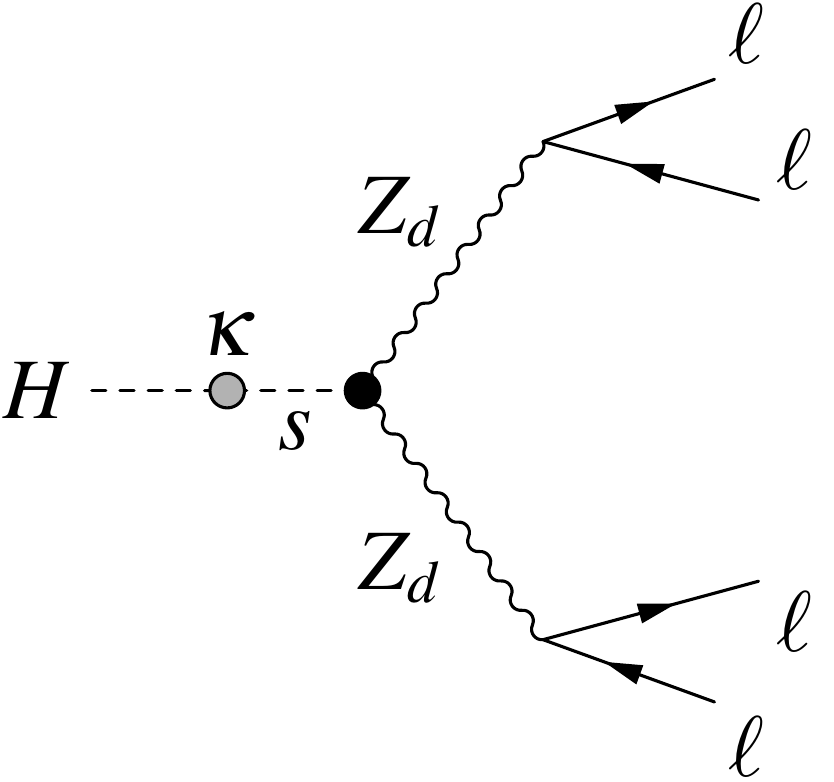}\label{fig:zd-diag}}
\hskip 0.1\textwidth
\subfloat[]{\includegraphics[width=0.25\textwidth]{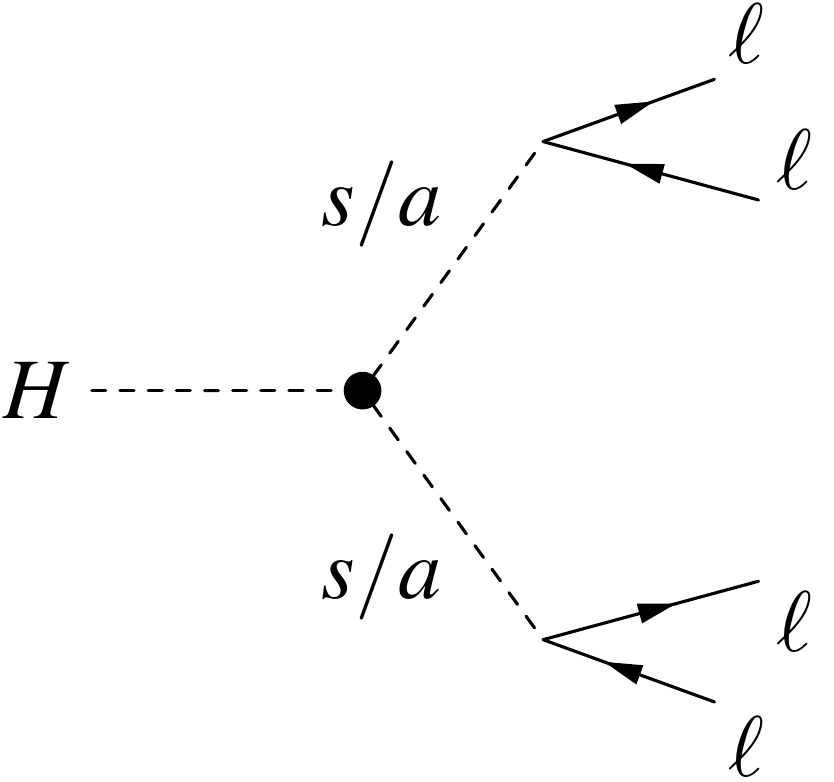}\label{fig:aa-diag}}
\caption{Exotic decays of the Higgs~boson into four leptons induced by intermediate
dark vector bosons via
(a)~the hypercharge portal (to which the ZX analysis is sensitive) and
(b)~the Higgs portal, where $s$ is a dark Higgs boson\ifnotpo{~\cite{Curtin:2014cca}}
(to which the HM and LM analyses are sensitive).
The $\Zd$ gauge boson decays into SM particles through kinetic mixing
with the hypercharge field (with branching ratios that are nearly
independent of $\epsilon$). The $HZ\Zd$
vertex factor is proportional to $\epsilon$ whereas the
$H\Zd\Zd$ vertex factor is proportional to
$\kappa$. (c)~illustrates the decay of a Higgs~boson into dark~Higgs scalars $s$
or pseudoscalars $a$ that couple to SM particles through
mixing with the SM~Higgs field in models
with an extended Higgs sector\ifnotpo{ (}\ifnotpo{\cref{sec:theory-haa}}\ifnotpo{)}.}
\label{fig:zd-diagrams}
\end{center}
\end{figure}

The processes probed in this paper that involve a SM Higgs~boson
decaying into $\Zd$ bosons are depicted in \cref{fig:zx-diag,fig:zd-diag}
and are included in the
Hidden Abelian Higgs Model (HAHM)~\cite{Curtin:2014cca}.
The decay $H\ra Z\Zd$ is sensitive to the parameters
$\epsilon$ and $m_{\Zd}$, but does not depend on $\kappa$.  However,
the presence of an irreducible background from the SM $H\ra ZZ^*$ process means
that this signal can be observed only as a peak in the dilepton
mass spectrum over the background.  The process $H\ra \Zd\Zd$, in contrast,
is much more easily separated from SM backgrounds and hence is potentially
sensitive to smaller values of the kinetic mixing $\epsilon$, where
it is only required that the mixing be large enough for the $\Zd$~boson
to decay promptly.  However, this process does require mixing
between the SM and dark-sector Higgs~bosons and thus depends on $\kappa$.
 
Limits on the kinetic mixing of $\epsilon \lesssim 0.03$ have been set
from precision electroweak measurements~\cite{Curtin:2014cca,Hook:2010tw,Pospelov:2008zw}
over the range $\SI{1}{\gev} < m_{\Zd} < \SI{200}{\gev}$.
Searches for dilepton resonances, $pp\ra \Zd\ra \ell\ell$, at the LHC
for $m_{\Zd} < m_Z$ imply that $\epsilon \lesssim 0.005$--$0.020$
for $\SI{20}{\gev} < m_{\Zd} < \SI{80}{\gev}$~\cite{Hoenig:2014dsa}.
Other searches rule out $\epsilon \gtrsim 10^{-3}$
for $\SI{10}{\mev} < m_{\Zd} < \SI{10}{\gev}$~\cite{Batley:2009aa,Lees:2012ra,PhysRevLett.59.755,PhysRevD.38.3375,PhysRevLett.67.2942,Adler:2004hp}.
The $H\ra XX\ra 4\ell$ analyses constrain the Higgs~mixing
parameter~$\kappa$,
while the $H\ra Z\Zd\ra 4\ell$ analysis provides information about the
kinetic mixing parameter~$\epsilon$.

\subsection{Extended Higgs sectors}
\label{sec:theory-haa}
 
Models containing two Higgs~doublets and an additional
scalar field (2HDM+S)~\cite{Curtin:2013fra,a0} are also relevant
for the search for $H\ra XX\ra 4\mu$.
Two-Higgs-doublet
models (2HDMs) generically contain two neutral scalars $H_{1,2}$,
two charged scalars $H^\pm$,
and one neutral pseudoscalar $A$.  The lighter of the neutral scalars $H_1$
is identified as the observed Higgs~boson~$H$, while the other states
are constrained to be heavy by existing data~\cite{Gunion:1997aq,Belanger:2013xza}.
Adding a complex scalar singlet that mixes
weakly with $H_{1,2}$ gives two additional states, a scalar $s$
and a pseudoscalar $a$.   If these are lighter than $m_H/2$, then
$H\ra aa$ and $H\ra ss$ decays are allowed (\cref{fig:aa-diag}).
This paper probes the process $H\ra aa\ra 4\mu$, but
limits on $H\ra aa\ra 4\mu$ also apply
to $H\ra ss\ra 4\mu$.
 
The decays of the scalar and pseudoscalar
into fermions are determined by their Yukawa couplings~\cite{Curtin:2013fra},
implying that the branching ratio to electrons is very small, and
that the branching ratio to muons is smaller than that of the $\Zd$ vector bosons
described previously.
Branching ratios for $H\ra aa$ and $a\ra \mu\mu$ can be significant
in the range $2m_\mu < m_a < 2m_\tau$, ranging
from $10^{-2}$ to $10^{-1}$ in some regions of
parameter space~\cite{Curtin:2013fra}.
In 2HDMs, there are several possible ways
for the Higgs sector to couple to fermions.  Of these,
type-III models (in which leptons and quarks couple to
different Higgs~doublets) at large $\tan\beta$
(where $\tan\beta$ is the ratio of the vacuum expectation values
of the two Higgs~doublets)
are particularly interesting for these analyses.
A light pseudoscalar can correspond to the $R$-symmetry limit of the
NMSSM~\cite{Ellis:1988er,Dermisek:2006wr},
which reduces the need for fine-tuning and addresses
the $\mu$-problem~\cite{Giudice:1988yz}.
Searches for exotic decays of the Higgs~boson into new light scalars
or pseudoscalars have been carried out for a variety of
mass ranges and final states with both
LHC~\cite{HIGG-2016-03,HIGG-2017-05,HIGG-2017-09,EXOT-2016-22,CMS-HIG-18-006,CMS-HIG-18-011,CMS-HIG-18-003,CMS-HIG-17-024,CMS-HIG-17-029,
HIGG-2014-02,EXOT-2013-15,EXOT-2013-24,CMS-HIG-13-010,CMS-HIG-16-015,CMS-HIG-14-019,CMS-HIG-16-015}
and Tevatron~\cite{Abazov:2009yi} data.
 
\section{ATLAS detector}
\label{sec:detector}

\newcommand{\AtlasCoordFootnote}{
ATLAS uses a right-handed coordinate system with its origin at the nominal interaction point (IP)
in the centre of the detector and the \(z\)-axis along the beam pipe.
The \(x\)-axis points from the IP to the centre of the LHC ring,
and the \(y\)-axis points upwards.
Cylindrical coordinates \((r,\phi)\) are used in the transverse plane,
\(\phi\) being the azimuthal angle around the \(z\)-axis.
The pseudorapidity is defined in terms of the polar angle \(\theta\) as \(\eta = -\ln \tan(\theta/2)\),
and the rapidity is defined in terms of energy \(E\) and momentum \(p\) as \(y = (1/2)[(E+p_z)/(E-p_z)]\).
Angular distance is measured in units of \(\Delta R \equiv \sqrt{(\Delta\eta)^{2} + (\Delta\phi)^{2}}\).}
 
The ATLAS detector~\cite{PERF-2007-01} at the LHC covers nearly the entire solid angle around the collision point.\footnote{\AtlasCoordFootnote}
It consists of an inner tracking detector surrounded by a thin superconducting solenoid, electromagnetic and hadron calorimeters,
and a muon spectrometer incorporating three large superconducting air-core toroidal magnets.
 
The inner-detector system (ID) is immersed in a \SI{2}{\tesla} axial magnetic field
and provides charged-particle tracking in the range \(|\eta| < 2.5\).
The high-granularity silicon pixel detector covers the vertex region and typically provides four measurements per track,
the first hit normally being in the insertable B-layer (IBL) installed before Run~2~\cite{ATLAS-TDR-19,PIX-2018-001}.
It is followed by the silicon microstrip tracker, which usually provides eight measurements per track.
These silicon detectors are complemented by the transition radiation tracker (TRT),
which enables radially extended track reconstruction up to \(|\eta| = 2.0\).
The TRT also provides electron identification information
based on the fraction of hits (typically 30 in total) above a higher energy-deposit threshold corresponding to transition radiation.
 
The calorimeter system covers the pseudorapidity range \(|\eta| < 4.9\).
Within the region \(|\eta|< 3.2\), electromagnetic calorimetry is provided by barrel and
endcap high-granularity lead/liquid-argon (LAr) calorimeters,
with an additional thin LAr presampler covering \(|\eta| < 1.8\)
to correct for energy loss in material upstream of the calorimeters.
Hadron calorimetry is provided by the steel/scintillator-tile calorimeter,
segmented into three barrel structures within \(|\eta| < 1.7\), and two copper/LAr hadron endcap calorimeters.
The solid angle coverage is completed with forward copper/LAr and tungsten/LAr calorimeter modules
optimized for electromagnetic and hadronic energy measurements respectively.
 
The muon spectrometer (MS) comprises separate trigger and
high-precision tracking chambers measuring the deflection of muons in a magnetic field generated by the superconducting air-core toroidal magnets.
The field integral of the toroids ranges between \num{2.0} and \SI{6.0}{\tesla\metre}
across most of the detector.
A set of precision chambers covers the region \(|\eta| < 2.7\) with three layers of monitored drift tubes,
complemented by cathode-strip chambers in the forward region, where the background is highest.
The muon trigger system covers the range \(|\eta| < 2.4\) with resistive-plate chambers in the barrel, and thin-gap chambers in the endcap regions.
 
Interesting events are selected by the first-level trigger system implemented in custom hardware,
followed by selections made by algorithms implemented in software in the high-level trigger~\cite{TRIG-2016-01}.
The first-level trigger accepts events from the \SI{40}{\MHz} bunch crossings at a rate below \SI{100}{\kHz},
which the high-level trigger reduces in order to record events to disk at about \SI{1}{\kHz}.
 
An extensive software suite~\cite{ATL-SOFT-PUB-2021-001} is used in the reconstruction and analysis of
real and simulated data, in detector operations, and in the trigger and data acquisition systems of the experiment.
 
 
\section{Data and simulated event samples}
\label{sec:DataAndMCSamples}
 
The results in this paper are based on \SI{139}{\per\femto\barn}
of $\sqrt{s}=\SI{13}{\tera\eV}$ proton--proton ($pp$) collision data collected with the ATLAS detector
over the period 2015--2018.
 
Monte Carlo (MC) simulation is used
to determine expected contributions from both
the signal processes and most background processes.
For most samples,
detector effects were included using a \GEANT~\cite{Agostinelli:2002hh} simulation of the ATLAS detector~\cite{SOFT-2010-01}.
The $H\to aa$, $H\to Z\Zd$, and $H\to Za$ signal samples, as well as a portion
of the $gg\to ZZ^*$ and triboson background samples, instead used
a fast simulation~\cite{SOFT-2010-01} which relies on a
parameterization of the calorimeter response~\cite{ATL-PHYS-PUB-2010-013}.
The effects of pile-up (additional $pp$ collisions
in the same or a neighbouring bunch crossing) are included in the simulation.
Weights are applied to the simulated events to correct for small differences
between data and simulation in the reconstruction, identification,
isolation, and impact parameter efficiencies for electrons and muons~\cite{EGAM-2018-01,PERF-2017-01,MUON-2018-03-pub}.
Further, the lepton momentum scales and resolutions in the simulation
are adjusted to match the data~\cite{EGAM-2018-01,PERF-2017-03,MUON-2018-03-pub}.
\cref{tab:MonteCarloOverview} summarizes the samples
and generators used, which include
\mgamc\ version 2.2.2~\cite{Alwall:2014hca},
\POWHEGBOX v2~\cite{Hamilton:2013fea,Hamilton:2015nsa,Alioli:2010xd,Nason:2004rx,Frixione:2007vw},
\PYTHIA\ 8.186~\cite{Sjostrand:2014zea}
(along with \EVTGEN 1.2.0~\cite{Lange:2001uf} to decay heavy-flavour hadrons),
and \sherpa\ 2.2.0, 2.2.1, and 2.2.2~\cite{Bothmann:2019yzt}.
 
\begin{table}[tbp]
\caption{Overview of the event generators used for the simulated
signal and background samples.  For each process, the table lists
the matrix element (ME) generator used along with the
parton distribution function (PDF), the model used
to implement parton showering (PS), the underlying event (UE),
and the decay of heavy-flavour hadrons (HF),
as well as the set of tuned parameters used to model the UE.
The text gives the full version numbers of the generators.}
\begin{center}{\small
\begin{tabular}{llllll}
\toprule
 
Process                      & ME generator
& ME PDF
& PS/UE/HF model
& UE tune \\
\hline
 
$ H \to \Zd\Zd$ / $Z\Zd$ \bigstrut[t] & \mgamc
& \NNPDF[2.3lo]~\cite{Ball:2012cx}
& \PYTHIA/\EVTGEN
& A14~\cite{ATL-PHYS-PUB-2014-021}\\
 
$ H \to aa$		\bigstrut[b]
& \POWHEGBOX
& \PDFforLHC[15nnlo]
& \PYTHIA/\EVTGEN
& AZNLO~\cite{STDM-2012-23} \\
 
&
& \quad~\cite{Butterworth:2015oua}
&
& \\
 
$ H \to Za$		\bigstrut[b]
& \POWHEGBOX
& \PDFforLHC[15nnlo]
& \PYTHIA/\EVTGEN
& AZNLO \\
 
\hline

ggF\bigstrut[t]              & \POWHEGBOX
& \PDFforLHC[15nnlo]
& \PYTHIA/\EVTGEN
& AZNLO \\
 
VBF                          & \POWHEGBOX
& \CT[10nlo]~\cite{Lai:2010vv}
& \PYTHIA/\EVTGEN
& AZNLO \\
 
$VH$                         & \PYTHIA
& \NNPDF[2.3lo]~\cite{Ball:2012cx}
& \PYTHIA/\EVTGEN
& A14 \\
 
$ggZH$                       & \POWHEGBOX
&   \NNPDF[3.0nlo]
& \PYTHIA/\EVTGEN
& AZNLO    \\
 
$\bbbar H$                   & \mgamc
&   \NNPDF[2.3lo]
& \PYTHIA/\EVTGEN
&   A14   \\
 
$\ttbar H$                 & \POWHEGBOX
&   \NNPDF[2.3lo]
& \PYTHIA/\EVTGEN
&   A14  \\
 
$ZZ$                         & \sherpa
& \NNPDF[3.0nnlo]
& \sherpa
& \sherpa default \\
 
$VVV$                        & \sherpa
& \NNPDF[3.0nnlo]
& \sherpa
& \sherpa default \\
 
$\ttbar Z$	             & \sherpa
&   \NNPDF[3.0nnlo]											& \sherpa					        &   \sherpa default \\
 
$Z+\textrm{jets}$          & \sherpa
&  \NNPDF[3.0nnlo]
& \sherpa
&  \sherpa default \\
 
$\ttbar$                   & \POWHEGBOX
&  \NNPDF[3.0nlo]
& \PYTHIA/\EVTGEN
&  A14 \\
 
$WZ$                        & \POWHEGBOX
& \CT[10nlo]
& \PYTHIA /\EVTGEN
&  A14 \\

\bottomrule
\end{tabular}
}
\label{tab:MonteCarloOverview}
\end{center}
\end{table}

 
Signal samples involving a \Zd vector boson were generated according to
the HAHM~\cite{wells2008find,gopalakrishna2008higgs,Curtin:2013fra, Curtin:2014cca}
implementation in \mgamc,
with the Higgs~bosons being produced via gluon--gluon fusion (ggF)
and the Higgs~boson mass set to $m_H = \SI{125}{\GeV}$.
For the $H\ra \Zd\Zd$ process, $\epsilon$ and $\kappa$ were both
set to $10^{-4}$ and samples were generated with $m_{\Zd} = \SI{0.5}{\GeV}$,
$\SI{1}{\GeV}$,
$\SI{2}{\GeV}$, and every $\SI{5}{\GeV}$ in the range
$\SI{5}{\gev} \le m_{\Zd} \le \SI{60}{\GeV}$.
For the $H\ra Z\Zd$ process, $\kappa$ was changed to $10^{-10}$,
and samples were generated every $\SI{5}{\GeV}$ in the range
$\SI{15}{\gev} \le m_{\Zd} \le \SI{55}{\GeV}$.
Final states with $\tau$-leptons were not included; the change in signal
region yield due to the omission of these decays was below~1\%
and thus neglected.
The much smaller production of signal events by vector-boson fusion (VBF),
$VH$, and $\ttbar H$ was also omitted.
 
Samples for $H\to aa$ were simulated using \POWHEGBOX at
next-to-next-to-leading order (NNLO) for arbitrary inclusive $gg\to H$
observables by reweighting the Higgs~boson rapidity to that
of HNNLO~\cite{Hamilton:2013fea,Hamilton:2012np,Campbell:2012am,Hamilton:2012rf,Catani:2007vq-fixed}.
Again, only the ggF production process was considered.
Higgs~boson decays into two scalars and thence into muons
were simulated at leading order (LO) using \PYTHIA. Samples were
generated for $m_a=0.5$, 1, 2, 2.5, 4, 6, 8, 10, 15, 30, 45, and $\SI{60}{\GeV}$.
Samples for $H\to Za$ were generated similarly, for
$m_a = 1$, 2, 4, 6, 8, 10, 15, 20, 25, and $\SI{30}{\GeV}$.

Prompt-lepton backgrounds are estimated directly from MC simulations.
They arise primarily from the SM $H\to ZZ^* \to 4\ell$
process along with the non-resonant $ZZ^* \to 4 \ell$ process.
Smaller leptonic backgrounds arise from triboson production
as well as $t\bar{t}+Z$ decays.
Decays involving $Z\ra\tau\tau$ were found to contribute negligibly
to the background yields and are thus not included in the simulation.
Backgrounds with jets misidentified as leptons
are estimated with data-driven methods,
detailed below in the individual analysis sections.
 
The $H\to ZZ^* \to 4\ell$ background process comprises various
Higgs~boson production modes.
The ggF process $gg\ra H$~\cite{ATL-PHYS-PUB-2016-006,Hamilton:2013fea,Hamilton:2015nsa,Alioli:2010xd,Nason:2004rx,Frixione:2007vw}
was simulated in the same way as the $H\ra aa$ signal sample described above.
The prediction was normalized to the
next-to-next-to-next-to-leading-order (N$^3$LO) cross section in QCD with
next-to-leading-order (NLO) electroweak corrections~\cite{
deFlorian:2016spz,Anastasiou:2016cez,Anastasiou:2015ema,Dulat:2018rbf,
Harlander:2009mq,Harlander:2009bw,Harlander:2009my,
Czakon:2021yub,Pak:2009dg,
Actis:2008ug,Actis:2008ts,Bonetti:2018ukf}.
The VBF process~\cite{
Nason:2009ai,Alioli:2010xd,Nason:2004rx,Frixione:2007vw}
was simulated using \POWHEGBOX at NLO.  The prediction was normalized
to an approximate-NNLO cross section in QCD with NLO electroweak
corrections~\cite{Ciccolini:2007jr,Ciccolini:2007ec,Bolzoni:2010xr}.
Associated production with a vector boson ($VH$)~\cite{Luisoni:2013kna,Ciccolini:2003jy,Brein:2003wg, Ferrera:2011bk,Brein:2011vx, Ferrera:2013yga, Ferrera:2014lca, Campbell:2016jau}
was simulated at LO, while $t\bar{t}$ and $b\bar{b}$ associated production ($\ttbar H$, $\bbbar H$)~\cite{ATL-PHYS-PUB-2016-005,Alwall:2014hca},
as well as loop-induced Higgs and $Z$~boson production ($ggZH$)~\cite{Kniehl:1990iva}, were simulated at NLO.
 
The non-resonant $\qqbar\to ZZ^* \to 4 \ell$
background process~\cite{ATL-PHYS-PUB-2017-005} was simulated
using \sherpa 2.2.2 at NLO for up to one additional parton and at LO for
up to three additional partons.  Matrix element calculations
were matched and merged with the \sherpa parton shower
based on the Catani--Seymour dipole
factorization~\cite{Gleisberg:2008fv,Schumann:2007mg} using the
MEPS@NLO prescription~\cite{Hoeche:2011fd,Hoeche:2012yf,Catani:2001cc,Hoeche:2009rj}.
The virtual QCD corrections are provided by the \openloops
library~\cite{Buccioni:2019sur,Cascioli:2011va,Denner:2016kdg}.
The gluon-initiated process ($gg\to ZZ^* \to 4 \ell$) was simulated in the
same manner,
except that it was at LO, and the s-channel $H$ diagrams
were omitted to avoid double-counting.  The gluon-initiated process
has a large QCD correction at NLO, so the cross section was scaled
by a NLO/LO $K$-factor of $1.70\pm0.15$~\cite{Caola:2015psa}.
Interference between the $gg\to ZZ^* \to 4 \ell$ and
$gg\to H\to  ZZ^* \to 4 \ell$ processes is neglected.
 
Higher-order electroweak processes include triboson production ($VVV$) and
vector-boson scattering (VBS).  These processes can yield final states
with four leptons along with two additional particles.  They
were generated with \sherpa 2.2.2 at NLO for the inclusive processes
and at LO for up to two additional partons, with the same treatment
as for $\qqbar\to ZZ^* \to 4 \ell$.  Higgs~boson production via VBF
was subtracted from these samples in order to avoid double-counting.
 
The process $\ttbar + (Z\to \ell\ell)$ was generated with \sherpa 2.2.0
at LO with up to one additional parton emission.
 
Other, reducible, backgrounds have fewer than four prompt leptons in the
final state, but can be accepted by the signal selection if there are
additional leptons from heavy-flavour decay or jets misidentified
as leptons.  The $Z+\textrm{jets}$ process was generated
with \sherpa 2.2.1 using NLO matrix elements for up to two partons and
LO matrix elements for up to four partons.  The $\ttbar$ process
was generated with \POWHEGBOX at NLO with the $h_\textrm{damp}$
parameter, which regulates the high-$\pt$ radiation against
which the $\ttbar$ system recoils,
set to $1.5m_{\textrm{top}} = \SI{258}{\GeV}$~\cite{ATL-PHYS-PUB-2016-020}.
The $WZ$ process was also generated with \POWHEGBOX at NLO with the
\CT[10nlo] PDF.

 
\section{Event reconstruction and selection}
 
\subsection{Lepton reconstruction}
\label{sec:LeptonReconstruction}
 
For the analyses considered in this paper, the final-state objects of interest
are electrons and muons.
 
Electrons are reconstructed and identified from charged-particle tracks
in the ID that match energy deposits in the calorimeters~\cite{EGAM-2018-01}.
The identification algorithm, based on a likelihood analysis,
corresponds to the `Loose' selection described in Ref.~\cite{EGAM-2018-01}.
The reconstruction and identification efficiency for electrons
from $Z\to ee$ decays is about 90\% per electron~\cite{EGAM-2018-01}.
 
Muon reconstruction~\cite{MUON-2018-03-pub} begins by independently finding
tracks in both the ID and MS.  These track candidates are combined in a second
step along with information from the calorimeters to form muon candidates
of different types.  Combined muons have matching tracks in both
the MS and ID.  Segment-tagged muons have an ID~track but only
a single-chamber track segment in the MS.  Calorimeter-tagged (CT) muons
have no MS~track but have a pattern of energy deposition in the calorimeters
consistent with a muon; this is used only in regions where the MS
is not fully instrumented ($|\eta| < 0.1$).  Finally, stand-alone (SA) muons
have an MS~track but no ID~track, and are used in regions beyond
the coverage of the ID, $2.5 < |\eta| < 2.7$.  Due to the reduced performance
of the latter two types, no more than one
CT or SA muon may be used in an event.  Muons are then identified by
imposing quality requirements, corresponding to the `Loose' selection
in Ref.~\cite{MUON-2018-03-pub}.
The reconstruction and identification efficiency for muons
from $W\to \mu\nu$ decays is greater than 98\%~\cite{MUON-2018-03-pub}.
 
To avoid identifying the same detector signature as multiple particles,
an electron candidate that has the
same ID track as a muon candidate is ignored, unless the muon
is only calorimeter-tagged, in which case the muon is ignored instead.
Electrons that have the same track or cluster as a higher-$\pt$
electron are also ignored.

\subsection{Invariant kinematic mass variables}
\label{sec:KinematicVariables}
 
All three analyses considered in this paper involve looking for mass resonances
in final states consisting of a quadruplet of two same-flavour opposite-sign (SFOS)
lepton pairs: $(e^+e^- + e^+e^-)$, ($e^+e^- + \mu^+\mu^-)$, or $(\mu^+\mu^- + \mu^+\mu^-)$.
The invariant masses of the two pairs are denoted by
$m_{12}$ and $m_{34}$, where $m_{12}$ is taken to be the one
closer in mass to the $Z$~boson, $|m_{12} - m_Z| < |m_{34}-m_Z|$.
 
If all four leptons have the same flavour, then for a given
$m_{12}$ and $m_{34}$ labelling, alternative SFOS pairings
can also be defined.  An invariant mass $m_{14}$ is constructed from the positively charged
lepton of the $m_{12}$ pair and the negatively charged lepton of the $m_{34}$
pair. The other alternative pairing $m_{23}$ is constructed analogously.

\subsection{Common event selection}
\label{sec:EventSelection}
 
The analyses all involve a Higgs~boson decaying into
a pair of new bosons $X$,
or into a new boson $X$ along with a $Z$~boson, which
in turn decay into pairs of leptons.  The $X$~bosons are presumed to be on-shell,
so the strategy is to search for resonances in the relevant dilepton
mass distributions.  Each analysis defines a signal region (SR) via a series
of selections on measured quantities which maximizes the sensitivity
to the signal.
 
All three analyses share a common preselection, but differ in the
subsequent steps of selecting the candidate final-state leptons,
forming them into quadruplets, selecting one of those quadruplets,
and applying further requirements to the selected quadruplet.
\cref{tab:eventpreselectionSummary} shows the event
selections of the different analyses.
\begin{table}
\caption{Summary of event selection requirements for the ZX, HM, and LM
analyses.  The quarkonia masses are taken to be
$m_{J/\Psi} = \SI{3.096}{\GeV}$, $m_{\Psi(2S)} = \SI{3.686}{\GeV}$,
$m_{\Upsilon(1S)} = \SI{9.461}{\GeV}$,
and $m_{\Upsilon(3S)} = \SI{10.355}{\GeV}$\pdgcite.
The text provides other definitions.
}

\centering
 
\begin{adjustbox}{width=1.0\textwidth}
 
\begin{tabular}{|c|c|>{\centering}m{5cm}|>{\centering}m{5cm}|>{\centering}m{5cm}|}
\cline{3-5} \cline{4-5} \cline{5-5}
\multicolumn{1}{c}{} &
& Single $Z$ (ZX) analysis \bigstrut[t]\\
$H\rightarrow ZX\rightarrow4\ell$ $(\ell=e,\mu)$ \bigstrut[b]
& High-mass (HM) analysis \bigstrut[t]\\
$H\rightarrow XX\rightarrow4\ell$ $(\ell=e,\mu)$ \bigstrut[b]
& Low-mass (LM) analysis\\
$H\rightarrow XX\rightarrow4\mu$ \tabularnewline
\hline
 
\multicolumn{2}{|c|}{Mass range\bigstrut}
& $\SI{15}{\GeV} < m_X < \SI{55}{\GeV}$ \bigstrut
& $\SI{15}{\GeV} < m_X < \SI{60}{\GeV}$ \bigstrut
& $\SI{1}{\GeV} < m_X < \SI{15}{\GeV}$ \bigstrut\tabularnewline
\hline
 
\multicolumn{2}{|c|}{Baseline electrons}
& \multicolumn{2}{>{\centering}m{10cm}|}{$\pt > \SI{7}{\GeV}$ and $|\eta| < 2.47$;
Loose identification with an IBL hit\bigstrut[t]\\
$|z_0 \sin \theta| < \SI{0.5}{\mm}$}\bigstrut[b]
& ---\tabularnewline
\hline
 
\multicolumn{2}{|c|}{Baseline muons}
& \multicolumn{3}{>{\centering}m{15cm}|}{$\pt > \SI{5}{\GeV}$ ($\SI{15}{\GeV}$ if calo-tagged) and $|\eta| < 2.7$; Loose identification\bigstrut[t]\\
$|z_0 \sin \theta| < \SI{0.5}{\mm}$ and $d_0 < \SI{1}{\mm}$ (except for stand-alone muons)}\bigstrut[b] \tabularnewline
\hline
 
\multicolumn{2}{|c|}{Quadruplet selection}
& \multicolumn{3}{>{\centering}m{15cm}|}{Trigger-matched $e^+e^-e^+e^-$, $e^+e^-\mu^+\mu^-$, or $\mu^+\mu^-\mu^+\mu^-$;
$\le 1$ SA+CT $\mu$\bigstrut[t] \\
Three leading-$\pt$ leptons satisfying $\pt> \SI{20}{\GeV}$, $\SI{15}{\GeV}$, $\SI{10}{\GeV}$\\
Define pairs $m_{12}$ and $m_{34}$ such that $|m_{12}-m_Z| < |m_{34} - m_Z|$  \bigstrut[b]} \tabularnewline
\cline{3-5}
 
\multicolumn{2}{|c|}{}
& $\SI{50}{\GeV} < m_{12} < \SI{106}{\GeV}$ \bigstrut[t]\\
$\SI{12}{\GeV} < m_{34} < \SI{115}{\GeV}$\\
$m_{14,32} > \SI{5}{\GeV}$ ($4e$/$4\mu$) \bigstrut[b]
& \multicolumn{2}{>{\centering}m{10cm}|}{---} \tabularnewline
\cline{3-5}
 
\multicolumn{1}{|c}{}  &
& \multicolumn{2}{>{\centering}m{10cm}|}{$\Delta R(\ell,\ell^{\prime})>0.10\,(0.20)$ for same-flavour (different-flavour) $\ell,\ell^{\prime}$\bigstrut}
& ---\bigstrut \tabularnewline
\hline
 
\multicolumn{2}{|c|}{Quadruplet ranking} \bigstrut
& In order $4\mu$, $2e2\mu$, $2\mu2e$, $4e$\bigstrut[t]\\
Smallest $|m_Z - m_{12}|$\\
Smallest $|m_Z - m_{34}|$ \bigstrut[b]
& \multicolumn{2}{>{\centering}m{10cm}|}{Select quadruplet with smallest $\Delta m_{\ell\ell}=|m_{12}-m_{34}|$}\tabularnewline
\hline
 
\multirow{10}{*}{\parbox{1.5cm}{Event\\ selection}} & \bigstrut[t]\parbox{3cm}{\centering Isolation \&\bigstrut[t]\\ impact parameter}
& \multicolumn{3}{>{\centering}m{15cm}|}{Track and calorimeter isolation,
excluding tracks/clusters from other leptons in the quadruplet\bigstrut[t]\\ 
$d_{0}/\sigma_{d_{0}}<5$ for electrons and $d_{0}/\sigma_{d_{0}}<3$  for muons}\bigstrut[b]\tabularnewline
\cline{2-5}
 
& $m_{4\ell}$ \bigstrut
& \multicolumn{2}{>{\centering}m{10cm}|}{$\SI{115}{\GeV} < m_{4\ell} < \SI{130}{\GeV}$}
& $\SI{120}{\GeV} < m_{4\ell} < \SI{130}{\GeV}$ \tabularnewline
\cline{2-5}

& $Z$-veto
& ---
& $\SI{10}{\GeV} < m_{12,34} < \SI{64}{\GeV}$\bigstrut[t] \\
For $4e$ and $4\mu$ channels: $\SI{5}{\GeV} < m_{14,23} < \SI{75}{\GeV}$ \bigstrut[b]
& ---
\tabularnewline
\cline{2-5}
 
& Heavy-flavour veto
& ---
& \multicolumn{2}{>{\centering}m{10cm}|}{Reject event if $m_{12,34,14,23}$ in:\bigstrut[t]\\
$(m_{J/\Psi}-\SI{0.25}{\GeV})$ to $(m_{\Psi(2S)}+\SI{0.30}{\GeV})$, or\\
$(m_{\Upsilon(1S)}-\SI{0.70}{\GeV})$ to $(m_{\Upsilon(3S)}+\SI{0.75}{\GeV})$}\bigstrut[b]
\tabularnewline
\cline{2-5}
 
& Signal region\bigstrut
& ---
& $m_{34}/m_{12} > 0.85 - 0.1125 f(m_{12})$
& $\SI{1.2}{\GeV} < m_{12,34} < \SI{20}{\GeV}$\bigstrut[t]\\
$m_{34}/m_{12}>0.85$\\
Reject event if $m_{12,34}$ in:\\
$\SI{2}{\GeV}$ to $\SI{4.4}{\GeV}$, or\\
$\SI{8}{\GeV}$ to $\SI{12}{\GeV}$
\bigstrut[b]
\tabularnewline
\hline
\end{tabular}
 
\end{adjustbox}
 
\label{tab:eventpreselectionSummary}
\end{table}
 
The common preselection requires that events were recorded
with the detector in good operating condition~\cite{Laycock:2017ieh}
and without excess calorimeter noise~\cite{ATLAS-CONF-2015-029}.
Each event must have an identified primary vertex with
at least two tracks~\cite{ATL-PHYS-PUB-2015-026} and at least
four lepton candidates.
Events were triggered by requiring either one or two
lepton candidates, where the candidates could be
either electrons or muons~\cite{TRIG-2018-05,TRIG-2018-01,CERN-EP-2021-076}.
The lepton candidates identified
offline must match candidates identified by the trigger.
The trigger \pT requirements range
from $\pT > \SI{7}{\GeV}$ to $\pT > \SI{60}{\GeV}$, depending
on lepton multiplicity and flavour. In either case, the trigger efficiency
is above \SI{95}{\percent} (relative to signal region events surviving
all other event selections).
 
Electron and muon candidates are reconstructed as described
in \cref{sec:LeptonReconstruction}.  Electrons must be within the central region
of the detector ($|\eta| < 2.47$), have $\pT > \SI{7}{\GeV}$,
have a longitudinal impact parameter $z_0$ that
satisfies $|z_0 \sin \theta| < \SI{0.5}{\mm}$ with respect to the primary vertex, and
have an additional associated hit in the insertable B-layer.
Muons must be within the acceptance of the muon spectrometer, $|\eta|<2.7$.
All muons must have $\pT > \SI{5}{\GeV}$, while CT muons
must pass the stronger requirement $\pT > \SI{15}{\GeV}$.
Lastly, all muon candidates that are associated with a vertex,
i.e.\ all except SA muons, must have a longitudinal impact parameter
with respect to the reconstructed primary vertex satisfying
$|z_0 \sin \theta| < \SI{0.5}{\mm}$ and a transverse
impact parameter with respect to the position of the beam
satisfying $d_0 < \SI{1}{\mm}$.
 
All possible quadruplets (\cref{sec:KinematicVariables}) are formed from
the selected leptons.  A quadruplet
may contain no more than one SA or CT muon,
and at least one lepton in the quadruplet
must correspond to a lepton found by one of the triggers satisfied
by the event.
The three highest-$\pt$ leptons must satisfy, respectively,
${\pt} > \SI{20}{\GeV}$, ${\pt} > \SI{15}{\GeV}$, and ${\pt} > \SI{10}{\GeV}$.
Except for the LM analysis, for which the angular separation
between leptons can be very small,
all pairs of same-flavour leptons in the quadruplet
must satisfy $\Delta R(\ell,\ell') > 0.1$, while different-flavour
pairs must satisfy $\Delta R(\ell,\ell') > 0.2$.
At least one quadruplet per event is required.
For the HM and LM analyses, if there is more than one quadruplet
passing these requirements, the one with the smallest mass difference
between the two pairs, $\Delta m_{\ell\ell} = |m_{12}-m_{34}|$,
is chosen.
The analogous procedure for the ZX analysis is described in \cref{sec:ZX_Selection}.
 
The leptons in the quadruplet must be isolated from other deposits in the
calorimeter or ID tracks~\cite{PERF-2017-01,MUON-2018-03-pub}.
This rejects backgrounds in which leptons arise from the decay
of heavy-flavour jets, or in which hadronic jets are misidentified
as leptons.
For each lepton, the sum of the transverse energies
of topological clusters~\cite{ATL-LARG-PUB-2008-002}
within a cone of $\Delta R = 0.2$ around
it (excluding energy attributed to the lepton itself)
must be less than 20\% of its \pt for electrons,
and less than 30\% of its \pt for muons.
The transverse momenta of tracks in a cone around the lepton are also
summed, and must be less than 15\% of its \pt.  The $\eta$--$\phi$ radius of the
cone depends on the momentum of the lepton.  For electrons,
the radius is $\Delta R = \min(0.2, \SI{10}{\GeV}/\pt$), while
for muons, it is $\Delta R = \min(0.3, \SI{10}{\GeV}/\pt$).
In both cases, tracks and energy clusters attributed
to other leptons in the quadruplet are also excluded from the sums.
This is particularly important for the LM analysis (\cref{sec:LM_Selection}),
where the angular separation between leptons may be very small.
 
In addition to the impact parameter requirements discussed earlier, each lepton
in the quadruplet must have transverse impact parameter significance
$d_0 / \sigma_{d_0} < 5$ for electrons and  $d_0 / \sigma_{d_0} < 3$ for muons
(with the exception of SA muons, which do not have an ID track),
where $\sigma_{d_0}$ is the estimated error in the reconstructed
transverse impact parameter $d_0$.

\subsection{HM event selection}
\label{sec:HM_Selection}
 
The high-mass analysis applies a set of kinematic requirements
to select events consistent
with $H\to XX\to 4\ell$ decays.  The invariant mass
of the four-lepton system must be consistent with the SM Higgs~boson:
$\SI{115}{\GeV} < m_{4\ell} < \SI{130}{\GeV}$.  Also, the lepton
pairs must not be consistent with the decays of $Z$~bosons
($Z$-veto): $\SI{10}{\GeV} < m_{12,34} < \SI{64}{\GeV}$.
For the $4e$ and $4\mu$ channels, it is possible that the leptons from
a single $X$ or $Z$ decay are not paired together, but rather
a lepton from one $Z/X$ decay may be paired with a lepton from
the other $Z/X$ decay.    Therefore, there are also requirements
on the alternative lepton pairings, $\SI{5}{\GeV} < m_{14,23} < \SI{75}{\GeV}$,
in order to suppress $ZZ^*$ background events in which the leptons are mispaired.
Events with pairs consistent
with $J/\psi$ or $\Upsilon$ decay are also rejected with requirements
on the lepton pair masses
(see \cref{tab:eventpreselectionSummary}).
 
The final requirement enforces consistency between $m_{12}$
and $m_{34}$: $m_{34}/m_{12} > 0.85 - 0.1125 f(m_{12}) $,
where the function $f(m_{12})$ is defined in the Appendix.
Together with the relation $|m_{12} -m_Z| < |m_{34} - m_Z|$,
this defines a wedge-shaped region in the $m_{12}$--$m_{34}$ plane,
as shown in \cref{fig:resultsDataScatterHighMass}.

\subsection{LM event selection}
\label{sec:LM_Selection}
 
The LM analysis is designed to be sensitive to the mass
range $\SI{1}{\GeV} < m_{X} < \SI{15}{\GeV}$.
For these low masses, the angular separation between the two leptons in the
$X\ra\ell\ell$ decay can become very small
($\Delta R(\ell,\ell)<0.1$ for $m_X=\SI{1}{\GeV}$).
In this case, the efficiency to select isolated
electrons is significantly smaller than that for muons, so this analysis
uses only the $4\mu$ final state.  Otherwise, the event selection
is very similar to that of the HM analysis
(\cref{sec:HM_Selection}), except that a few kinematic criteria differ.
The $\Delta R$ requirements between final-state leptons are removed, and the
$Z$-veto requirement is not relevant.
In addition to the HM heavy-flavour veto, the two lepton pair
masses $m_{12}$ and $m_{34}$ must not be in the ranges
$2$--$\SI{4.4}{\GeV}$ or
$8$--$\SI{12}{\GeV}$.
The $m_{4\ell}$ requirement is narrowed to
$\SI{120}{\GeV} < m_{4\ell} < \SI{130}{\GeV}$, because muons have smaller
radiative losses than electrons, and both lepton pairs
must satisfy
$\SI{1.2}{\GeV} < m_{12,34} < \SI{20}{\GeV}$.
Also, the final requirement for the signal region is simplified
to $m_{34}/m_{12} > 0.85 $.

\subsection{ZX event selection}
\label{sec:ZX_Selection}
 
The selection for the ZX analysis differs from those of the
HM and LM analyses as it is selecting
a $Z$~boson along with a new $X$~boson.  It is, however, very similar
to the selection used for the ATLAS SM $H\to ZZ^*\to 4\ell$
analysis~\cite{HIGG-2016-22}.
In addition to the common criteria described in \cref{sec:EventSelection},
each quadruplet must satisfy
$\SI{50}{\GeV} < m_{12} < \SI{106}{\GeV}$,
$\SI{12}{\GeV} < m_{34} < \SI{115}{\GeV}$, and,
for the $4e$ and $4\mu$ channels, the alternative pairings must satisfy
$m_{14,23} > \SI{5}{\GeV}$.  The latter requirement suffices to remove
mispaired $J/\psi$ events.  Backgrounds from $\Upsilon$ decays
were found to be negligible after all selections.
If there is more than one such quadruplet,
quadruplets are ranked by the following criteria, applied in sequence:
\begin{itemize}
\item Rank by the flavours of the two lepton pairs according to the
reconstruction efficiencies of the leptons.
The reconstruction efficiency for muons is higher
than that for electrons, especially at lower
lepton momenta.
Therefore, the final state lepton pairs in order of decreasing
reconstruction efficiency are $4\mu$, $2e2\mu$, $2\mu2e$, and $4e$.
\item Choose the quadruplet with the smallest $|m_Z - m_{12}|$.
\item Choose the quadruplet with the smallest $|m_Z - m_{34}|$.
\end{itemize}
This is strictly applied to all quadruplets, even same-flavour ones,
and thus  differs slightly from the prescription used in the analysis
of Ref.~\cite{HIGG-2016-22}, where
the lower-ranked alternative pairing of a same-flavour
quadruplet is prevented from being considered in the quadruplet selection.
For this analysis, the alternative pairing is treated as a separate quadruplet
and participates in the ranking and quadruplet selection.
 
Following the selection of the quadruplet, the tracks associated
with all four leptons are required to be consistent with originating
from a common vertex: $\chi^2/N_\textrm{dof} < 9$ (tightened
to $<6$ for the $4\mu$ channel), where these upper bounds were chosen
to give an efficiency of 99.5\%.  This removes additional reducible backgrounds,
mainly $Z+\textrm{jets}$ and $\ttbar$.
(These backgrounds are already very small
for the HM and LM analyses, so this requirement is not applied in those cases.)
Finally, the total invariant
mass is required to be consistent with the decay of a Higgs~boson, in the same
manner as in the HM analysis: $\SI{115}{\GeV} < m_{4\ell} < \SI{130}{\GeV}$.
None of the other requirements of the HM analysis ($Z$~boson/heavy-flavour veto
and signal region requirements) are applied here.

 
\section{Systematic uncertainties}
\label{sec:systematics}
 
Many systematic uncertainties are common to all the analyses
considered here.  The dominant ones include:
\begin{itemize}
\item \textbf{Luminosity and pile-up}: The uncertainty in the integrated
luminosity is 1.7\%~\cite{ATLAS-CONF-2019-021},
obtained using the LUCID-2 detector~\cite{LUCID2}
for the primary luminosity measurements.  Uncertainty due to pile-up
arises from differences between the predicted and measured
inelastic cross sections, as well as from the reweighting procedure
described in \cref{sec:DataAndMCSamples}.
This uncertainty is approximately~1\%.
 
\item \textbf{Lepton-related uncertainties}: The efficiency for events
to pass the selection depends on the reconstruction and identification
efficiencies for leptons, as well as the determination of their momentum
scale.
Tag-and-probe techniques are applied to
the dilepton resonances $Z\to \ell^+\ell^-$,
$J/\psi\to \ell^+\ell^-$, and $\Upsilon\to \mu^+\mu^-$
in order to measure the efficiencies and momentum scales
and resolutions for electrons and muons.
This leads to corrections, usually of the order of up to a percent,
to account for differences observed
between data and simulation, as well as an estimate of the
residual uncertainty~\cite{EGAM-2018-01, MUON-2018-03-pub}.
As there are four leptons in the final state, small single-lepton
uncertainties can result in larger uncertainties in the final
yields, which range up to 15\%, dominated by the uncertainty
in the electron reconstruction and identification efficiency.
 
\item \textbf{Theoretical uncertainties}: Uncertainties in the
modelling of the simulated signal and
background processes are estimated by varying the
parton distribution functions, the factorization,
renormalization, and QCD scales, and the modelling of hadronization
and the underlying event.  The total uncertainty in the acceptance
of the signal is around 3\%,
and the uncertainty in the background yield is
3\%--9\% for the $H\to ZZ^*\to 4\ell$ process~\cite{LHCHiggsCrossSectionWorkingGroup:2011ti}
and about 5\% for $ZZ^*\to 4\ell$~\cite{Caola:2015psa,ATL-PHYS-PUB-2017-005,Gleisberg:2008fv, Cascioli:2011va, Schumann:2007mg, Hoeche:2012yf}.
 
\end{itemize}
 
Uncertainties related to data-driven background estimates are discussed
in the analysis-specific sections below.
 
Each source of systematic uncertainty is considered to be uncorrelated
with others: in the statistical description of the data, each source
of systematic uncertainty is parameterized by several nuisance parameters
that are constrained by Gaussian probability density distributions.
The luminosity and lepton-related uncertainties are completely correlated
among all Monte Carlo samples.
 
\section{HM analysis: $H\to XX\to 4\ell$ ($\SI{15}{\GeV} < m_X < \SI{60}{\GeV}$)}
\label{sec:HM_analysis}
 
The high-mass analysis searches for decays of a SM Higgs~boson
into a pair of new bosons $X$, where $X$ could be $\Zd$, $a$, or $s$,
which in turn decay into pairs of electrons
or muons (see \cref{fig:zd-diagrams}).
The event selection (detailed in \cref{sec:HM_Selection})
seeks two same-flavour opposite-sign pairs of leptons
of similar invariant mass that are consistent with the decay of
a SM Higgs~boson and inconsistent with the subsequent decay of $Z$~bosons.

\subsection{Background estimate}
 
Backgrounds with four prompt leptons are estimated from
simulation (see \cref{sec:DataAndMCSamples}) and validated using
data from background-dominated control samples.  The dominant backgrounds are
$H\to ZZ^*\to 4\ell$ (about 72\% of the total background)
and $ZZ^*\to 4\ell$ (about 24\% of the total background).
Other such processes include $\ttbar Z\to 4\ell$ and processes
with three gauge bosons.  These are found to be negligible.
 
Reducible backgrounds include those from processes with leptons originating
from the decay of heavy-flavour jets, or with jets misidentified
as leptons.
The background from the $Z+\textrm{jets}$ process is estimated using data.
Control regions enriched in misidentified leptons are defined by
selecting quadruplets with one or two of its subleading leptons satisfying
`inverted' criteria but otherwise passing the signal region selection.
`Inverted' electrons fail either the
Loose electron selection or the isolation requirement, but not both.
`Inverted' muons fail either the isolation requirement or
the transverse impact parameter significance requirement
($d_0 / \sigma_{d_0} < 3$), or both.
Two samples are defined, both requiring
two leptons consistent with the decay of a $Z$~boson.  The `good' sample
requires at least one extra lepton passing the nominal selection, while
the `inverted' sample requires at least one extra lepton passing
the `inverted' selection.   Since both samples are highly enriched
in $Z\to\ell\ell$ decays, the extra leptons originate mostly
from jets misidentified as leptons.  Transfer factors are
defined as the ratio of the number of extra leptons passing the `good' selection
to the number passing the `inverted' selection.
These transfer factors are applied to events in the `inverted' control
regions in order to extrapolate to the signal region.
The systematic uncertainties in this procedure are estimated by propagating
the statistical uncertainties in the transfer factors as well as comparing
the results from several different definitions of `good' and `inverted' leptons.
This yields an estimate of the background due to the $Z+\textrm{jets}$ process
in the signal region compatible with zero.
 
Other reducible backgrounds are estimated from simulation.  The dominant
contribution is from $\ttbar$, with about 3\% of the total background.
Other such backgrounds, including those from diboson production
and heavy-flavour processes, are found to be negligible.

\subsection{Background validation}
 
The background estimates are validated using four dedicated
background-enriched validation regions, defined so that they
do not overlap with the HM signal region:
 
\begin{itemize}
\item VR1:  The $Z$-veto requirement on the alternative pairings
is inverted, requiring $m_{14,23} \ge \SI{75}{\GeV}$,
and the compatibility requirement on $m_{34}/m_{12}$ is removed.
This produces a sample enriched in the $H\to ZZ^*\to 4\ell$ process
as well as the non-resonant  $ZZ^*\to 4\ell$ process.
Only the $4e$ and $4\mu$ final states contribute to this region.
 
\item VR2: The requirements on the four invariant mass pairings are removed
and replaced with $m_{12} \ge \SI{64}{\GeV}$, and the compatibility
requirement  on $m_{34}/m_{12}$ is removed.
This sample is also enriched in both the $H\to ZZ^*\to 4\ell$
and $ZZ^*\to 4\ell$ processes.
All final states contribute to this region.
 
\item VR3:  Both the Higgs~boson mass window requirement
($\SI{115}{\GeV} < m_{4\ell} < \SI{130}{\GeV}$) and the final
$m_{34}/m_{12}$ compatibility requirement are inverted,
producing a sample dominated by $ZZ^*\to 4\ell$.
 
\item VR4: The final $m_{34}/m_{12}$ compatibility requirement is inverted,
and all four dilepton mass requirements are changed
to $m_{\ell\ell} < \SI{55}{\GeV}$.  This sample mainly
consists of $H\to ZZ^*\to 4\ell$, but has a significant contribution
from $ZZ^*\to 4\ell$.
 
\end{itemize}
 
Although these validation regions are constructed so that they do not
overlap with the signal region for the HM~analysis, there is some overlap
of VR1 and VR2 with the signal region of the ZX~analysis.
However, given the cross-section limits found for the
$H\ra Z\Zd\ra 4\ell$ process (\cref{fig:totsig_zzd}),
the contribution of ZX~signal events to either of these regions is
less than 5\% of the SM~background expectation.

\begin{figure}[!htbp]
\begin{center}
\subfloat[]{\includegraphics[width=0.47\textwidth]{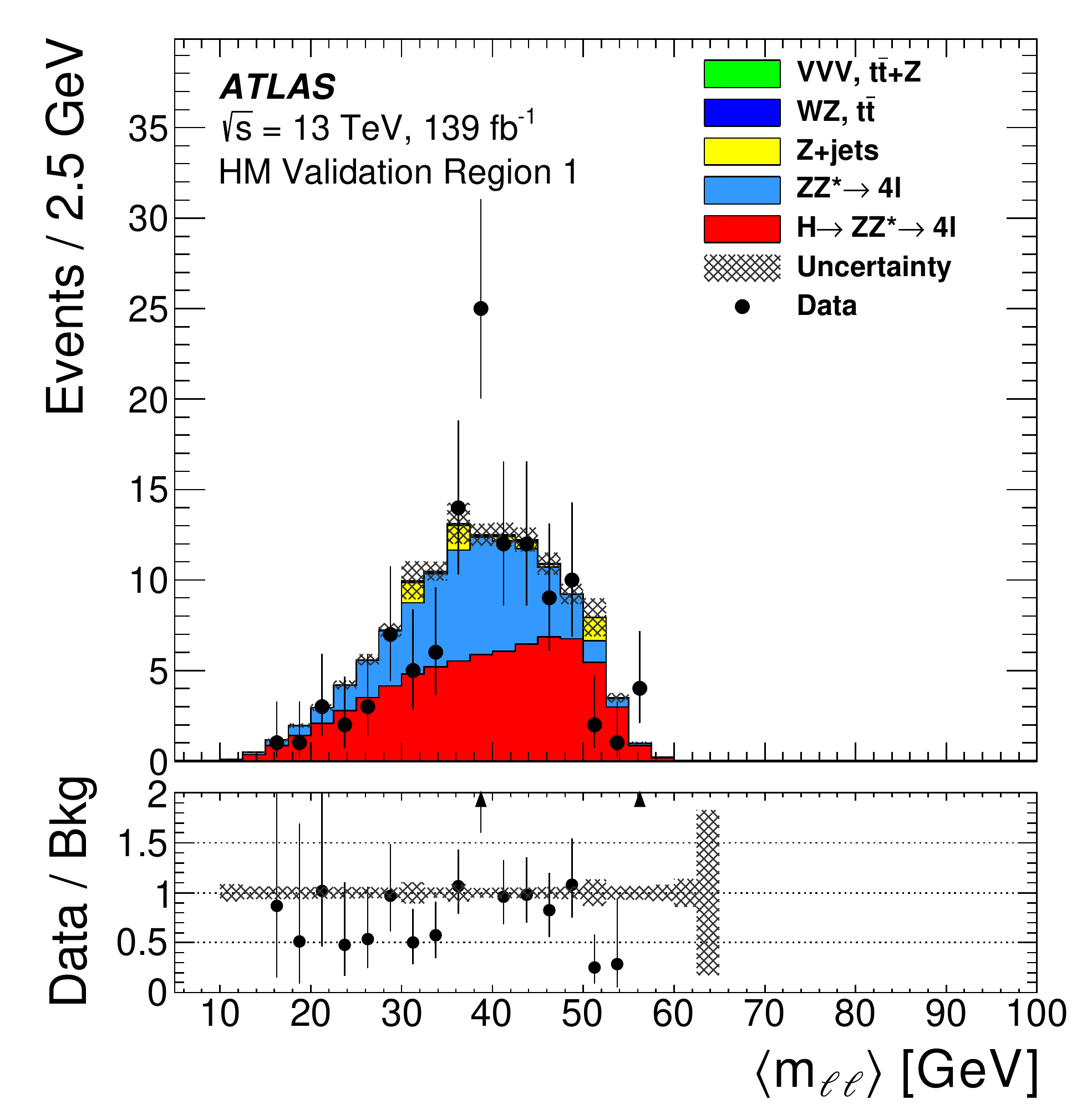}}
\subfloat[]{\includegraphics[width=0.47\textwidth]{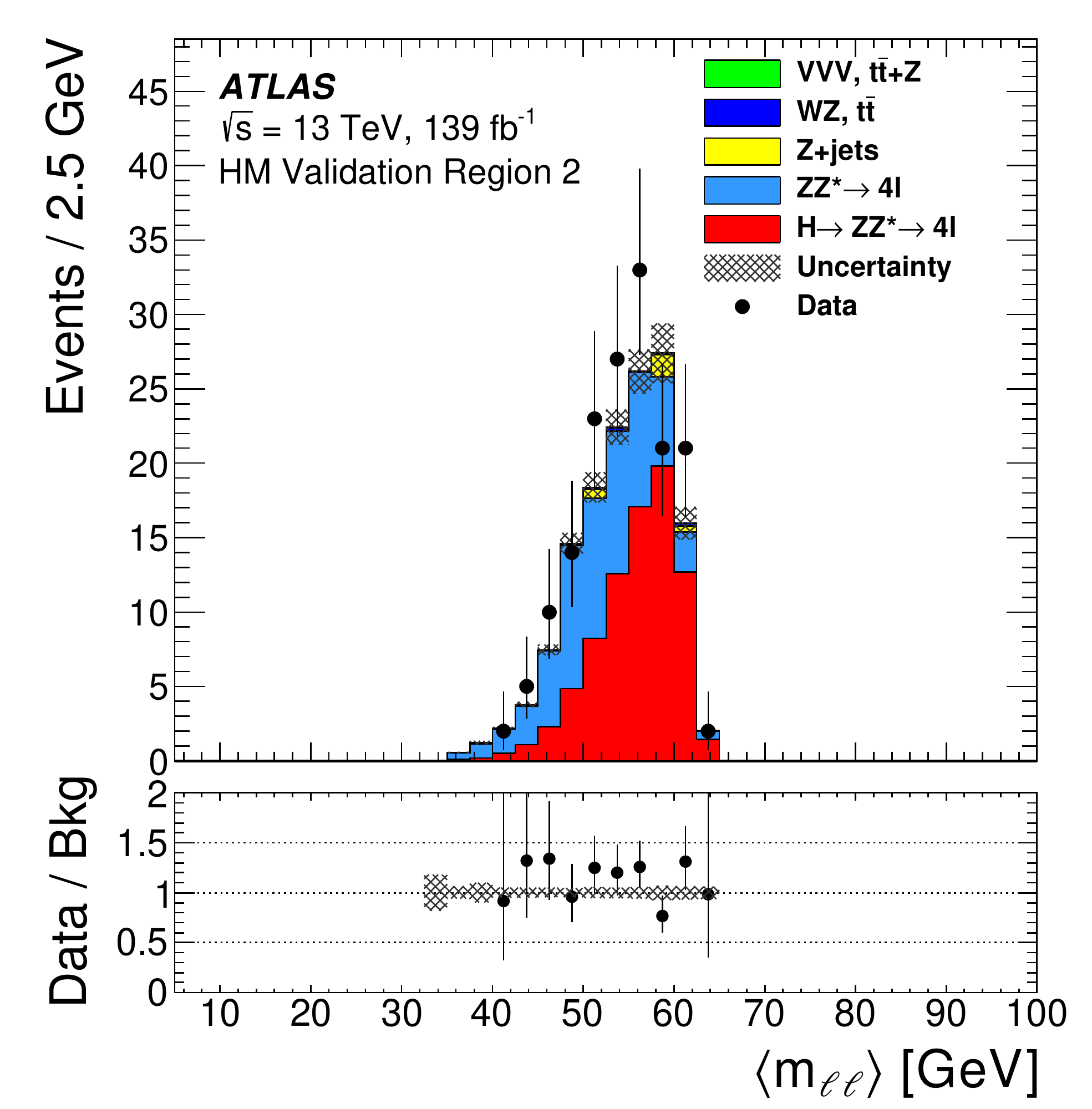}}\\
\subfloat[]{\includegraphics[width=0.47\textwidth]{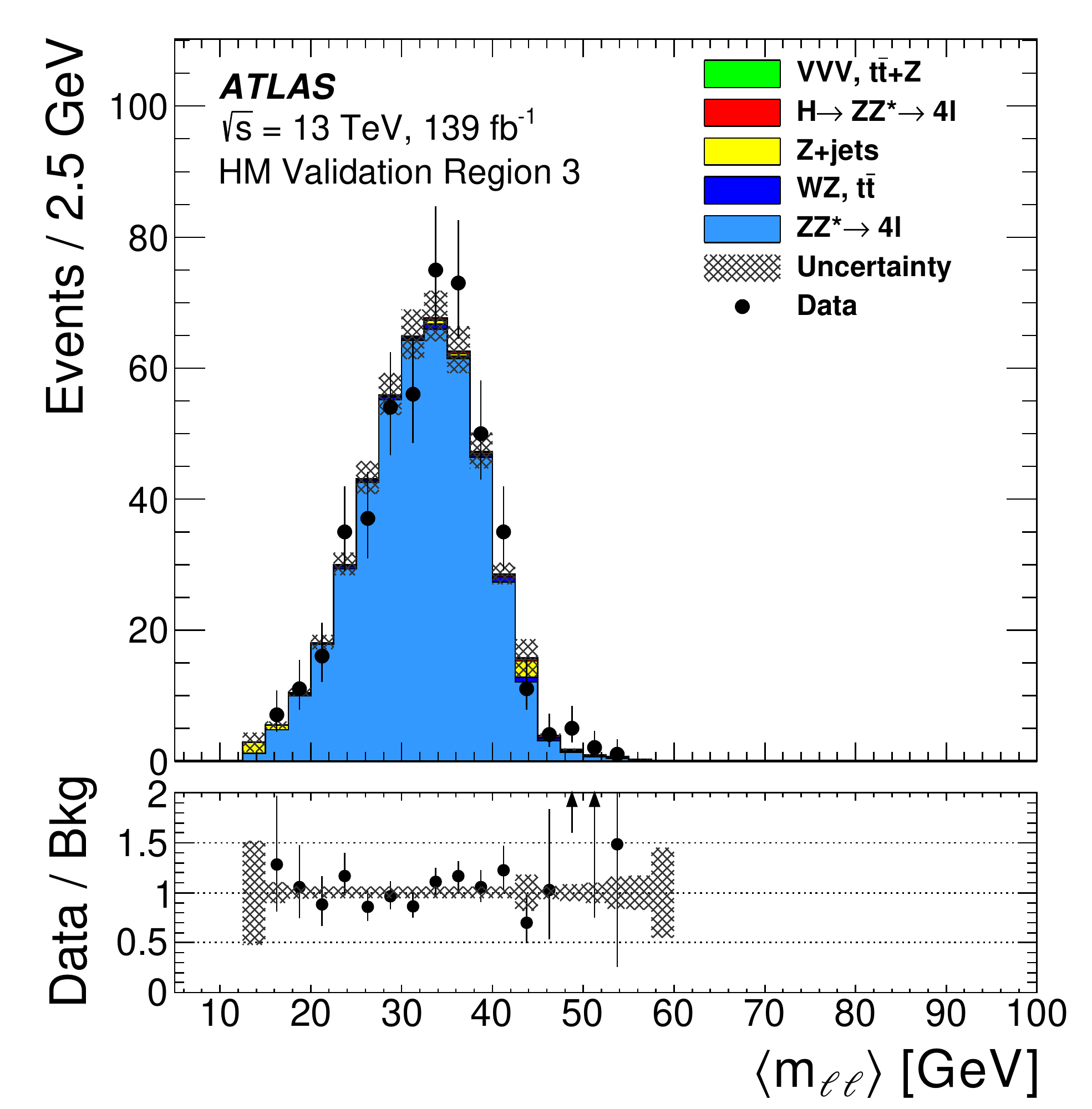}}
\subfloat[]{\includegraphics[width=0.47\textwidth]{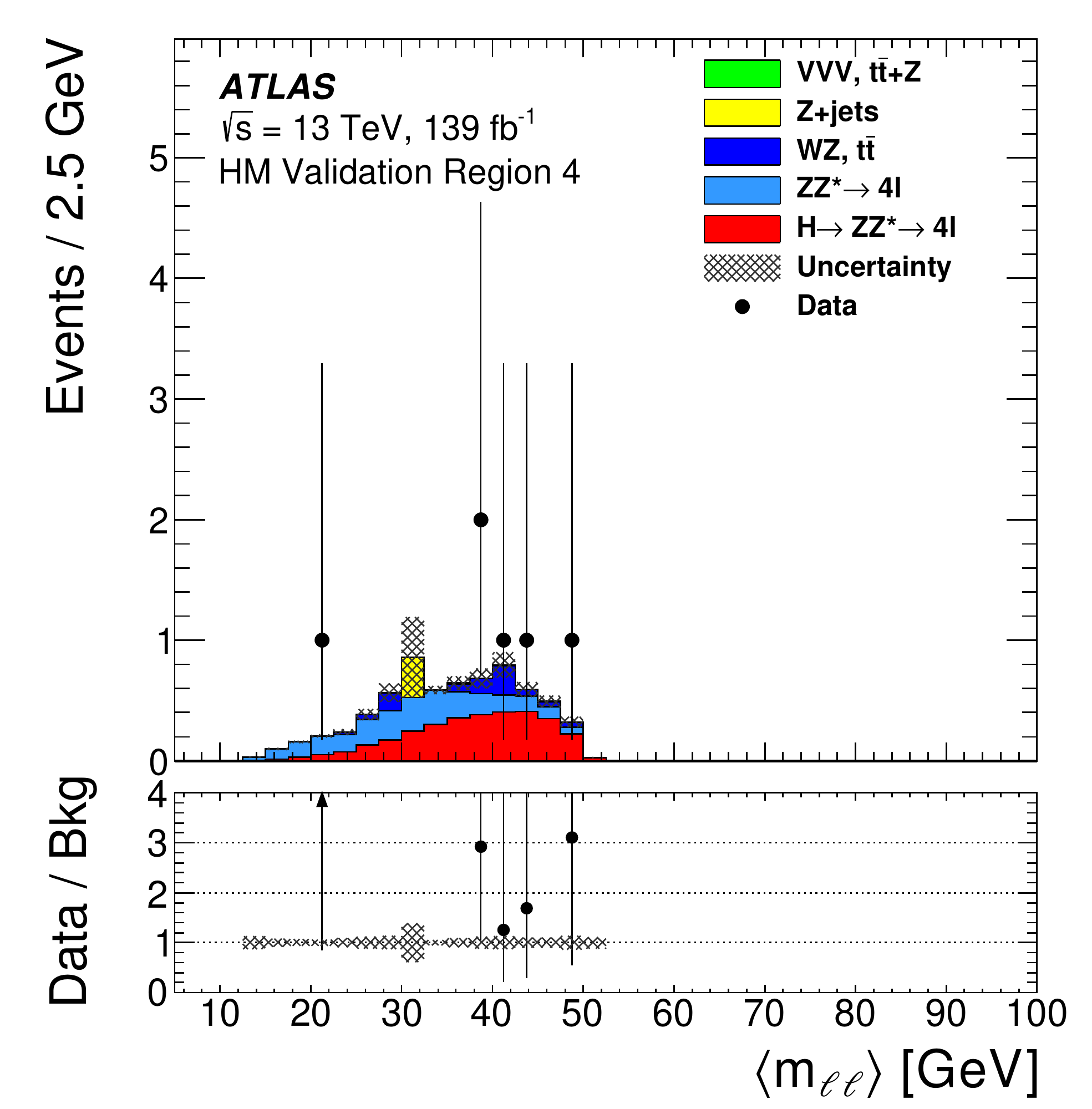}}\\
\caption{Distributions of $\langle m_{\ell\ell}\rangle= \half(m_{12}+m_{34})$ in
the validation regions for the HM $H\to XX\to 4\ell$ ($\SI{15}{\GeV} < m_X < \SI{60}{\GeV}$)
analysis: for (a)~VR1, (b)~VR2, (c)~VR3, and (d)~VR4
(see text for definitions). The signal contribution to these regions is negligible.
The shaded band represents the total uncertainty of the prediction.
The lower panels show the ratio of the observed data to the
(pre-fit) MC~predictions; the arrows at the upper edge indicate data points that fall outside of the  $y$-axis range.
The uncertainties of the plotted data are asymmetric and are calculated
using
Eqs.~(40.76) \ifnotpo{of Ref.~\cite{pdg20}}\ifpo{from the 2020 Review of Particle Properties}.
}
\label{fig:HighMassVRs}
\end{center}
\end{figure}

\cref{fig:HighMassVRs} compares the predicted backgrounds in these
regions with the data for the variable $\langle m_{\ell\ell}\rangle= \frac{1}{2}(m_{12}+m_{34})$.
Good agreement is found in all cases.
In these validation regions, the $Z+\textrm{jets}$ background is estimated
from MC~simulations, while for the signal region it is estimated from data.

\subsection{Results}
 
\begin{figure}[!htbp]
\begin{center}
\subfloat[]{\includegraphics[width=0.43\textwidth]{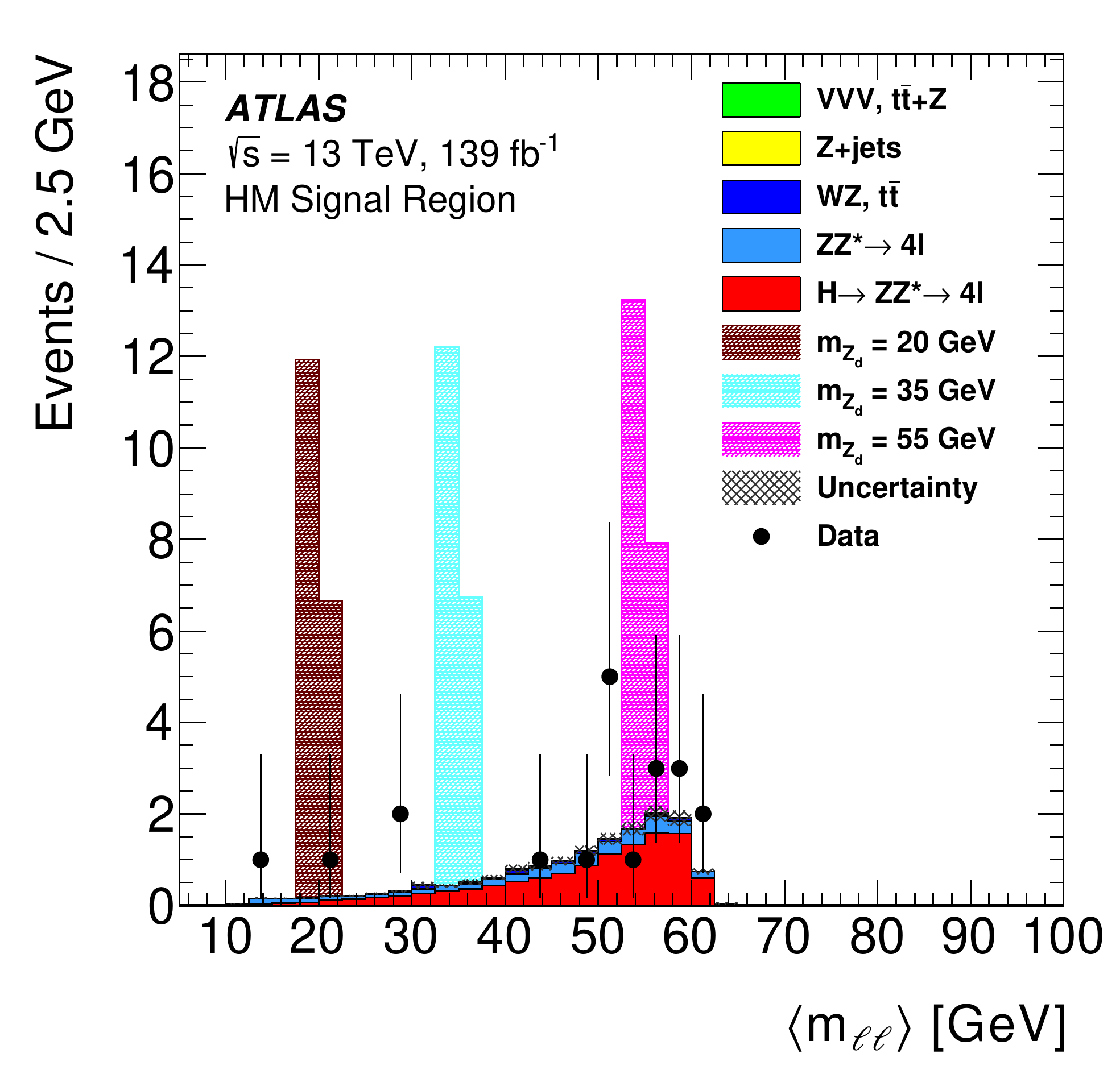}
\label{fig:resultsDataAvgMHighMass}}
\subfloat[]{\includegraphics[width=0.57\textwidth]{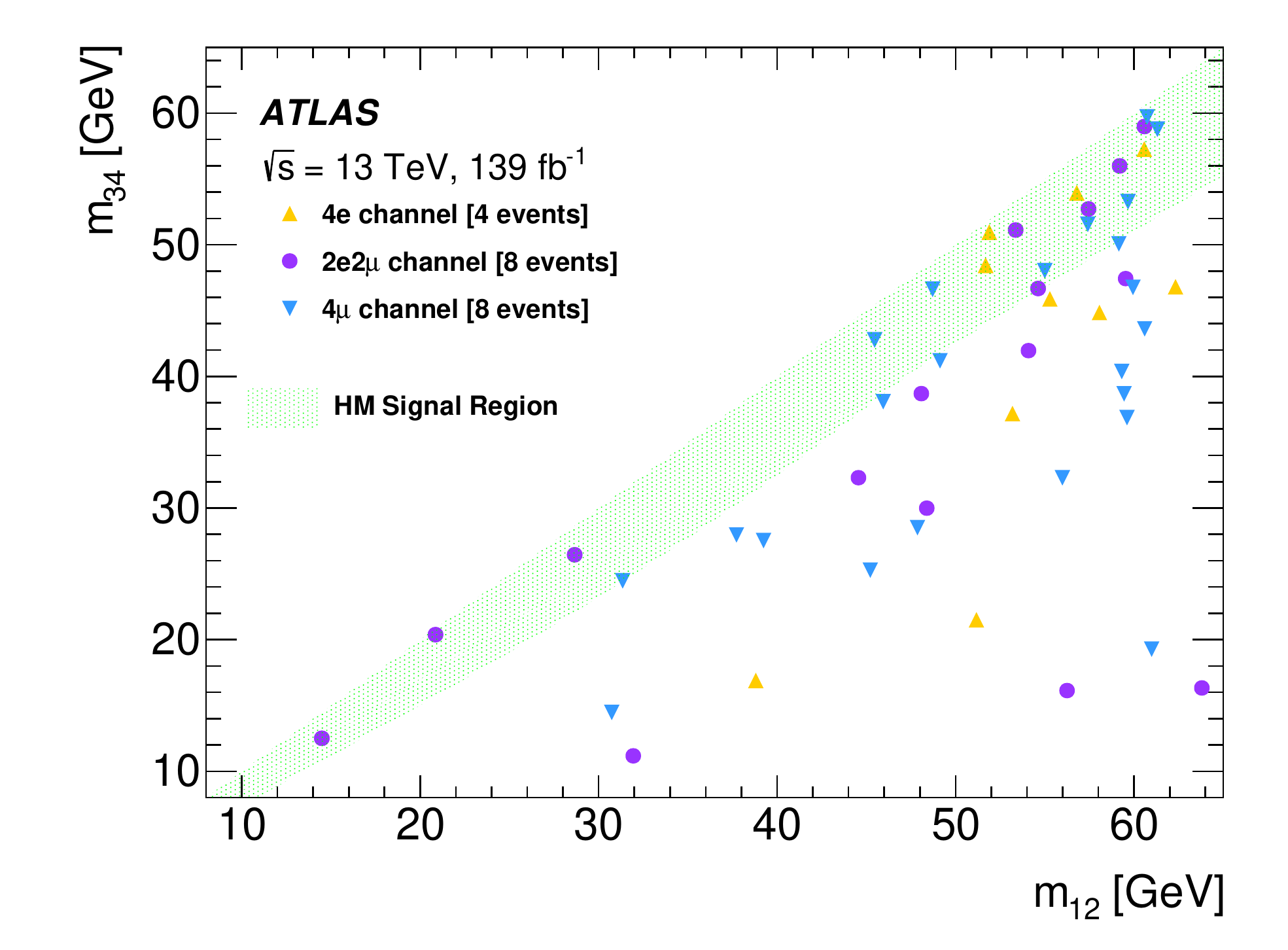}
\label{fig:resultsDataScatterHighMass}}
\caption{Distribution of (a)~$\langle m_{\ell\ell}\rangle$
and (b)~$m_{34}$ vs~$m_{12}$,
for events selected in the
HM
$H\to XX \to 4\ell$ $(\SI{15}{\GeV} < m_{X} < \SI{60}{\GeV})$
analysis.  In the
$\langle m_{\ell\ell}\rangle$
distribution (a), the (pre-fit) background expectations are also
shown; the hatched band contains the statistical
and systematic uncertainties.
The expectations for the signal are also shown,
for several masses.  The signal histograms are stacked on top
of the background histograms, and expected yields are normalized
with
$\sigma(pp\to H\to \Zd\Zd\to 4\ell) =
\tenth\sigma_{\textrm{SM}}(pp\to H\to ZZ^{*}\to 4\ell) = \SI{0.60}{\fb}$ (ggF process only).
The uncertainties of the plotted data are asymmetric and are calculated
using
Eqs.~(40.76) \ifnotpo{of Ref.~\cite{pdg20}}\ifpo{from the 2020 Review of Particle Properties}.
For the
$m_{34}$ vs~$m_{12}$
distribution (b), each marker corresponds to an event that passed
the Higgs~boson window requirement and $Z$~boson veto.
The markers (differentiated by channel) that fall inside
the green shaded area correspond to the events of the signal
region.}
\end{center}
\end{figure}

\begin{figure}[!htbp]
\begin{center}{\includegraphics[width=0.65\textwidth]{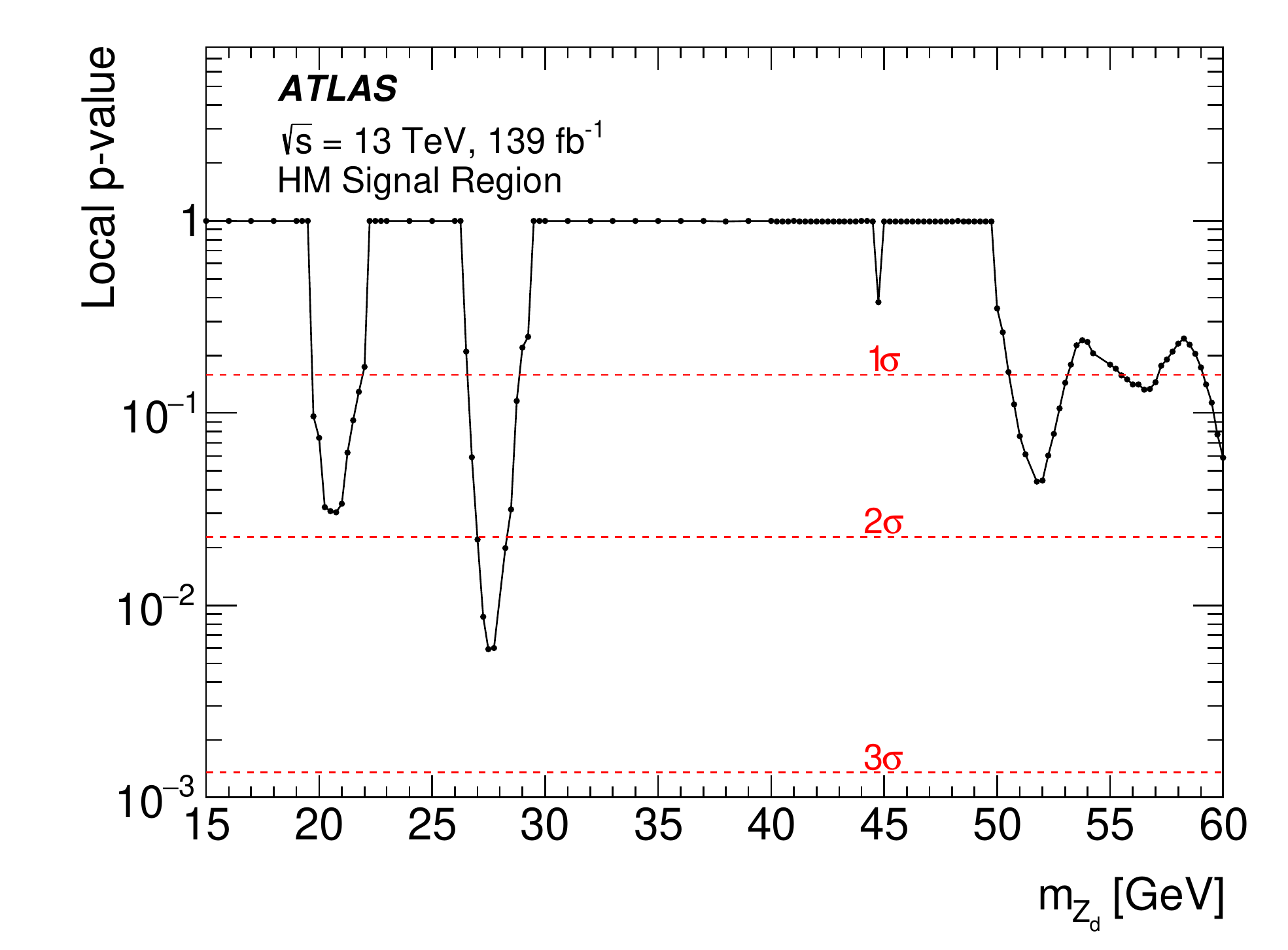}}
 
\caption{Observed local $p$-values under the background-only hypothesis for
the process $H\to XX\to 4\ell$ in the high-mass range. For the limit
determination, the distributions of $\langle m_{\ell\ell}\rangle$  in the
signal region are binned with a width of $\SI{1}{\GeV}$. The  $p$-values
are plotted in steps of $\SI{0.25}{\GeV}$ in the vicinity of observed
data and $\SI{1}{\GeV}$ elsewhere. The most significant excess
corresponds to a local significance
of~$2.5\sigma$ at $m_{\Zd} = \SI{28}{\GeV}$.}
\label{fig:ZdZdHighMassAnalysis_p0sig}
\end{center}
\end{figure}

The resulting $\langle m_{\ell\ell}\rangle$ distribution for this analysis
is shown in \cref{fig:resultsDataAvgMHighMass},
while \cref{tab:ZdZdHighMassAnalysis_yields} summarizes the final yields
and uncertainties in the signal region as defined in \cref{tab:eventpreselectionSummary}.
A total of 20 events are observed,
with a total predicted background of $15.6\pm1.3$ events.
The $p$-values for the background-only hypothesis as a function of $m_X$
are shown in \cref{fig:ZdZdHighMassAnalysis_p0sig}.
The profile-likelihood ratio
($-2\log[L(\mu=0,\hat{\hat{\theta}})/L(\hat{\mu},\hat{\theta})]$)
is used as the test statistic, and the likelihood used is described
in \cref{sec:Limits}.
Different final states are not distinguished in the fit; distributions
used are summed over all channels.
The largest deviation from SM expectations occurs around
$m_{\Zd}=\SI{28}{\GeV}$, corresponding to the two events with
$\langle m_{\ell\ell} \rangle \approx\SI{28}{\GeV}$, with a local significance of $2.5\sigma$.
Following procedures fixed before the data in the signal region were examined,
the one event with $\langle m_{\ell\ell}\rangle < \SI{15}{\GeV}$
and the two with $\langle m_{\ell\ell}\rangle > \SI{60}{\GeV}$
are not considered when setting limits and
do not affect \cref{fig:ZdZdHighMassAnalysis_p0sig}.
The distribution of $m_{34}$ versus $m_{12}$ for the selected events
is shown in \cref{fig:resultsDataScatterHighMass}.

\begin{table}[htbp!]
\caption{
Expected event yields of the SM background processes and data yield
for the HM
$H\to XX \to 4\ell$ ($\SI{15}{\GeV} < m_{X} < \SI{60}{\GeV}$)
selection.
Three of the 20 observed events are outside the
range
$\SI{15}{\GeV} < \langle m_{\ell\ell}\rangle < \SI{60}{\GeV}$
and are thus not considered when setting limits.
The systematic uncertainties of the background estimates
are highly correlated between the different sources of background
(see \ifnotpo{\cref{sec:systematics}}\ifpo{text}).
}
\begin{center}
\begin{small}
\begin{tabular*}{0.4\textwidth}{@{\extracolsep{\fill}}ll}
\toprule
Process & Yield ($\pm \mathrm{stat.}\pm\mathrm{syst.}$)\bigstrut \\
\hline
$H\to ZZ^*\to 4\ell$\bigstrut[t]                & $11.1~~\pm0.1~~\pm1.0$   \\
$ZZ^*\to 4\ell$                                 & ~~$3.38\pm0.05\pm0.25$ \\
$\ttbar$                                        & ~~$0.47\pm0.13\pm0.09$\\
$Z+\textrm{jets}$                               & ~~$0.43\pm0.39^{~+~0.17}_{~-~0.01}$\\
$Z+\ttbar\to 4\ell$                             & ~~$0.09\pm0.02\pm0.02$\\
$WZ$                                            & ~~$0.05\pm0.03^{~+~0.05}_{~-~0.00}$\\
$VVV$/VBS                                         & \qquad Negligible \\
Heavy flavour\bigstrut[b]                       & \qquad Negligible \\
\hline
Total\bigstrut                                  & $15.6~~\pm0.4~~\pm1.2$   \\
\hline
Data\bigstrut                                   & 20             \\
\bottomrule
\end{tabular*}
\label{tab:ZdZdHighMassAnalysis_yields}
\end{small}
\end{center}
\end{table}

\FloatBarrier
 
\section{LM analysis: $H\to XX\to 4\mu$ ($\SI{1}{\GeV} < m_X < \SI{15}{\GeV}$)}
\label{sec:LM_analysis}

The LM analysis extends the HM analysis to the region
$\SI{1}{\GeV} < m_X < \SI{15}{\GeV}$, where $X = \Zd$, $a$, or $s$.
Only the $4\mu$ final state is considered
for this analysis.  The event selection is detailed
in \cref{sec:LM_Selection} and is similar to that of the HM analysis,
with some adjustments for the different kinematic region.

\subsection{Background estimate}
 
Backgrounds involving four prompt leptons are estimated directly
from MC simulations (see \cref{sec:DataAndMCSamples}).
The $H\to ZZ^*\to 4\mu$ and $ZZ^*\to 4\mu$ processes together
comprise about two thirds of the total background estimate.
Higher-order electroweak processes, including triboson production
and vector-boson scattering, are found to be negligible.

The remaining backgrounds involve non-prompt leptons,
primarily from decays of heavy-flavour hadrons in events
with multiple $b$-quarks such as \bbbar.  A leading part of this
contribution comes from double semileptonic decays,
where a $b$-hadron decays into a muon and a $c$-hadron, which further
decays into another muon and light hadrons.  Resonances produced
in the $b$-hadron decay chain (i.e.,~$\omega$, $\rho$, $\phi$, $J/\psi$)
are also an important background but
are almost completely suppressed by the heavy-flavour vetoes
on dilepton masses required as part of the LM event selection.
There is also a small contribution from $\bbbar\bbbar$, where each muon originates
from an independent $b$-quark.
As the muons selected here are all isolated, $b$-jet tagging is not useful
for reducing these backgrounds.
The backgrounds from these processes are estimated together using a
data-driven method~\cite{EXOT-2016-22,CMS-HIG-13-010}.
 
The first step is to find the shape of the background in the
$m_{12}$--$m_{34}$ plane.
The invariant mass distribution of each muon pair is modelled separately
to account for the different kinematic selections imposed
on the leading, subleading, and remaining muons.  Two distinct control
samples are used, each of which contains an opposite-sign muon pair
plus a third muon.
The first sample, used to model $m_{12}$, requires
a muon pair with ${\pt}_1 > \SI{20}{\GeV}$ and ${\pt}_2 > \SI{10}{\GeV}$
satisfying a dimuon trigger, and a third muon with ${\pt}_3 > \SI{5}{\GeV}$.
The second sample, used to model $m_{34}$, requires
a muon pair with ${\pt}_{1,2} > \SI{5}{\GeV}$
and a third muon with ${\pt}_3 > \SI{27}{\GeV}$, satisfying a single-muon
trigger.  In both cases, the muons must pass the
same isolation and quality requirements as for the signal region.
Ninety-seven percent of signal events pass both these selections, with the
$m_{12}$ and $m_{34}$ pairs passing the requirements of the muon pair
in the first and second samples, respectively.  The invariant masses
of the muon pairs are taken from the two control samples and used
to form a 2D template in the $m_{12}$--$m_{34}$ plane as the direct product
of the two distributions.
 
A correction to the $m_{12}$--$m_{34}$ template is made
to account for a correlation between the kinematics
of the two muon pairs, which is introduced by the Higgs~boson mass requirement.
Another control sample is defined by inverting the
isolation and vertex requirements on the muons in the signal event
selection, defining a sample enriched in events with muons from
heavy-flavour quark decays.  Comparing
the distributions of muon pair invariant masses before and after
the Higgs~boson mass requirement yields the correction to the background
shape as a function of $m_{12}$ and $m_{34}$.
 
\begin{figure}[!htbp]
\begin{center}
\subfloat[]{\includegraphics[width=0.49\textwidth]{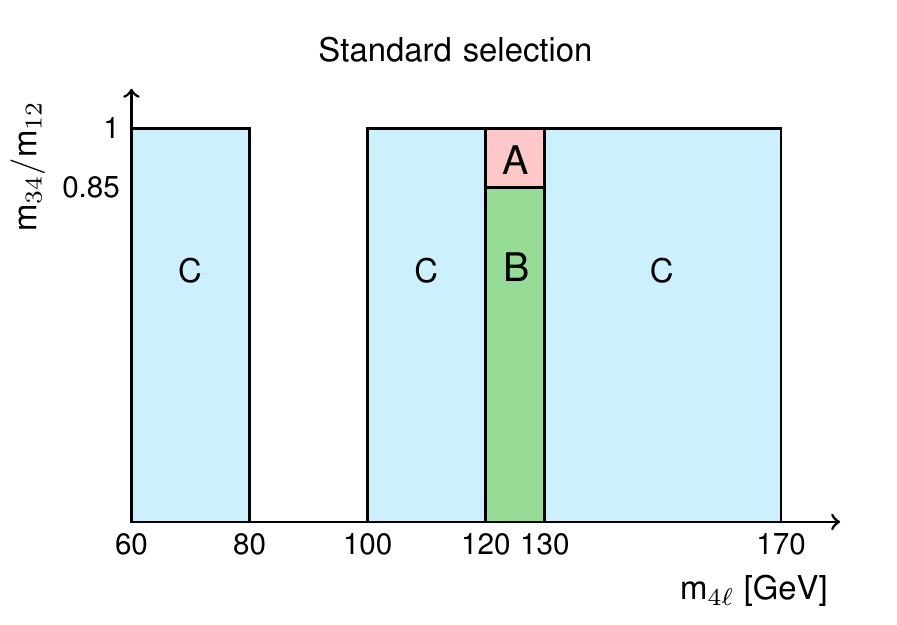}}
\subfloat[]{\includegraphics[width=0.49\textwidth]{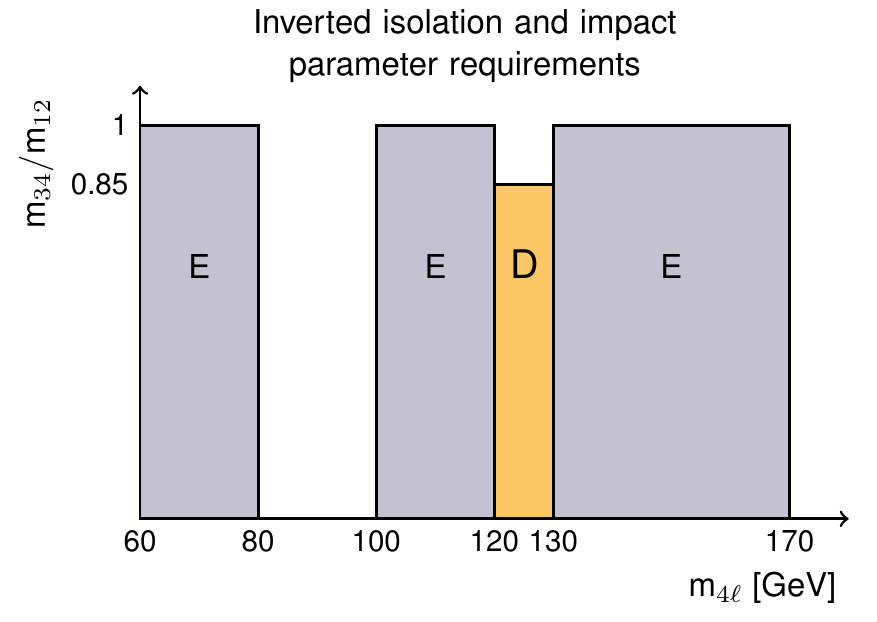}}
\caption{Definition of regions used in the normalization of the heavy-flavour
background in the LM analysis.
(a)~Region~A is the signal region.  The $m_{34}/m_{12} > 0.85$ compatibility
requirement is inverted in region~B, and the Higgs~boson mass requirement
is inverted in region~C. The isolation and impact parameter requirements are inverted in
regions~D and~E shown in~(b).}
\label{fig:lm-bg-regions}
\end{center}
\end{figure}

Finally, the overall normalization for the background from non-prompt
leptons is determined from data in regions defined by inverting
several selection criteria. As shown in \cref{fig:lm-bg-regions},
region~B is defined by inverting the compatibility requirement
$m_{34}/m_{12} > 0.85$. In order to improve the statistical precision
of the background prediction, additional regions are defined by
inverting the Higgs~boson mass requirement (region~C in
\cref{fig:lm-bg-regions}) and also the isolation and vertex
requirements (regions~D and E in \cref{fig:lm-bg-regions}). The
regions with $\SI{81}{\GeV} < m_{4\ell} < \SI{101}{\GeV}$ are
excluded in order to reduce contributions from $Z$~bosons. The
contribution with prompt muons, mostly $ZZ^*$ in regions~B and C, is
subtracted from the data. The background with non-prompt leptons in
region~B is then estimated using $B = C\cdot D/E$.
The 2D template is then used to scale from the estimate
in region~B to  the signal region satisfying the
compatibility requirement, region~A.

The uncertainty in the heavy-flavour background estimate is found by varying
each parameter of the background shape model up and down
by $2\sigma$ and taking the largest change in yield for each bin,
giving an uncertainty of 38\%.
Statistical uncertainties in the normalization of the signal region
are also propagated to the heavy-flavour background yield,
giving an uncertainty of 33\%.
Adding these uncertainties in quadrature gives a total systematic uncertainty
of 50\% in the heavy-flavour background yield.

\subsection{Results}
 
\begin{figure}[!htbp]
\begin{center}
\subfloat[]{\includegraphics[width=0.41\textwidth]{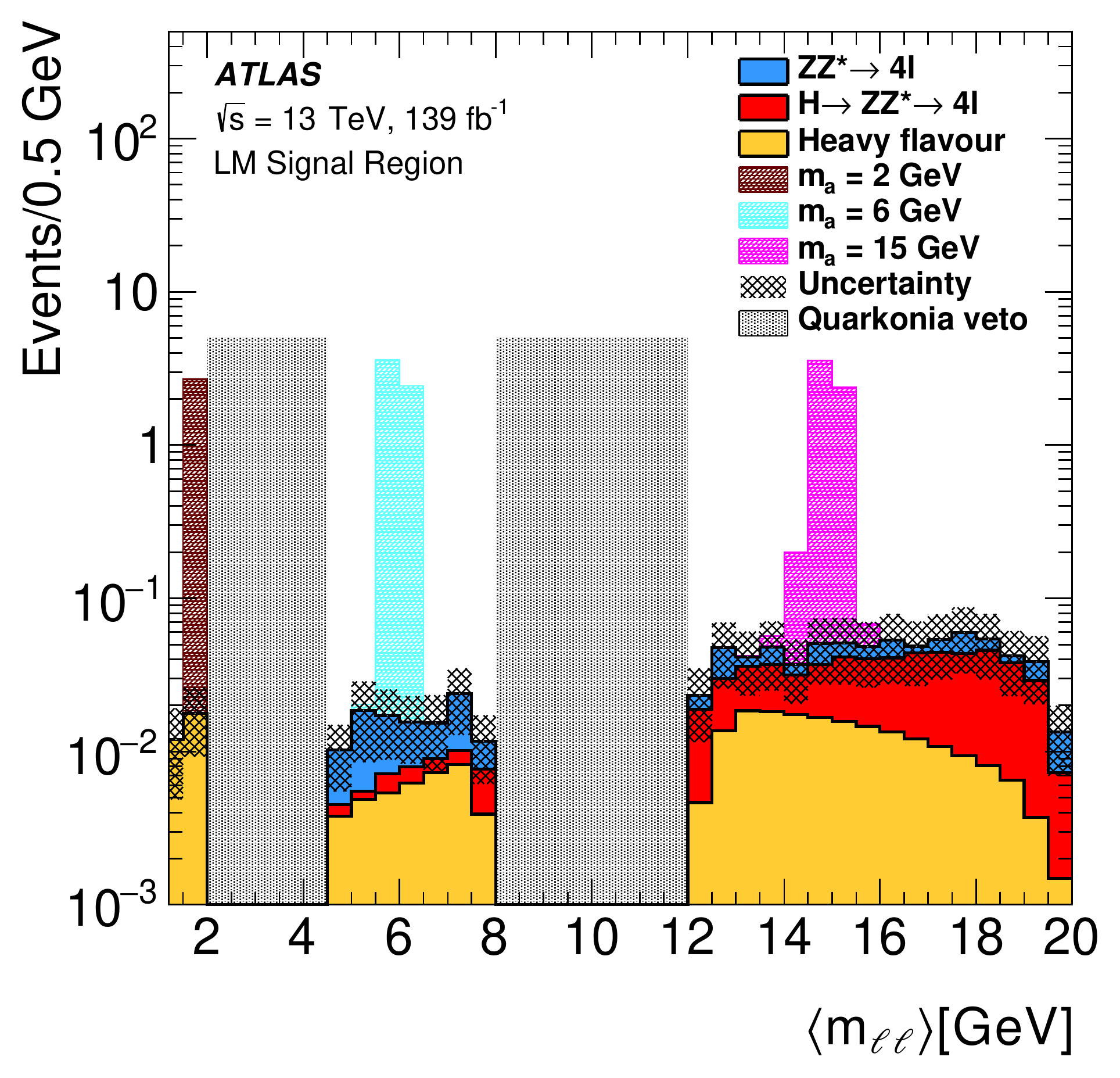}
\label{fig:resultsDataAvgMLowMass}}
\subfloat[]{\includegraphics[width=0.59\textwidth]{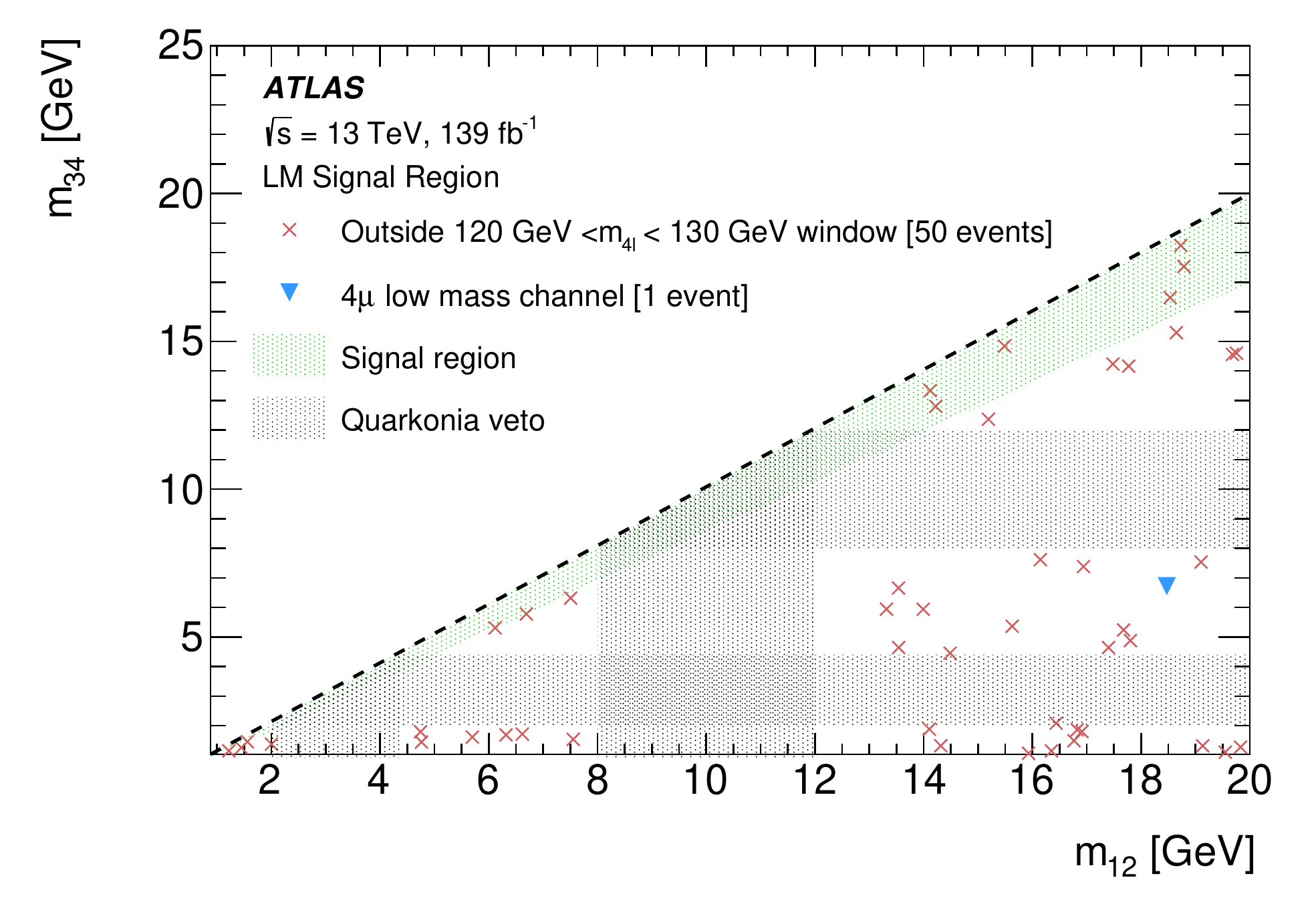}
\label{fig:resultsDataScatterLowMass}}
 
\caption{Distribution of (a)~$\langle m_{\ell\ell}\rangle$ and
(b)~$m_{34}$ vs~$m_{12}$,  for events selected in the LM
$H\to XX \to 4\mu$ $(\SI{1}{\GeV} < m_{X} < \SI{15}{\GeV})$ analysis.
No data events pass this selection. The expectation for a
$H\to aa\to 4\mu$  signal is also shown, for several masses.
The signal histograms are stacked on top of the (pre-fit)
background histograms,
and expected yields are normalized with
$\sigma(pp\to H\to aa\to 4\mu) = \tenth\sigma_{\textrm{SM}}(pp\to H\to ZZ^{*}\to 4\mu) = \SI{0.15}{\fb}$ (ggF process only).
The shaded band represents the total uncertainty of the prediction.
The crossed-through points in~(b) correspond to the 50 events that
are outside the  $m_{4\ell}$  mass window
of $\SI{120}{\GeV} < m_{4\ell} < \SI{130}{\GeV}$.
The events outside the green signal region are events that fail the
$m_{34}/m_{12} > 0.85$  requirement and include one event within the
$m_{4\ell}$  mass window.}
\end{center}
\end{figure}

The $\langle m_{\ell\ell}\rangle$ distribution in the LM signal region
is shown in \cref{fig:resultsDataAvgMLowMass}.
The distribution of $m_{12}$ vs~$m_{34}$ is shown
in \cref{fig:resultsDataScatterLowMass},
while \cref{tab:ZdZdLowMassAnalysis_yields} summarizes the final yields
and uncertainties.  No events are observed,
with a total background prediction of $0.89\pm0.15$ events.

\begin{table}[htbp!]
\caption{
Expected event yields of the SM background processes and data yield
for the LM $H\to XX \to 4\mu$ ($\SI{1}{\GeV} < m_{X} < \SI{15}{\GeV}$)
selection.
The systematic uncertainties of the background estimates
are highly correlated between the different sources of background
(see \ifnotpo{\cref{sec:systematics}}\ifpo{text}).
}
\begin{center}
\begin{small}
\begin{tabular*}{0.4\textwidth}{@{\extracolsep{\fill}}ll}
\toprule
Process & Yield ($\pm \mathrm{stat.}\pm\mathrm{syst.}$)\bigstrut \\
\hline
$H\to ZZ^*\to 4\mu$\bigstrut[t]               & $0.41\pm 0.01\pm0.03$   \\
$ZZ^*\to 4\mu$                                & $0.22\pm0.04\pm0.04$ \\
$VVV$/VBS                                             & \qquad Negligible \\
Heavy flavour\bigstrut[b]                      & $0.26\pm0.09\pm0.10$ \\
\hline
Total\bigstrut                                 & $0.89\pm0.10\pm0.11$   \\
\hline
Data\bigstrut                                  & 0             \\
\bottomrule
\end{tabular*}
\label{tab:ZdZdLowMassAnalysis_yields}
\end{small}
\end{center}
\end{table}

\section{ZX analysis: $H\to ZX\to 4\ell$ ($\SI{15}{\GeV} < m_X < \SI{55}{\GeV}$)}
\label{sec:ZX_analysis}
 
The ZX analysis searches for decays of a SM Higgs boson into a $Z$~boson
along with a new boson~$X$, where both bosons in turn decay into pairs
of electrons or muons.  The event selection is detailed
in \cref{sec:ZX_Selection}.  Like the previous analyses,
it involves finding two same-flavour opposite-sign lepton pairs
with an overall invariant mass consistent with the decay
of a SM~Higgs boson.  Unlike the other analyses, the leading
pair must be broadly consistent with the decay of a $Z$~boson,
and the analysis then searches for a peak in the invariant mass
distribution of the other pair.
For $m_{\Zd} > \SI{55}{\GeV}$,
the invariant mass distribution for $\Zd\rightarrow\ell\ell$ starts
to overlap significantly with that for $Z\rightarrow\ell\ell$.
Since this analysis accepts events with the leading lepton pair
consistent with the decay of a $Z$~boson,
it relies much more on the invariant mass distribution of the other
lepton pair to reject
the $ZZ\rightarrow 4\ell$ background than does the HM analysis.
Therefore, the upper search range for this
analysis is limited to $\SI{55}{\GeV}$, rather than $\SI{60}{\GeV}$
as for the HM analysis.

\subsection{Background estimate}
 
The dominant backgrounds in this analysis are
$H\to ZZ^*\to 4\ell$ (about 65\% of the total)
and non-resonant $ZZ^*\to 4\ell$ (about 33\% of the total).
Additional prompt backgrounds include the triboson
processes $ZZZ$, $WZZ$, and $WWZ$.  These are estimated
from simulation (see \cref{sec:DataAndMCSamples}), but the
$ZZ^*\to 4\ell$ background estimate is checked using background-enriched
validation samples.
 
Other, reducible, backgrounds, such as those from $Z+\textrm{jets}$, $\ttbar$,
and $WZ$ processes, contain either additional non-isolated leptons
from heavy-flavour decay or objects misidentified as leptons
and constitute only a few percent of the background.
The procedure used to estimate the total yield of these
backgrounds is identical to that of the
ATLAS SM $H\to ZZ^*\to 4\ell$ analysis~\cite{HIGG-2016-22,HIGG-2016-25}.
 
The reducible background is estimated separately for the cases where
the second lepton pair ($m_{34}$) decays into muons ($\ell\ell\mu\mu$)
and those where it decays into electrons ($\ell\ell ee$).
For the $\ell\ell\mu\mu$ case, a number of mutually exclusive control regions
are defined by inverting or relaxing some of the lepton identification
requirements, including the isolation and impact parameter requirements
for the subleading muon pair.  A fit to the $m_{12}$ distribution is then performed to
estimate the amount of background due to each of $\ttbar$, $Z+\textrm{heavy-flavour}$
(having $b$- or $c$-quark content), and $Z+\textrm{light-flavour}$.
Transfer factors derived from simulation are then used to extrapolate
the fitted yield of each background in the control regions
to the signal region.  The contribution from $WZ$ production
is estimated using simulation.
 
The $\ell\ell ee$ background from $Z+\textrm{jets}$, $\ttbar$, and $WZ$
production is classified into processes with jets being misidentified
as electrons ($f$), electrons from photon conversions ($\gamma$),
and electrons from semileptonic decay of heavy-flavour hadrons ($q$).
The $q$ component is estimated from simulation.  The other two components
are estimated from a control region in which the identification requirements
of the lowest-$\pT$ electron are relaxed.  Further, to suppress the
$ZZ^*$ contribution, the two subleading electrons must have the same sign.
The expectations for the $f$ and $\gamma$ components are obtained by fitting to
the distribution of the number of inner pixel detector hits
associated with the track of the lowest-$\pT$ electron.
The estimated yields of all three components are then extrapolated
to the signal region using transfer factors derived from simulation.
 
Finally, the shape of the $m_{34}$ distribution for the reducible
background is taken from simulation.  An additional 10\% systematic uncertainty
is assigned to the reducible background estimate
to account for differences
in the lepton isolation requirements between this analysis and
that of Refs.~\cite{HIGG-2016-22,HIGG-2016-25}.

\subsection{Background validation}
 
The estimate of the non-resonant $ZZ^*\to 4\ell$ background is further validated
in control samples that are enriched in this process.  Two validation
regions are defined by replacing the requirement
$\SI{115}{\GeV} < m_{4\ell} < \SI{130}{\GeV}$ with either
$m_{4\ell} < \SI{115}{\GeV}$ (VR5) or
$\SI{130}{\GeV} < m_{4\ell} < \SI{170}{\GeV}$ (VR6).
The latter validation region also requires $|m_{12}-m_Z| < \SI{6}{\GeV}$.
For consistency with the ATLAS SM $H\to ZZ^*\to 4\ell$ analysis~\cite{HIGG-2016-22,HIGG-2016-25},
the requirement on $m_{34}$ is also changed for both validation regions:
\begin{itemize}
\item $m_{34} > \SI{5}{\GeV}$ for $m_{4\ell} < \SI{100}{\GeV}$;
\item $m_{34} > 1.4 (m_{4\ell} - \SI{100}{\GeV}) + \SI{5}{\GeV}$ for $\SI{100}{\GeV} < m_{4\ell} < \SI{105}{\GeV}$;
\item $m_{34} > \SI{12}{\GeV}$ for $\SI{105}{\GeV} < m_{4\ell} < \SI{140}{\GeV}$;
\item $m_{34} > 0.76 (m_{4\ell}-\SI{140}{\GeV})+\SI{12}{\GeV}$ for $ \SI{140}{\GeV} < m_{4\ell} < \SI{170}{\GeV} $.
\end{itemize}
 
\begin{figure}[!htbp]
\begin{center}
\subfloat[]{\includegraphics[width=0.49\textwidth]{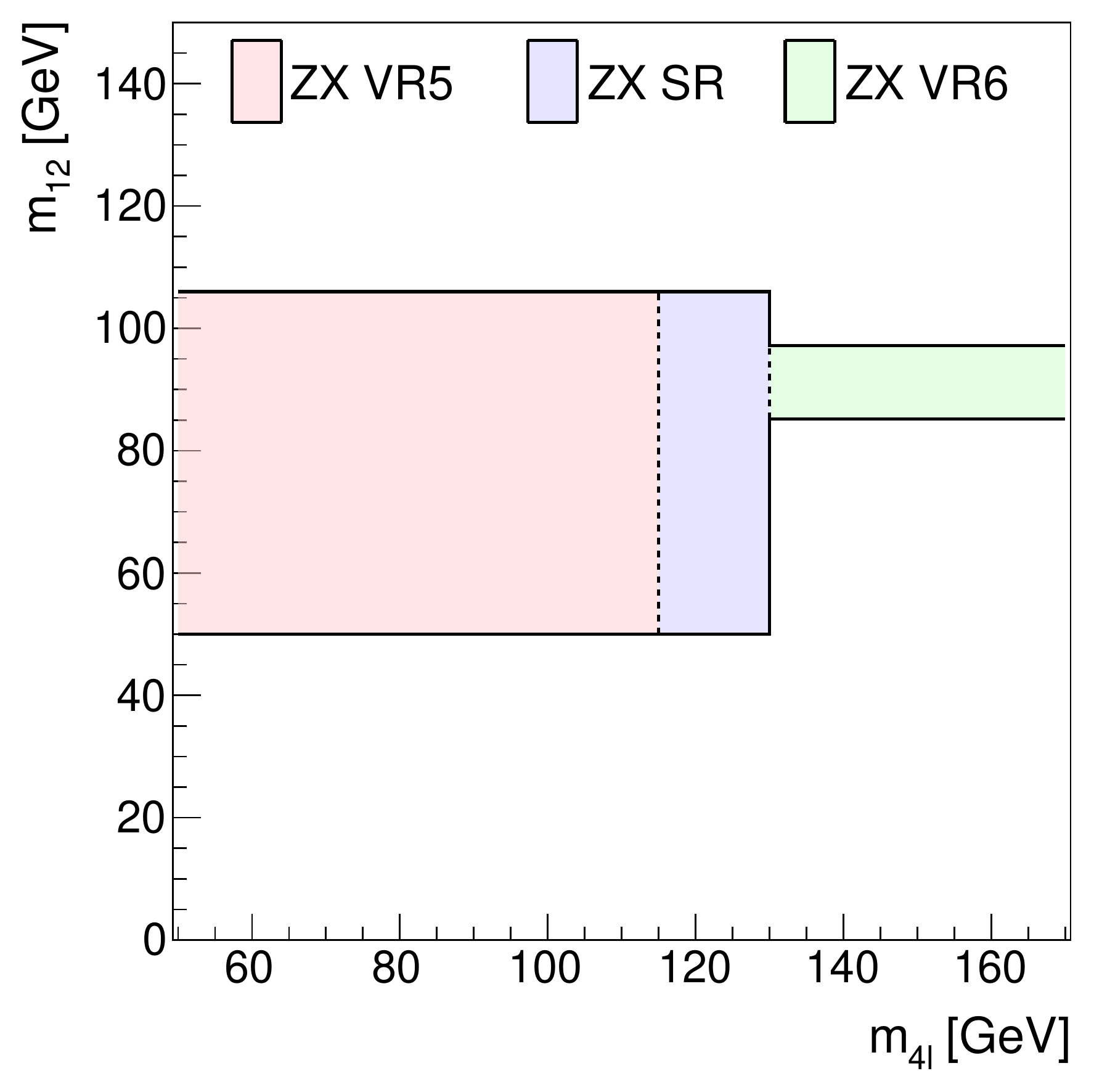}}
\subfloat[]{\includegraphics[width=0.49\textwidth]{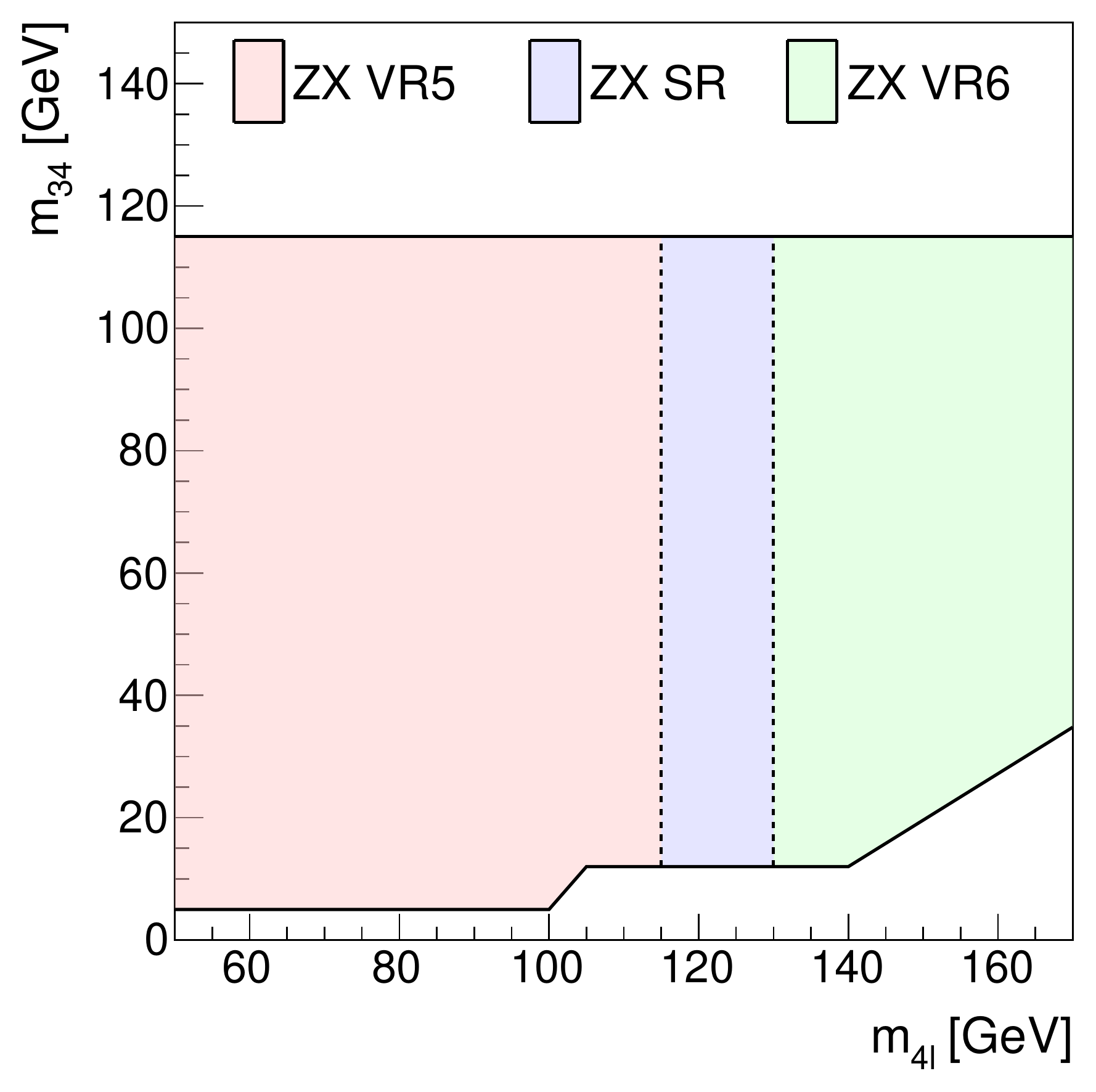}}
 
\caption{
Illustration of the validation region definitions for
the $H\ra ZX\ra 4\ell$ analysis.  Shown are the selections
in the
(a)~$m_{12}$~vs~$m_{4\ell}$ and
(b)~$m_{34}$~vs~$m_{4\ell}$ planes for the two validation regions
as well as the signal region.  Details of the selections are given in the text.
}
\label{fig:zxvalidation-regions}
\end{center}
\end{figure}

\begin{figure}[!htbp]
\begin{center}
\subfloat[]{\includegraphics[width=0.49\textwidth]{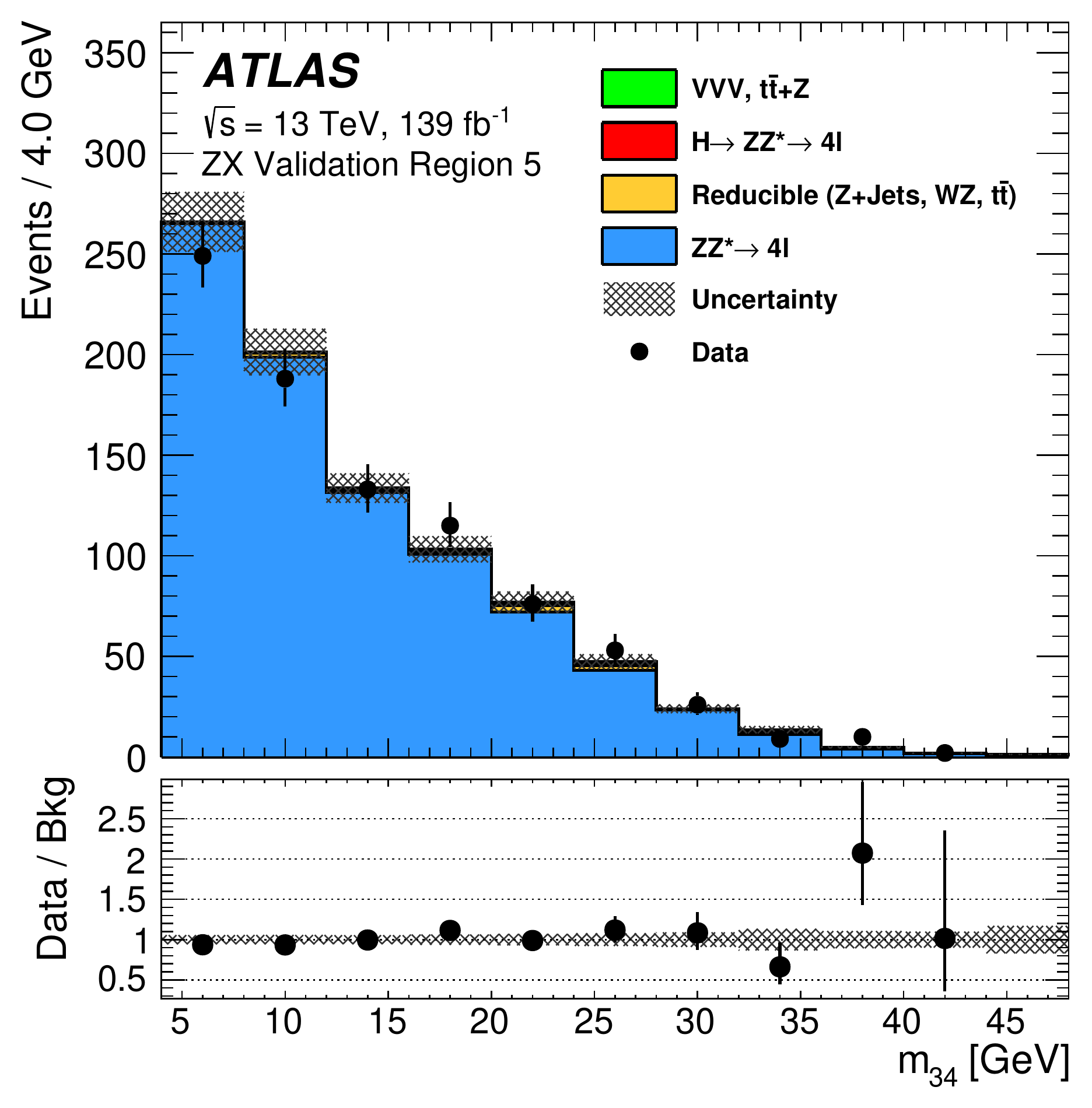}}
\subfloat[]{\includegraphics[width=0.49\textwidth]{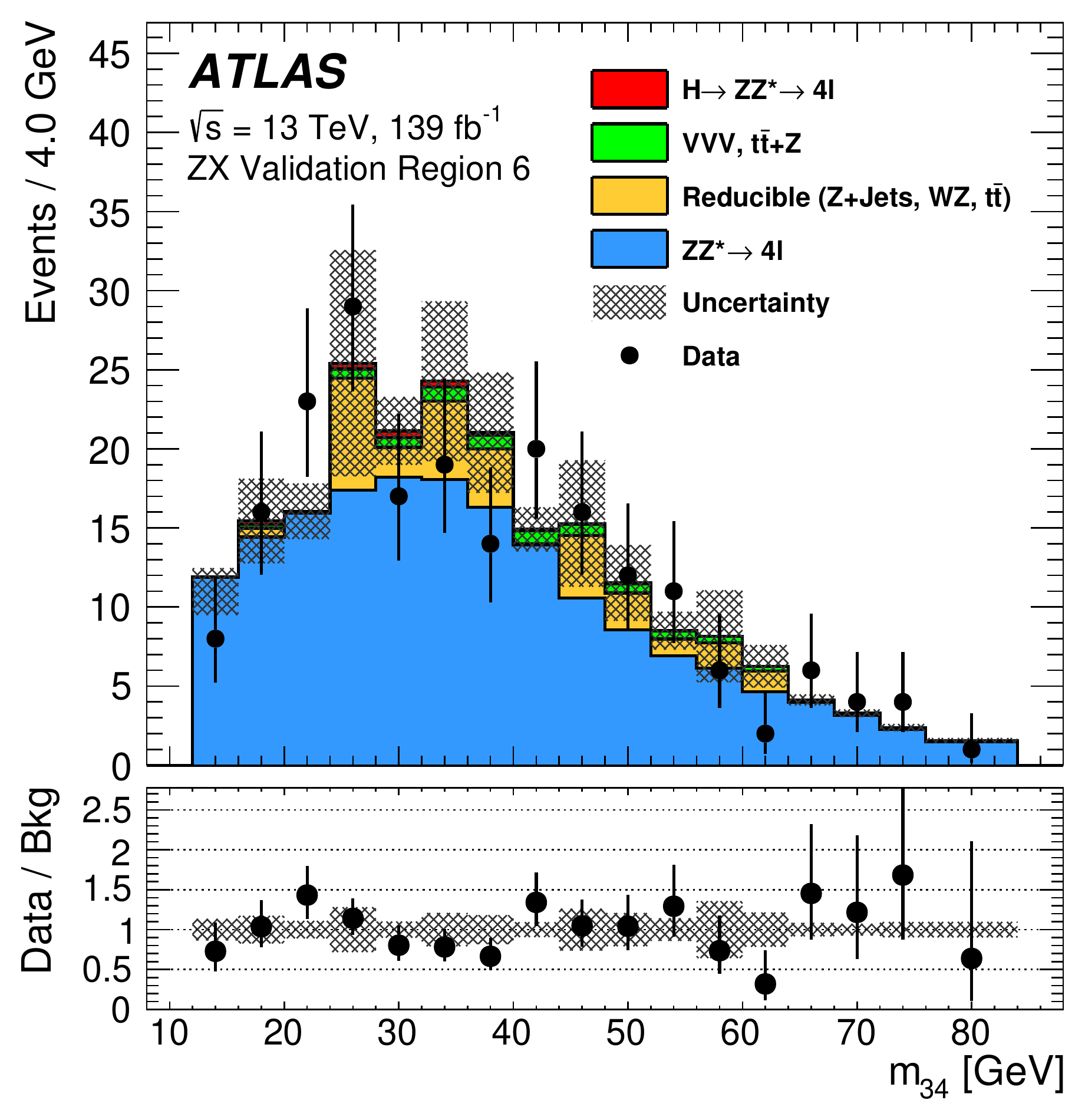}}
 
\caption{Distributions of $m_{34}$
in the two validation regions
for the $H\to ZX\to 4\ell$
analysis.
(a)~VR5: $m_{4\ell} < \SI{115}{\GeV}$;
(b)~VR6: $\SI{130}{\GeV} < m_{4\ell} < \SI{170}{\GeV}$.
The shaded band represents the total uncertainty of the (pre-fit) prediction.
The lower panels show the ratio of the observed data to the
MC~predictions.
The uncertainties of the plotted data are asymmetric and are calculated
using
Eqs.~(40.76) \ifnotpo{of Ref.~\cite{pdg20}}\ifpo{from the 2020 Review of Particle Properties}.
}
\label{fig:zxvalidation}
\end{center}
\end{figure}

These requirements are illustrated in \cref{fig:zxvalidation-regions}.
Distributions of $m_{34}$ for the two validation regions are shown
in \cref{fig:zxvalidation}.  Good agreement is found with background
expectations.

\subsection{Results}
 
\begin{figure}[!htbp]
\begin{center}
\includegraphics[width=0.7\textwidth]{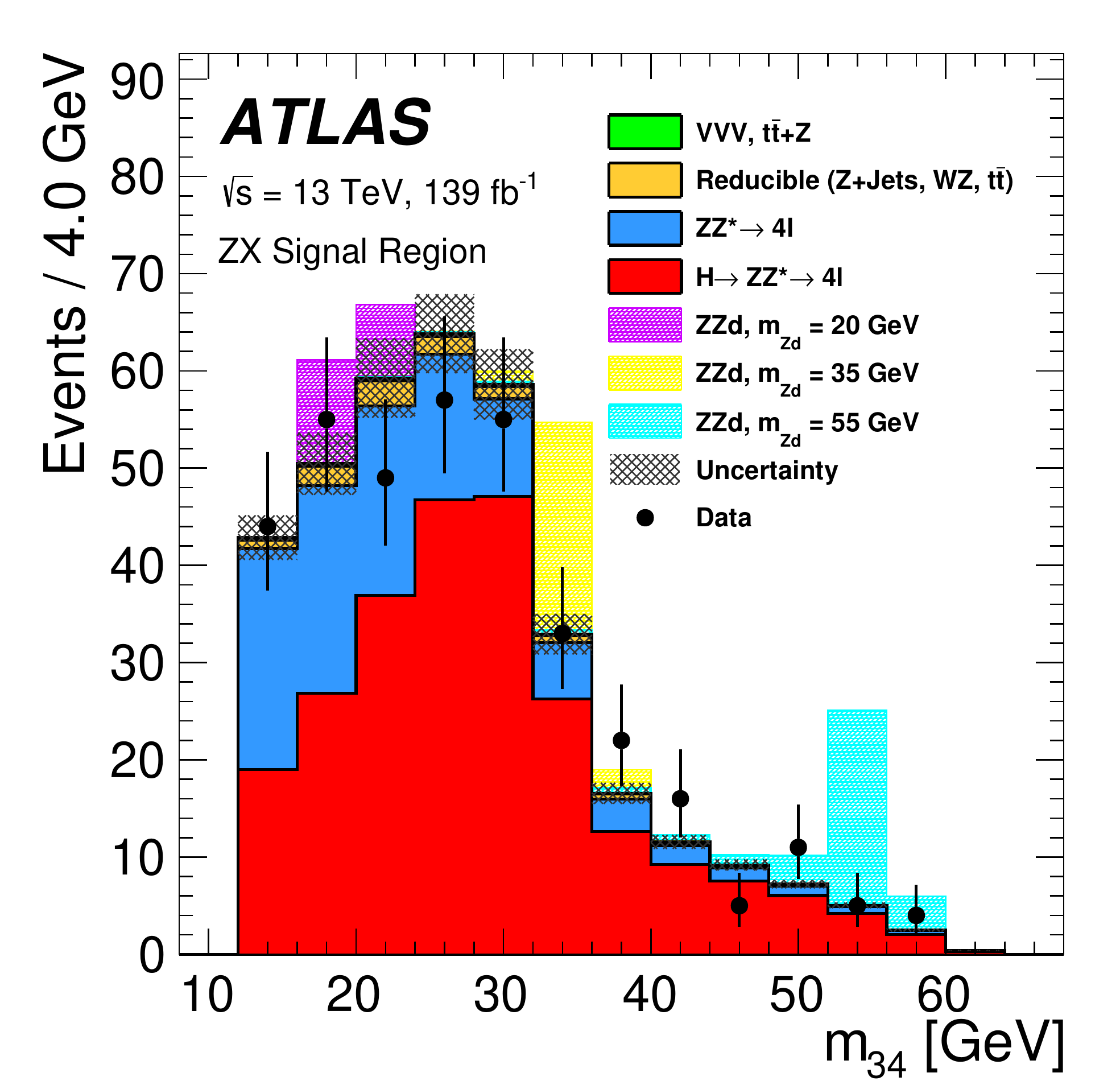}
\caption{Distribution of $m_{34}$ for data and background events
in the mass range $\SI{115}{\GeV} < m_{4\ell} < \SI{130}{\GeV}$
after the $H\to ZX\to 4\ell$~selection. The background normalization is taken from the fit (see text);
the shaded band represents the total uncertainty of the background
prediction.
Three signal points for the $H\to ZZ_d\to 4\ell$~model are shown,
stacked on top of the background histograms.
The signal yields are normalized
with
$\sigma(pp\to H\to Z\Zd\to 4\ell) =
\tenth\sigma_{\textrm{SM}}(pp\to H\to ZZ^{*}\to 4\ell) = \SI{0.69}{\fb}$.
The uncertainties of the plotted data are asymmetric and are calculated
using
Eqs.~(40.76) \ifnotpo{of Ref.~\cite{pdg20}}\ifpo{from the 2020 Review of Particle Properties}.}
\label{fig:zxm34}
\end{center}
\end{figure}

\begin{figure}[!htbp]
\begin{center}{\includegraphics[width=0.65\textwidth]{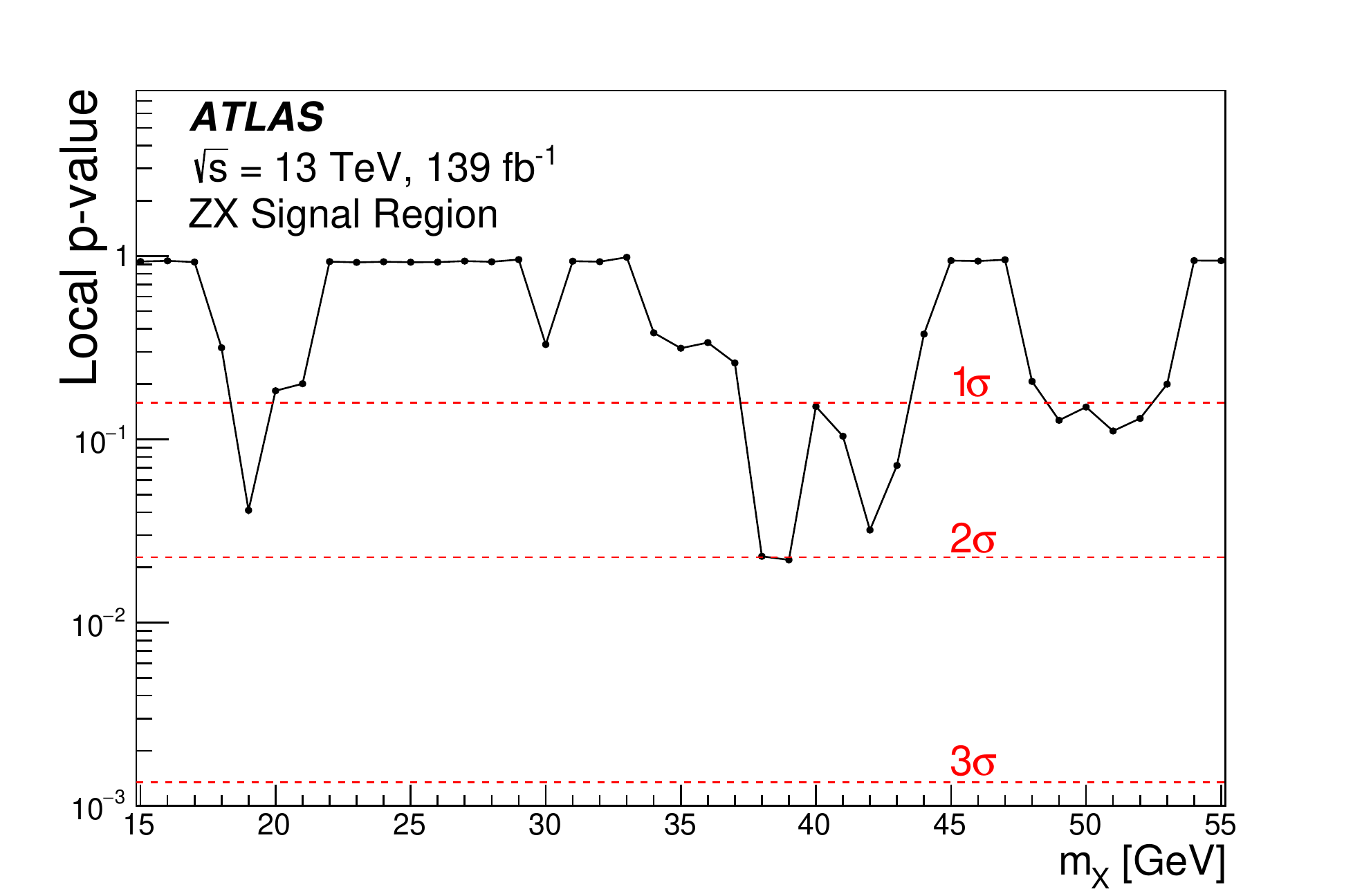}}
 
\caption{Observed local $p$-values under the background-only hypothesis
for the process $H\to Z X\to 4\ell$.}
\label{fig:zxp0sig}
\end{center}
\end{figure}

The final $m_{34}$ distribution for this analysis is shown
in \cref{fig:zxm34}, while \cref{tab:zxyields} summarizes the
final yields and uncertainties.
The dominant systematic uncertainty in final states
that contain electrons arises from the modelling of the
electron identification efficiency.  For the $4\mu$ channel,
the dominant systematic uncertainty arises from the modelling of muon isolation.
A total of 356 events are observed
with an expected background of $320 \pm 17$.
\cref{fig:zxp0sig} shows the observed local $p$-values
for the background-only hypothesis.
The profile-likelihood ratio is again used as the test statistic.
Different final states are not distinguished in the fit; distributions
used are summed over all channels.
The normalization of the
$H\to ZZ^*$ background is allowed to float (as an unconstrained
nuisance parameter, see \cref{sec:zxlimit}), with a resulting normalization
of $1.2\pm0.16$.  The largest excess,
with a local significance of around $2\sigma$, is at about $m_X=\SI{39}{\GeV}$.
 
These results are slightly different from the corresponding results
from the ATLAS SM $H\to ZZ^*\to 4\ell$ analysis~\cite{HIGG-2018-29},
which observed 310~events and found a signal strength
of $\sigma_{\textrm{fid}} / \sigma_{\textrm{fid,SM}} = 0.96\pm 0.11$.
The difference is largely due to the differences in quadruplet
handling mentioned in \cref{sec:ZX_Selection}, and also due to
differences in the definitions of the isolation and impact parameter
selections.  When this analysis is repeated using the quadruplet
definition of Ref.~\cite{HIGG-2018-29}, the resulting normalization
of the $H\to ZZ^*$ background is $1.12\pm 0.15$.

\begin{table}[htbp]
\caption{Expected and observed numbers of events in each channel after the $H\to ZX\to 4\ell$~
event selection defined by the mass range $\SI{115}{\GeV} < m_{4\ell} < \SI{130}{\GeV}$.
The background normalization is prior to the fit (see text).
The systematic uncertainties of the background estimates
are highly correlated between the different sources of background
(see \ifnotpo{\cref{sec:systematics}}\ifpo{text}).
}
\newcommand{\hd}{\hphantom{0}}
\newcommand{\hp}{\hphantom{.}}
 
\begin{center}
\begin{tabular}{llll}
\toprule
Process   & \multicolumn{3}{c}{Yield ($\pm \mathrm{stat.}\pm\mathrm{syst.}$)} \\
&   \qquad\quad $2\ell 2\mu$    &    \qquad\quad $2\ell2e$   &  \qquad\quad Total \\
\hline
\bigstrut[t]
$H\to ZZ^{*}\to 4\ell$ & $127.9\pm 0.1\pm 3.6$ & ~~$76~~~\pm 0.1\pm 10$ & $204~~~\pm 0.2\pm 12$ \\
$ZZ^{*}\to 4\ell$      & ~~$ 70.2\pm 0.2\pm 1.9$ & ~~$33.0\pm 0.2\pm~~3.6$  & $103.3\pm 0.3\pm~~4.6$ \\
Reducible              & ~~~~$  4.9\pm 0.1\pm 0.3$ & ~~~~$5.8\pm 0.3\pm~~0.6$   & ~~$ 10.7\pm 0.3\pm~~1.0$ \\
$VVV, t\bar{t}+Z$      & ~~~~$  1.1\pm 0.1\pm 0.04$ & ~~~~$0.7\pm 0.1\pm~~0.1$   & ~~~~$ 1.8\pm 0.1\pm~~0.1$ \\
 
\hline
Total\bigstrut         & $204.1\pm 0.3\pm 5.5$ & $116~~~\pm 0.5\pm 14$ & $320~~~\pm 0.5\pm 17$ \\
\hline
Data\bigstrut[t] &  237 & 119 & 356 \\
\bottomrule
\end{tabular}
\
\label{tab:zxyields}
\end{center}
 
\end{table}

\FloatBarrier
 
\section{Limits and interpretation}
\label{sec:Limits}
 
No significant excess is observed above SM background predictions for any of the
analyses considered.  Therefore, the results are interpreted in terms
of exclusion limits.
Firstly, model-independent limits are placed on fiducial
cross sections.  Model-dependent exclusion limits are then set for
the benchmark models described in \cref{sec:theory}.
 
For the HM and LM $H\ra XX \ra 4\ell$ analyses, evaluating the limits
entails parameterizing the signal distribution as a function of both
$\langle m_{\ell\ell}\rangle$
and
$m_X$, while the $H\to ZX\to 4\ell$ analysis requires
the parameterization to be a function
of $m_{34}$ and $m_X$.
Since simulated events are generated only at
discrete values of $m_X$, the signal templates are
interpolated between $m_X$ values.  For the
HM and ZX analyses,
this is done
using moment morphing~\cite{Baak:2014fta}.  The distributions at the generated
values of $m_X$ are used as templates, and the normalization is
determined from interpolation of the simulated signal yields.
For the LM analysis, Gaussian distributions are fit to the
$\langle m_{\ell\ell}\rangle$ distributions at each generated $m_X$,
and the fit parameters are interpolated in $m_X$.
 
The data are described statistically by a likelihood function consisting of a
Poisson factor for each histogram bin, summed over each channel, along with a
Gaussian constraint for each nuisance parameter~\cite{cranmer2012histfactory}:
\begin{equation}\nonumber{\cal L} (N, \alpha) = \prod_i
\textrm{Pois}\left(\sum_j N_{ij}; \sum_j \mu S_{ij}(\alpha) + B_{ij}(\alpha) \right)
\prod_k \textrm{Gaus} (\alpha_k; s_k, \sigma_k),
\end{equation}
where $N_{ij}$ is the number of observed events observed in bin $i$
for channel $j$, $\alpha$ is the set of nuisance parameters,
$S_{ij}(\alpha)$ and $B_{ij}(\alpha)$ are the predicted
numbers of signal
and background events for each bin and channel, $\mu$ is the signal strength,
and $s_k$ and $\sigma_k$ are mean and width of the Gaussian constraint
for nuisance parameter $\alpha_k$.  Systematic uncertainties are modelled
via nuisance parameters which are profiled in the calculation
of the test statistic; the effect of systematic uncertainties on the limits
is small.

\subsection{Limits on fiducial and total cross sections}
 
Model-independent cross-section limits for the HM, LM, and ZX analyses
are derived in fiducial regions defined using generator-level quantities.
These fiducial selections,
shown in \cref{tab:fiducial_selection}, are designed to mimic the signal region
selection requirements.  In order to account for the effects of quasi-collinear
electromagnetic radiation from the leptons within the detector resolution,
the four-momenta of prompt photons
close to a lepton ($\Delta R < 0.1$) are added to the four-momentum
of that lepton~\cite{ATL-PHYS-PUB-2015-013}.
 
\begin{table}
\caption{Summary of the fiducial phase-space definitions,
appropriate for $H \rightarrow XX \rightarrow 4\ell$
or $H \rightarrow ZX \rightarrow 4\ell$, where $X$ is a  promptly decaying, on-shell, narrow resonance.
}

\centering
 
\begin{adjustbox}{width=1.0\textwidth}
\small
 
\begin{tabular}{|c|c|>{\centering}m{5cm}|>{\centering}m{5cm}|>{\centering}m{5cm}|}
\cline{3-5} \cline{4-5} \cline{5-5}
\multicolumn{1}{c}{} &
& Single $Z$ (ZX) analysis \bigstrut[t]\\
$H\rightarrow XZ\rightarrow4\ell$ $(\ell=e,\mu)$ \bigstrut[b]
& High-mass (HM) analysis \bigstrut[t]\\
$H\rightarrow XX\rightarrow4\ell$ $(\ell=e,\mu)$ \bigstrut[b]
& Low-mass (LM) analysis\\
$H\rightarrow XX\rightarrow4\mu$ \tabularnewline
\hline
 
\multicolumn{2}{|c|}{Mass range}
& $\SI{15}{\GeV} < m_X < \SI{55}{\GeV}$ \bigstrut
& $\SI{15}{\GeV} < m_X < \SI{60}{\GeV}$ \bigstrut
& $\SI{1}{\GeV} < m_X < \SI{15}{\GeV}$
\tabularnewline
\hline
 
\multicolumn{2}{|c|}{Electrons}
& \multicolumn{3}{>{\centering}m{15cm}|}{$\pt > \SI{7}{\GeV}$ \quad\quad\quad $|\eta| < 2.5$\bigstrut
} \tabularnewline
\hline
 
\multicolumn{2}{|c|}{Muons}
& \multicolumn{3}{>{\centering}m{15cm}|}{$\pt > \SI{5}{\GeV}$ \quad\quad\quad $|\eta| < 2.7$\bigstrut
} \tabularnewline
\hline
 
\multicolumn{2}{|c|}{Quadruplet}
& \multicolumn{3}{>{\centering}m{15cm}|}{Three leading-$\pt$ leptons satisfying $\pt > \SI{20}{\GeV}$, $\SI{15}{\GeV}$, $\SI{10}{\GeV}$} \bigstrut \tabularnewline
\cline{3-5}
\multicolumn{1}{|c}{}  &
& \multicolumn{2}{>{\centering}m{10cm}|}{$\Delta R>0.10\,(0.20)$ between same-flavour (different-flavour) leptons\bigstrut}
& ---\bigstrut \tabularnewline
\cline{3-5}
\multicolumn{1}{|c}{}  &
& ---\bigstrut
& $m_{34}/m_{12} > 0.85 - 0.1125 f(m_{12})$ \bigstrut
&   $m_{34}/m_{12} > 0.85$\bigstrut \tabularnewline
\cline{3-5}
\multicolumn{1}{|c}{}  &
& $\SI{50}{\GeV} < m_{12} < \SI{106}{\GeV}$\bigstrut[t]\newline
$\SI{12}{\GeV} < m_{34} < \SI{115}{\GeV}$ \newline
$m_{14,23} > \SI{5}{\GeV}$ ($4e$/$4\mu$)\bigstrut[b]
& $\SI{10}{\GeV} < m_{12,34} < \SI{64}{\GeV}$\bigstrut[t]\newline
For $4e$ and $4\mu$ channels:\newline
$\SI{5}{\GeV} < m_{14,23} < \SI{75}{\GeV}$\bigstrut[b]
& $\SI{1.2}{\GeV} < m_{12,34} < \SI{20}{\GeV}$  \tabularnewline
\cline{3-5}
\multicolumn{1}{|c}{}  &
& ---
& \multicolumn{2}{>{\centering}m{10cm}|}{Reject event if $m_{12,34,14,23}$ in either:\bigstrut[t]\newline
$(m_{J/\psi}-\SI{0.25}{\GeV})$ to $(m_{\psi(2S)}+\SI{0.30}{\GeV})$, or \newline
$(m_{\Upsilon(1S)}-\SI{0.70}{\GeV})$ to
$(m_{\Upsilon(3S)}+\SI{0.75}{\GeV})$\bigstrut[b]}
\tabularnewline
\cline{3-5}
\multicolumn{1}{|c}{}  &
& ---
& ---
&
{Reject event if $m_{12,34}$ in either\bigstrut[t]\newline
$\SI{2}{\GeV}$ to $\SI{4.4}{\GeV}$, or\newline
$\SI{8}{\GeV}$ to $\SI{12}{\GeV}$\bigstrut[b]} \tabularnewline
\cline{3-5}
\multicolumn{1}{|c}{}  &
& $\SI{115}{\GeV} < m_{4\ell} < \SI{130}{\GeV}$ \bigstrut
& ---
& --- \tabularnewline
\hline
\end{tabular}
 
\end{adjustbox}
 
\label{tab:fiducial_selection}
\end{table}

The fiducial selections are used to factorize the effects of the event
selection into a largely model-independent `efficiency' and a model-dependent
`acceptance'.  The efficiency for a given
channel is defined as the fraction of events passing the fiducial
selection (using generator-level quantities) that also pass the
full event selection (using reconstructed quantities).  This mostly
depends on the lepton reconstruction, but not on the model used.
Systematic uncertainties relevant to the reconstruction of leptons
are propagated to the efficiency.
For a given theoretical signal model, the acceptance
for a channel $c$ is defined as
\begin{equation}\nonumber
\alpha_c = \frac{N^c_{\textrm{fid}}}{N^c_{\textrm{tot}}} ,
\end{equation}
where $N^c_{\textrm{fid}}$ is the yield for channel $c$ within the
fiducial region (at generator level) and $N^c_{\textrm{tot}}$
is the total yield for channel $c$ (simply the total generator-level
yield for the channel).  The efficiency may thus be used to find
a model-independent fiducial cross-section limit, which may then
be converted to a model-dependent total cross-section limit
using the acceptance.

\subsubsection{HM and LM limits}
 
\begin{figure}[!htbp]
\begin{center}
\subfloat[]{
\includegraphics[width=0.49\textwidth]{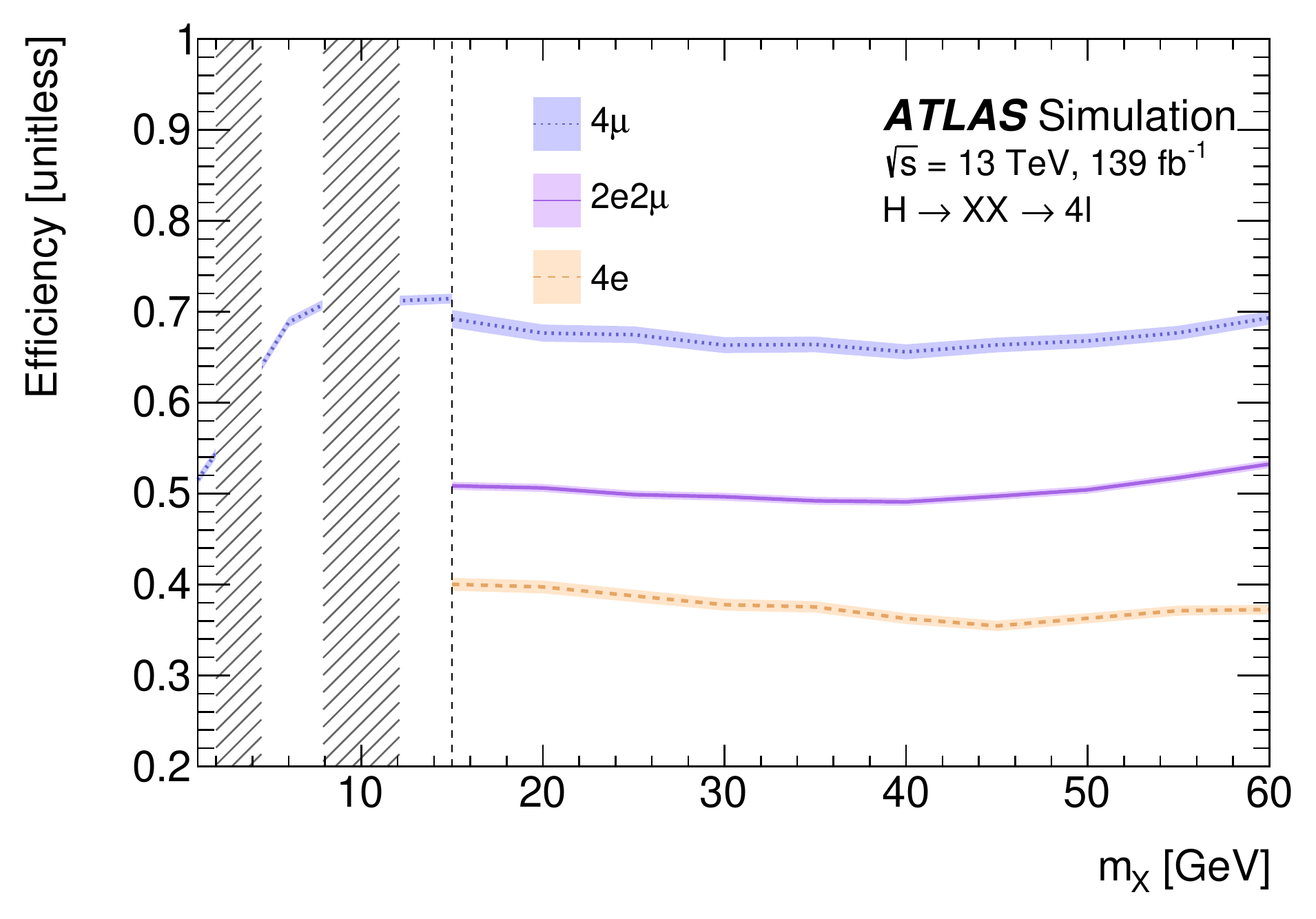}
\label{fig:efficiency_ZdZd}}
\hskip 0.03\textwidth
\subfloat[]{\includegraphics[width=0.47\textwidth]{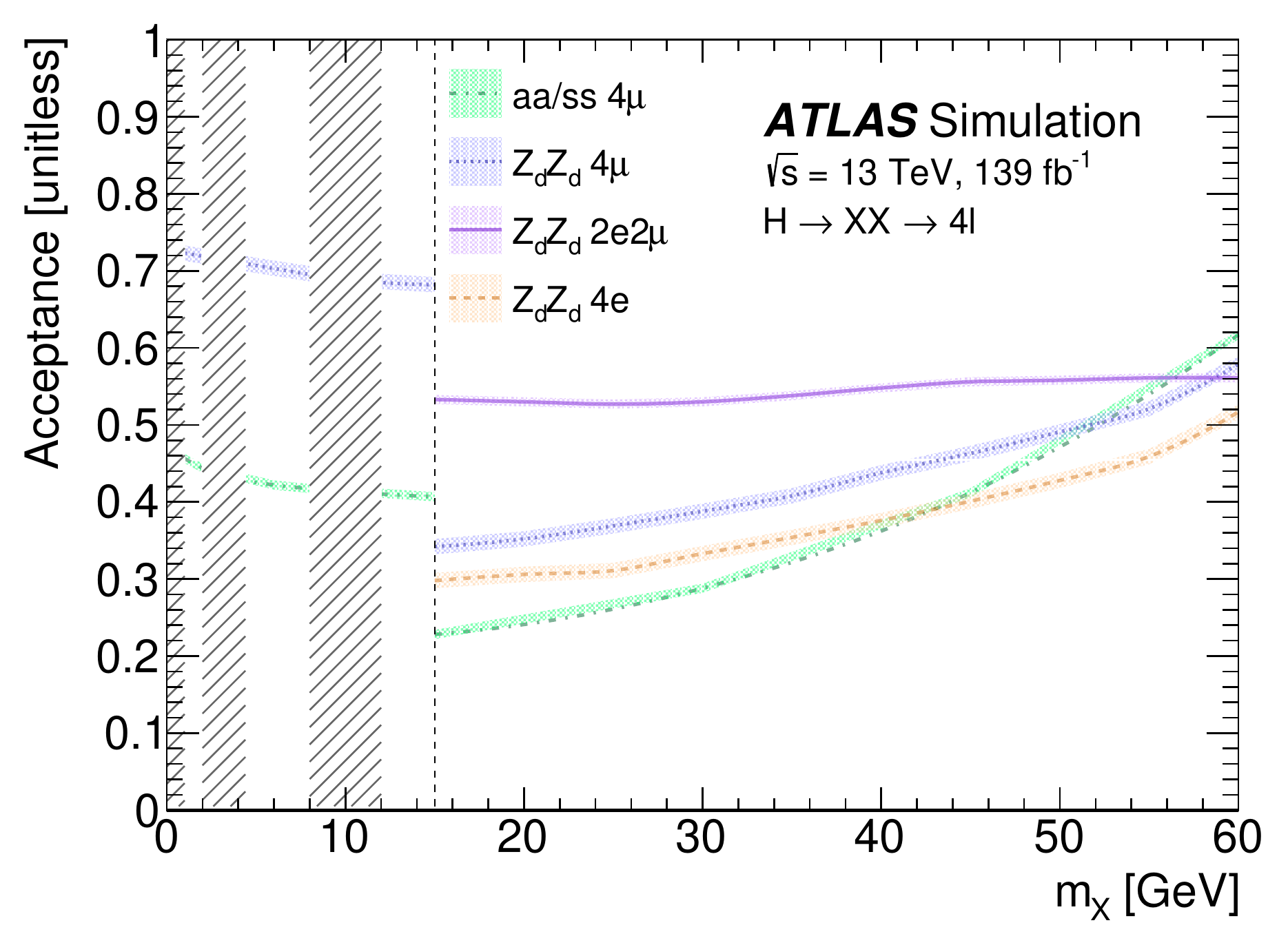}
\label{fig:acceptance_ZdZd}}
\caption{(a)~Model-independent per-channel efficiencies $\epsilon_{c}$
calculated in the fiducial volumes described in the $\SI{1}{\GeV} < m_{X} < \SI{15}{\GeV}$
and $\SI{15}{\GeV} < m_{X} < \SI{60}{\GeV}$
columns of
\ifpo{the fiducial region definition table} \ifnotpo{\cref{tab:fiducial_selection}}
(i.e.\ separate phase spaces are
defined for $m_{X}$ above and below $\SI{15}{\GeV}$).
(b)~Model-dependent per-channel fiducial region acceptances
for the $H\to \Zd\Zd\to 4\ell$ and $H\to aa\to 4\mu$ processes.
The shaded areas are the
quarkonia veto regions.}
\end{center}
\end{figure}

\begin{figure}[!htbp]
\begin{center}
\subfloat[]{\includegraphics[width=0.47\textwidth]{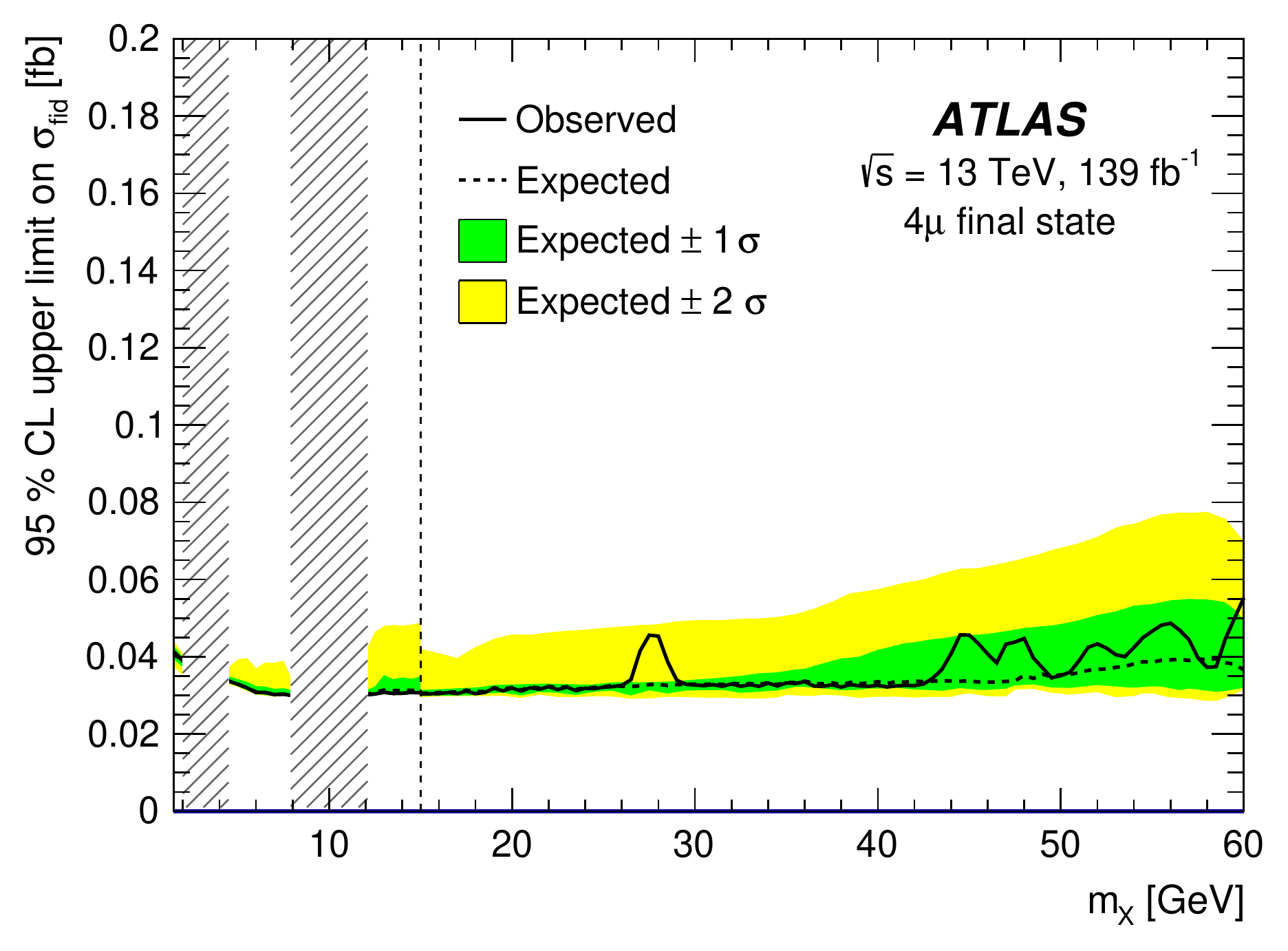}}
\subfloat[]{\includegraphics[width=0.47\textwidth]{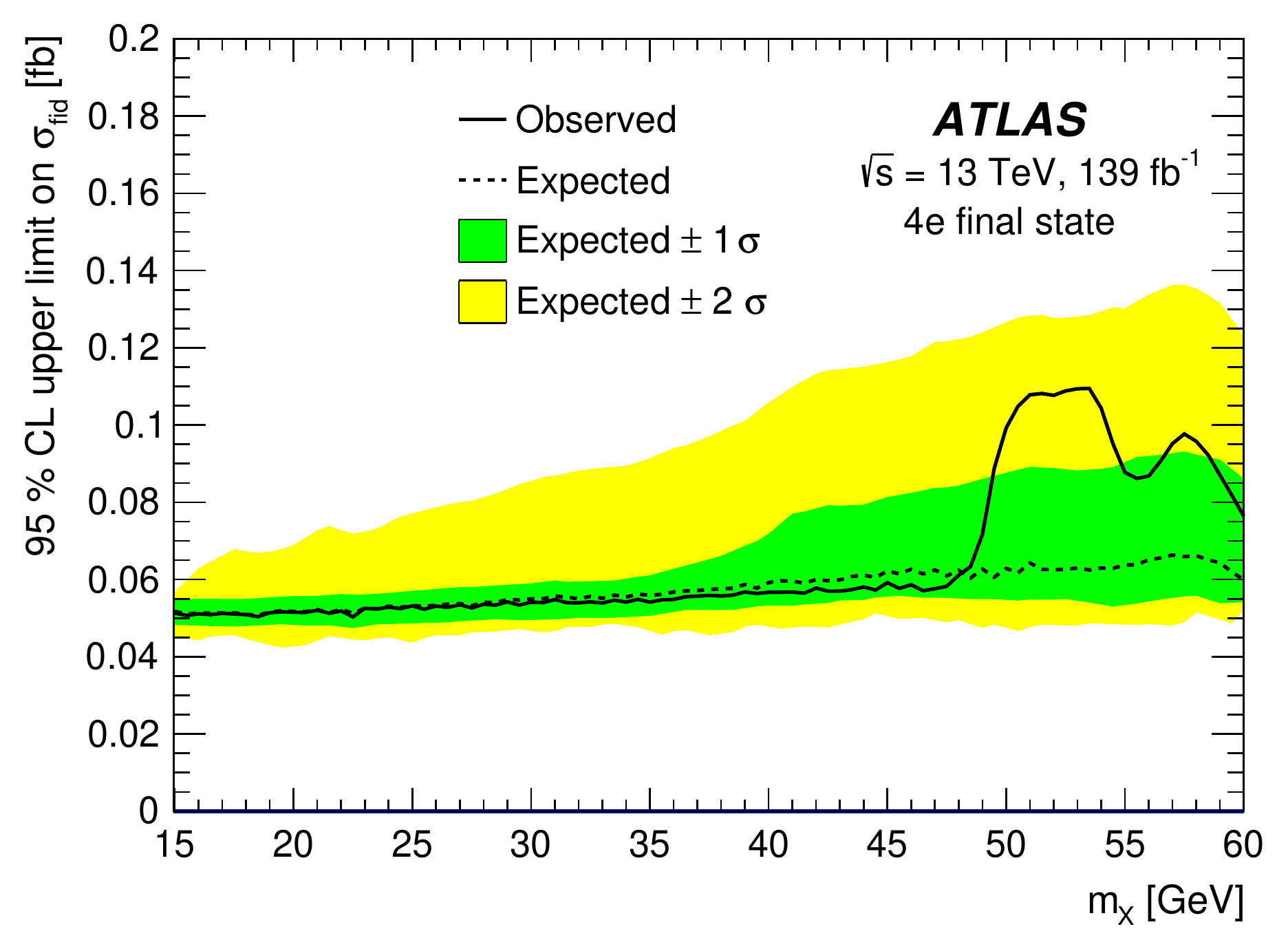}}\\%
\subfloat[]{\includegraphics[width=0.47\textwidth]{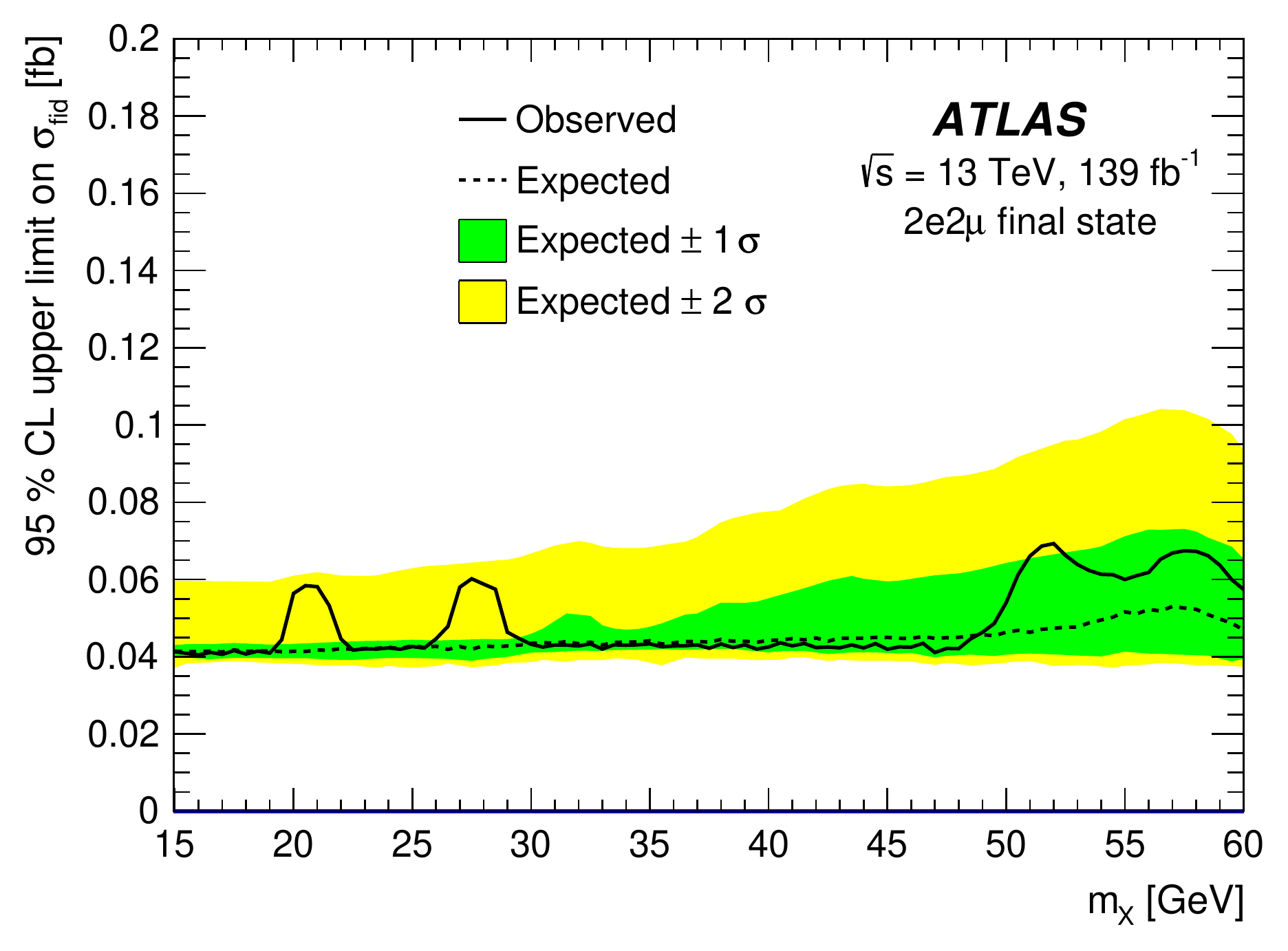}}
\caption{Per-channel upper limits at 95\popercent CL on fiducial
cross sections for the $H\rightarrow XX \rightarrow 4\ell$
process, for the (a)~$4\mu$, (b)~$4e$, and (c)~$2e2\mu$ final states.
The step change
in the $4\mu$ channel at $m_X=\SI{15}{\GeV}$ is due to the
change in efficiency caused by the change in fiducial
phase-space definition. The shaded areas are the
quarkonia veto regions.}
\label{fig:fiducialLimits_ZdZd}
\end{center}
\end{figure}

\begin{figure}[!htbp]
\begin{center}
\includegraphics[width=0.6\textwidth]{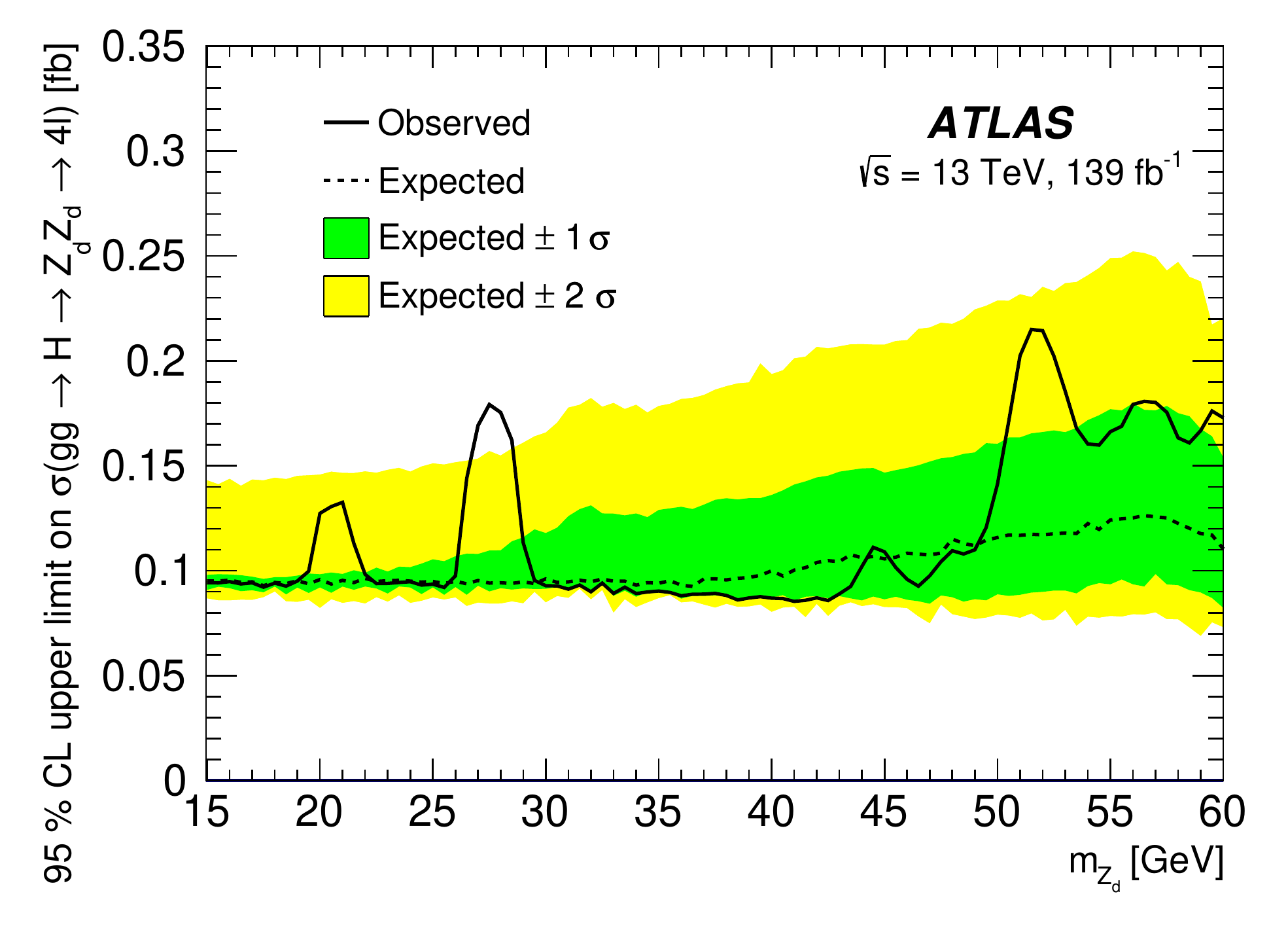}
\caption{Observed and expected upper limits at 95\popercent CL
for the cross section of the $H\to \Zd\Zd\to 4\ell$ process,
assuming SM Higgs~boson production via the gluon--gluon fusion
process.  All final states are combined.
HAHM parameters were set to $\kappa=\epsilon=10^{-4}$.}
\label{fig:Limit_xs_ZdZd_4l}
\end{center}
\end{figure}

\begin{figure}[!htbp]
\begin{center}
\subfloat[]{
\includegraphics[width=0.49\textwidth]{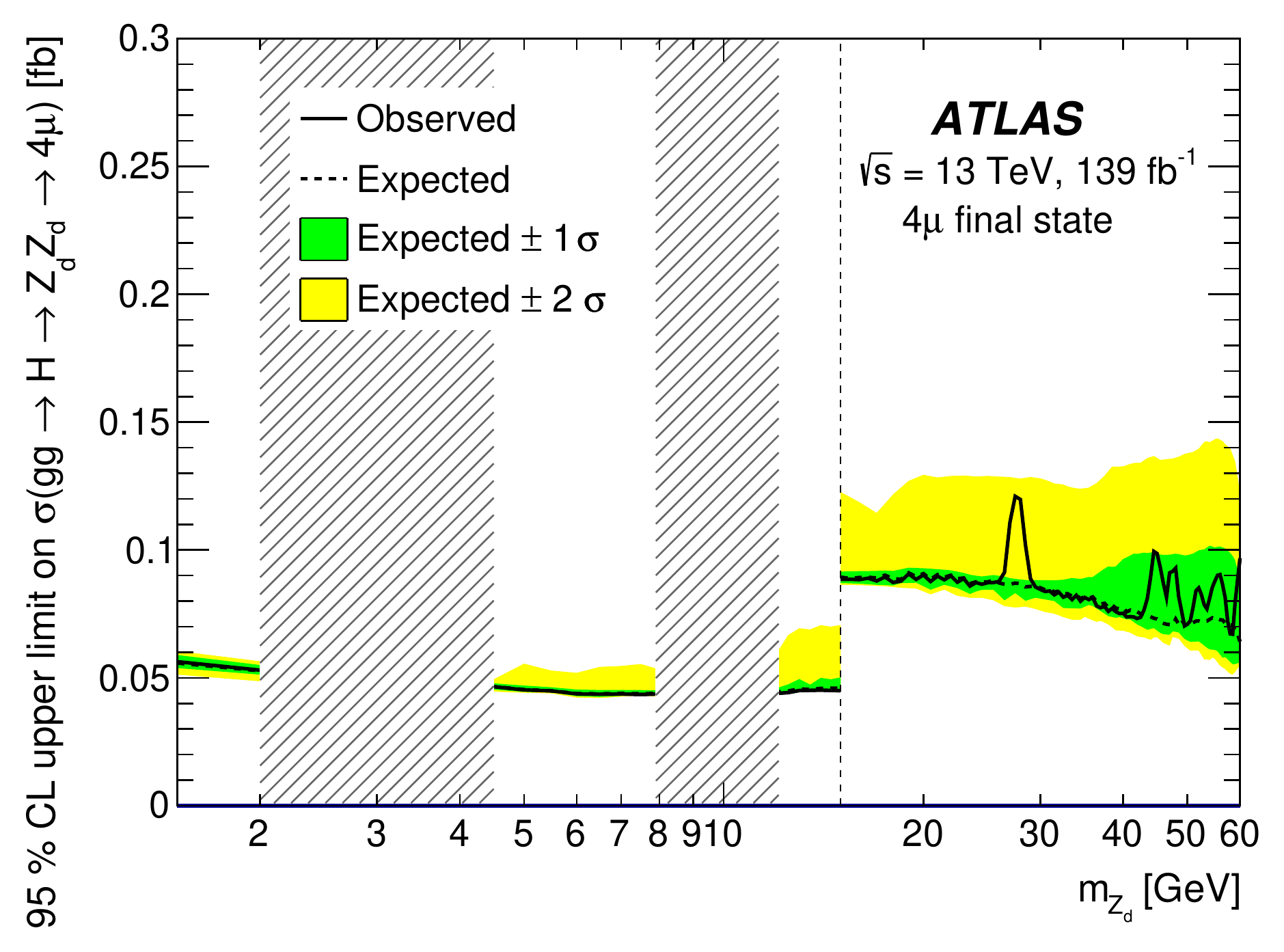}}
\subfloat[]{
\includegraphics[width=0.49\textwidth]{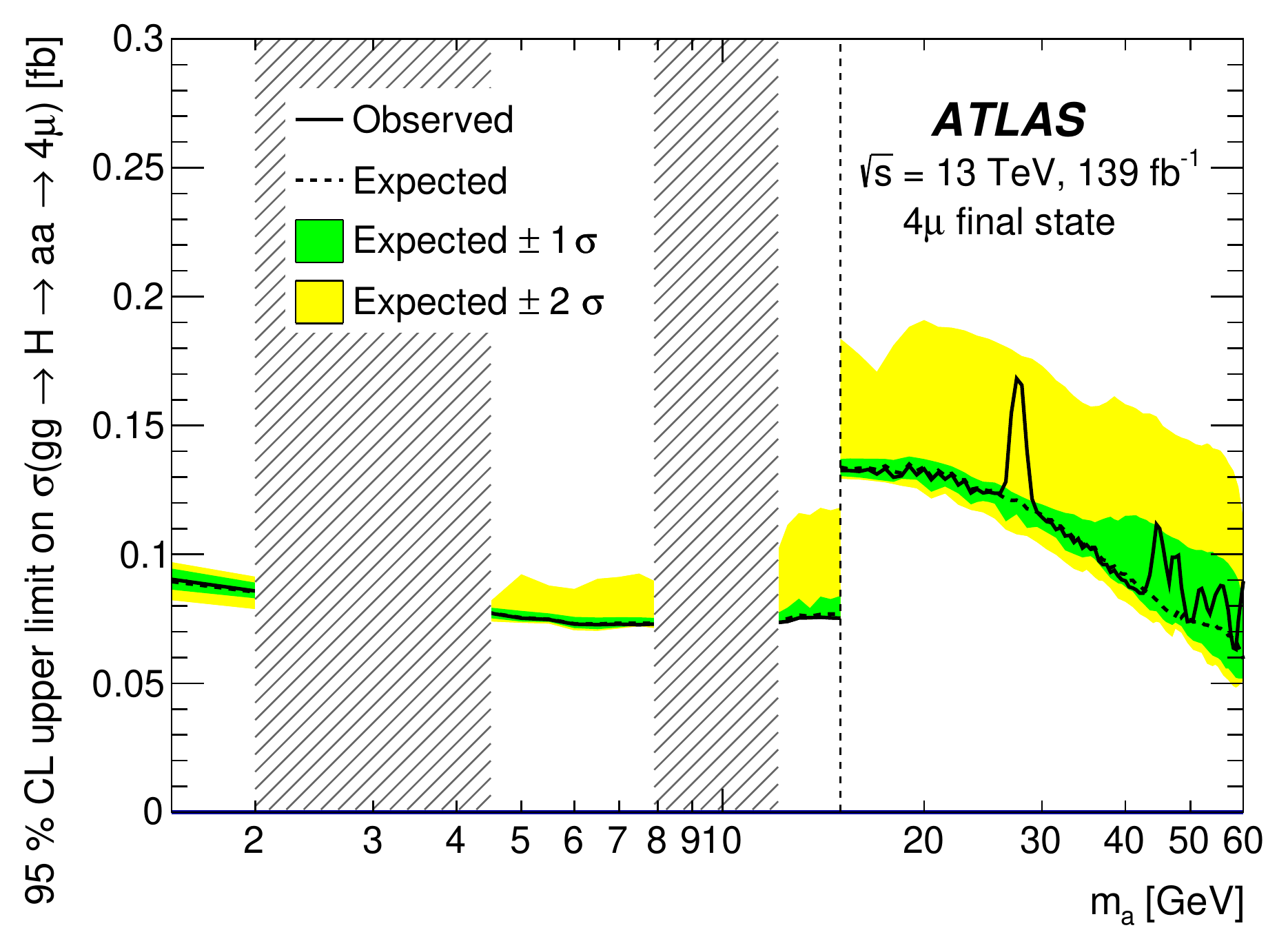}
\label{fig:Limit_xs_aa_4mu}}
\caption{Observed and expected upper limits at 95\popercent CL
for the cross sections of the (a)~$H\to \Zd\Zd\to 4\mu$ and (b)~$H\to aa\to 4\mu$
processes, assuming SM Higgs~boson production via the gluon--gluon fusion process.
The shaded areas are the quarkonia veto regions.
HAHM parameters were set to $\kappa=\epsilon=10^{-4}$.
The step changes at $m_{\Zd}=\SI{15}{\GeV}$ are due
to the change in selection from the LM to the HM analysis.}
\label{fig:Limit_xs_xx_4mu}
\end{center}
\end{figure}

The efficiencies within the fiducial regions for the HM and LM analyses are shown
in \cref{fig:efficiency_ZdZd}.  These were calculated using the
benchmark $H\to \Zd\Zd$ model, but the efficiencies
are mostly model-independent: for $H\to aa\to 4\mu$
over the range $\SI{1}{\GeV} < m_a < \SI{15}{\GeV}$ the efficiencies
are the same as for $H\to \Zd\Zd\to 4\mu$ to
within a relative difference of $3\%$.
The difference in efficiency between different final states is mainly
due to the fact that the efficiencies for reconstruction, identification,
and selection are lower for electrons than for muons.
These efficiencies are used to compute 95\% confidence level (CL) upper limits on the
cross section within the fiducial region, using the CL$_\text{s}$ frequentist
formalism~\cite{Read:2002hq} with the profile-likelihood-ratio
test statistic~\cite{Cowan:2010js-witherratum}.
The resulting limits are shown in \cref{fig:fiducialLimits_ZdZd}.  These limits
should be applicable to any models of the SM Higgs~boson decaying
into four leptons via two intermediate bosons that are narrow, on-shell,
and that decay promptly.
The model-dependent acceptances for the HM and LM analyses are shown
in \cref{fig:acceptance_ZdZd} for the $H\to \Zd\Zd$ and $H\to aa\to 4\mu$
models.  The resulting
upper limit on the product of the total cross section and decay branching
ratio for the benchmark model, $\sigma(gg\to H\to \Zd\Zd\to 4\ell)$,
for the HM analysis is shown in \cref{fig:Limit_xs_ZdZd_4l},
while \cref{fig:Limit_xs_xx_4mu} shows upper limits on
$\sigma(gg\to H\to \Zd\Zd\to 4\mu)$ and
$\sigma(gg\to H\to aa\to 4\mu)$ for both the HM and LM analyses.
These results are independent of assumptions about the decay branching
ratios of the $\Zd$~and $a$~bosons.  In particular, \cref{fig:Limit_xs_aa_4mu}
also applies to the scalar case $\sigma(gg\to H\to ss\to 4\mu)$.

\subsubsection{ZX limits}
\label{sec:zxlimit}
 
\begin{figure}[!htbp]
\begin{center}
\subfloat[]{\includegraphics[width=0.49\textwidth]{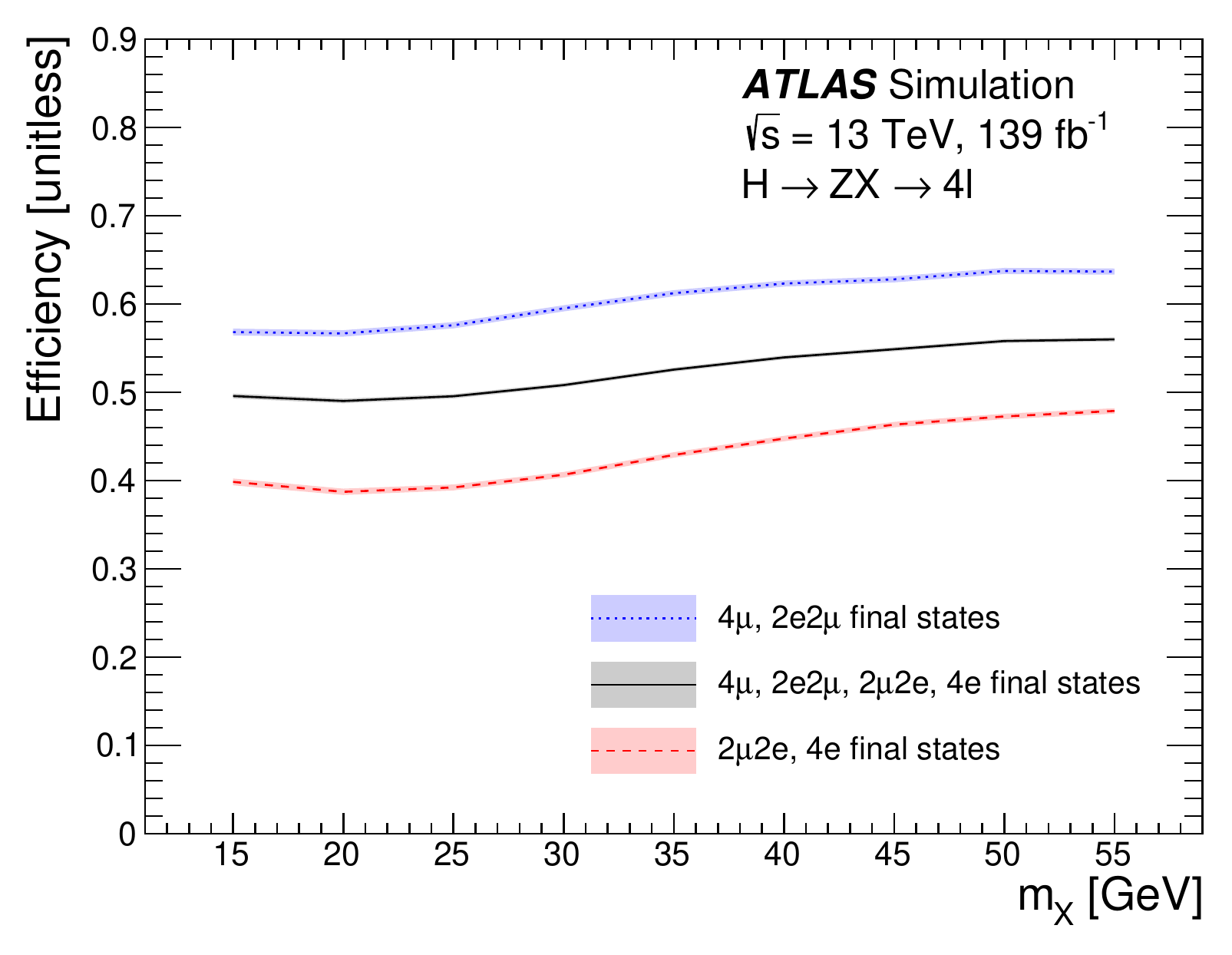}
\label{fig:zx-efficiency}}
\subfloat[]{\includegraphics[width=0.49\textwidth]{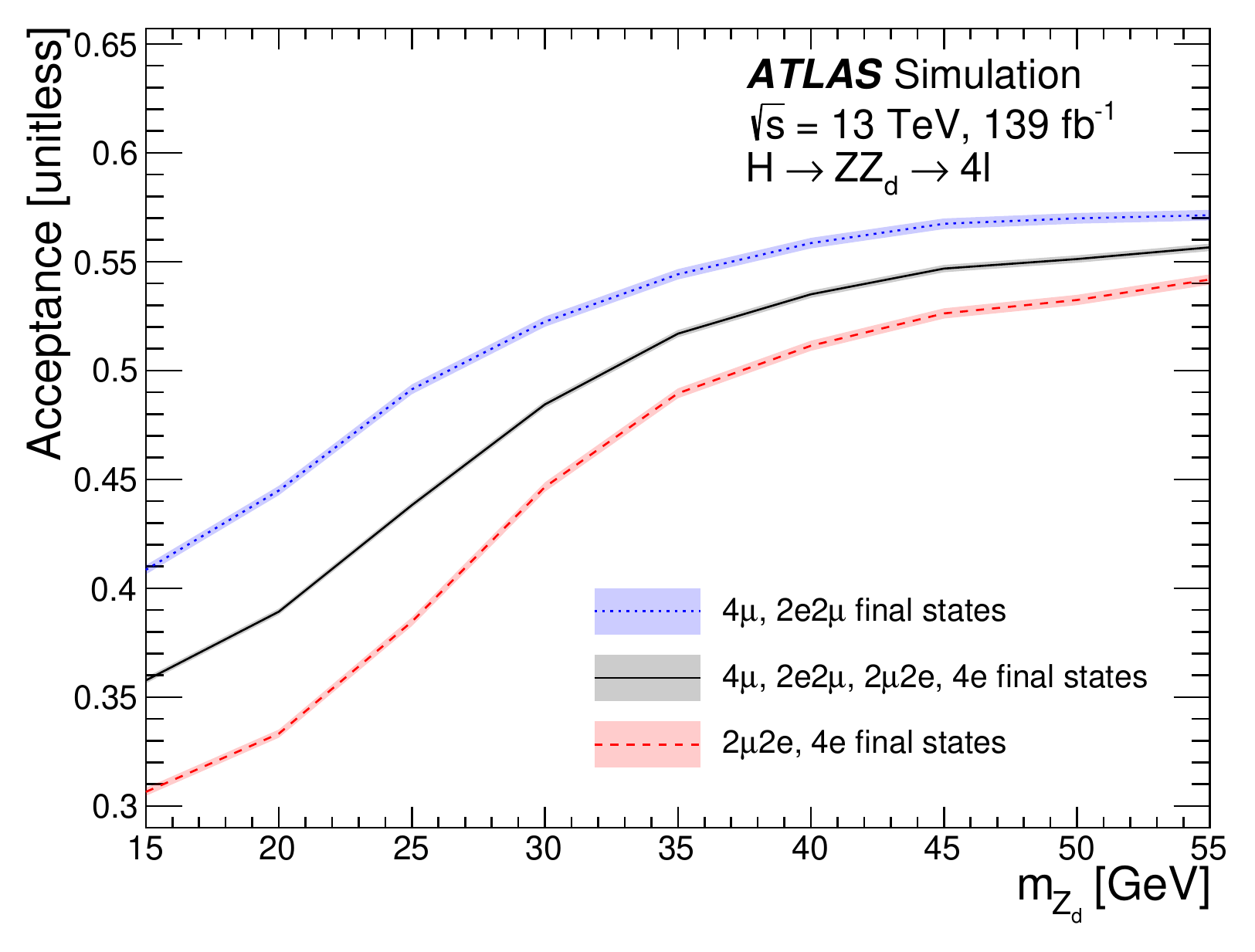}
\label{fig:zx-acceptance}}
\caption{
(a)~Model-independent efficiencies $\epsilon_{c}$
for the $H\to ZX$ process for different combinations of the final states
calculated in the fiducial volumes described in
\ifpo{the fiducial region definition table}\ifnotpo{\cref{tab:fiducial_selection}}.
(b)~Model-dependent per-channel fiducial region acceptances
for the $H\to Z\Zd\to 4\ell$ process for different combinations
of the final states.
}
\end{center}
\end{figure}

\begin{figure}[!htbp]
\begin{center}
\includegraphics[width=0.7\textwidth]{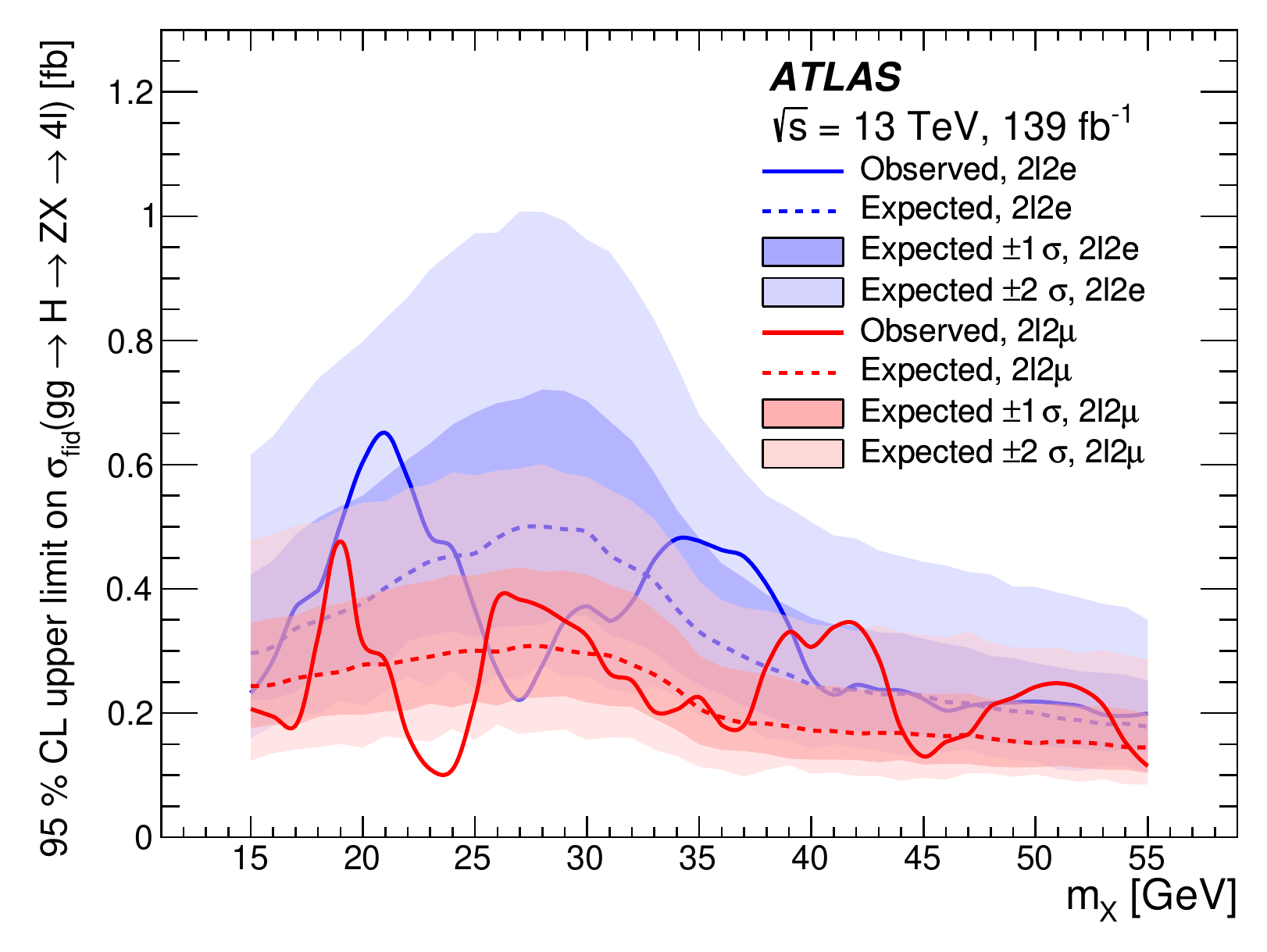}
\caption{
Per-channel upper limit at 95\popercent CL on the fiducial cross section
for the $H\rightarrow ZX \rightarrow 4\ell$
process.
}
\label{fig:zx-fid-limit}
\end{center}
\end{figure}

\begin{figure}[!htbp]
\begin{center}
\subfloat[]{\includegraphics[width=0.48\textwidth]{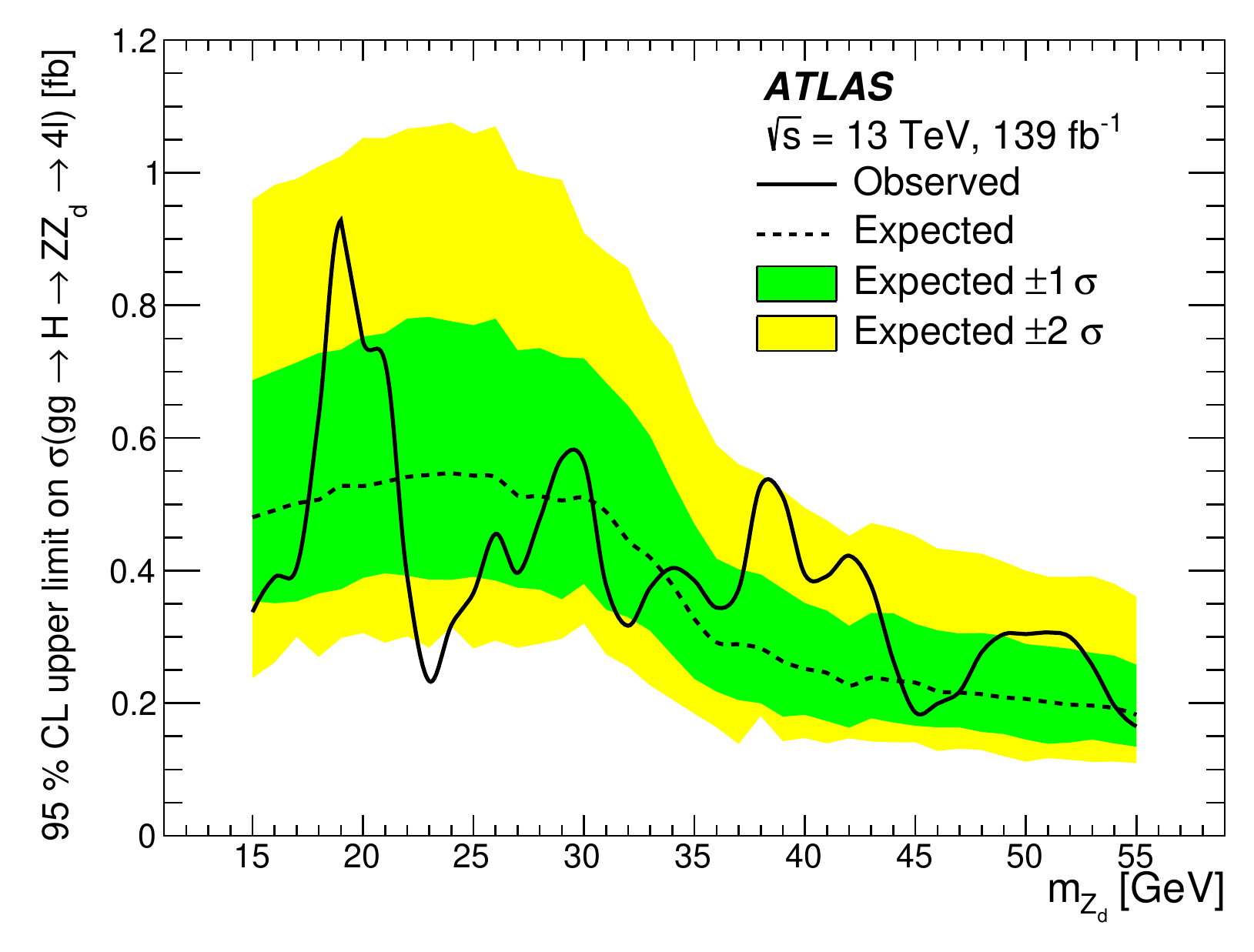}
\label{fig:totsig_zzd}}
\hskip 0.04\textwidth
\subfloat[]{\includegraphics[width=0.48\textwidth]{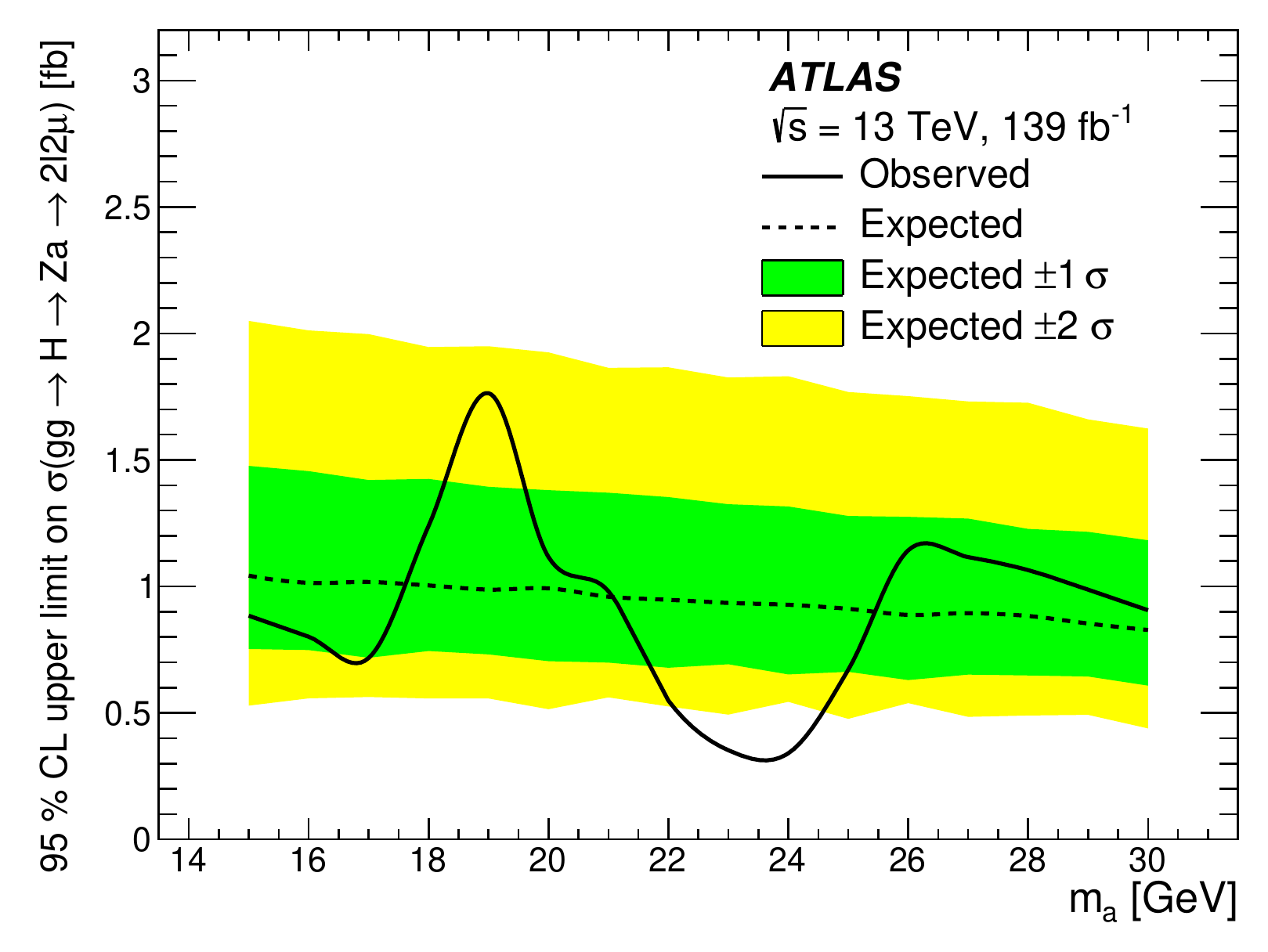}}
\caption{Observed and expected upper limits at 95\popercent CL
for the cross sections of the (a)~$H\to Z\Zd\to 4\ell$ and (b)~$H\to Za\to 2\ell2\mu$
processes, assuming SM Higgs~boson production via the gluon--gluon fusion process.
All final states are combined.
HAHM parameters were set to $\epsilon=10^{-4}$ and $\kappa=10^{-10}$.
}
\label{fig:ZX_limits_All}
\end{center}
\end{figure}

For limits involving ZX processes,
the normalization of the non-resonant $ZZ^*\to 4\ell$ background is
validated using control samples, but the normalization of the remaining
significant background,
$H\to ZZ^*\to 4\ell$, is allowed to float in the limit determination
as an unconstrained nuisance parameter.
The model-independent efficiency within the fiducial region is shown
in \cref{fig:zx-efficiency}, and
the resulting 95\% CL upper limit on the fiducial region cross section
is shown in \cref{fig:zx-fid-limit}.  The fiducial region acceptance
for the $H\ra Z\Zd\ra 4\ell$ process is shown in \cref{fig:zx-acceptance},
and the upper limits on the product of the
total cross section and decay branching ratio for the benchmark models,
$\sigma(gg\to H\to Z\Zd\to 4\ell)$ and
$\sigma(gg\to H\to Za\to 2\ell2\mu)$,
are shown in \cref{fig:ZX_limits_All}.

\subsection{Limits on branching ratios}
 
\begin{figure}[!htbp]
\begin{center}
\subfloat[]{\includegraphics[width=0.49\textwidth]{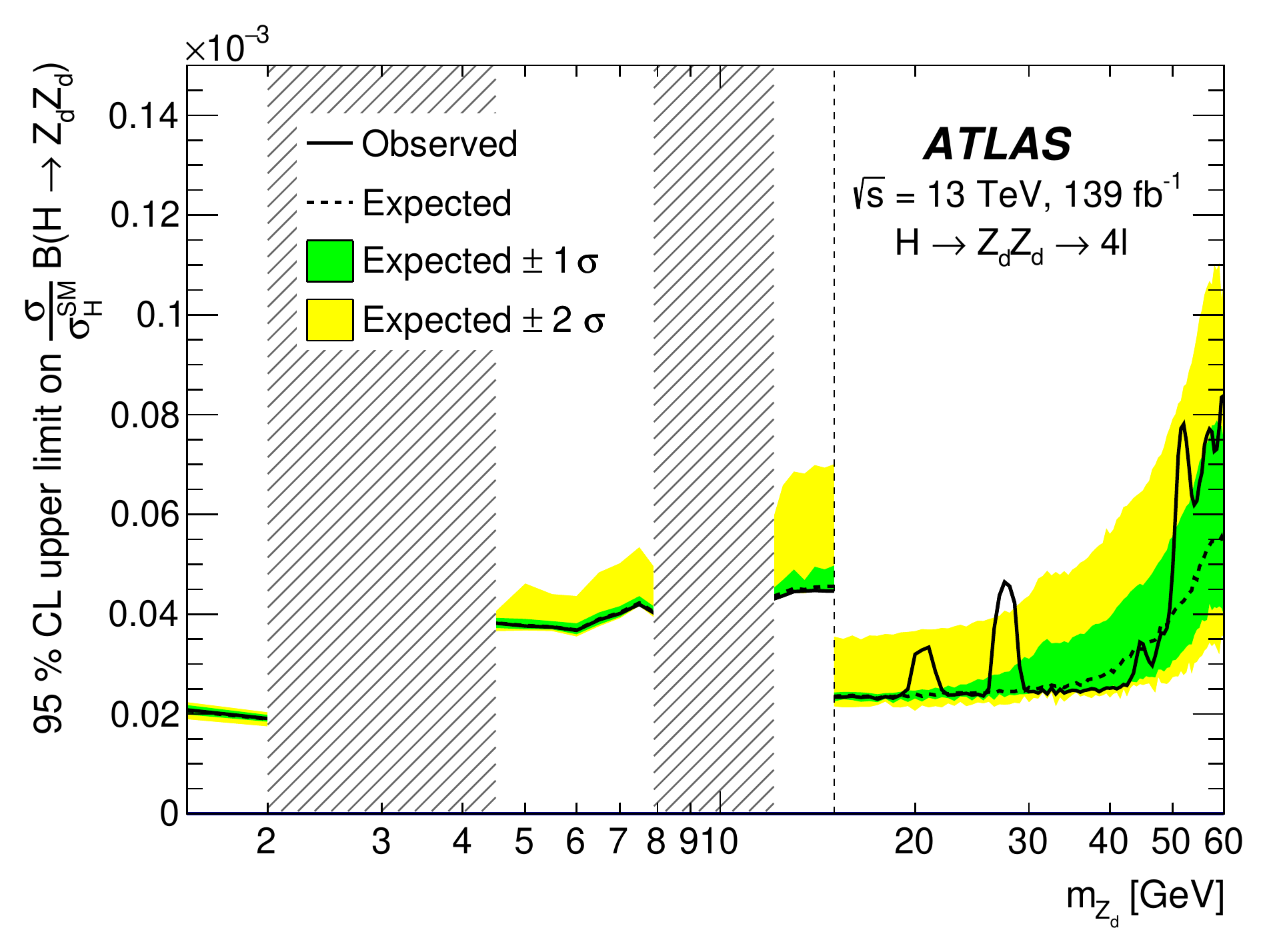}}
\subfloat[]{\includegraphics[width=0.49\textwidth]{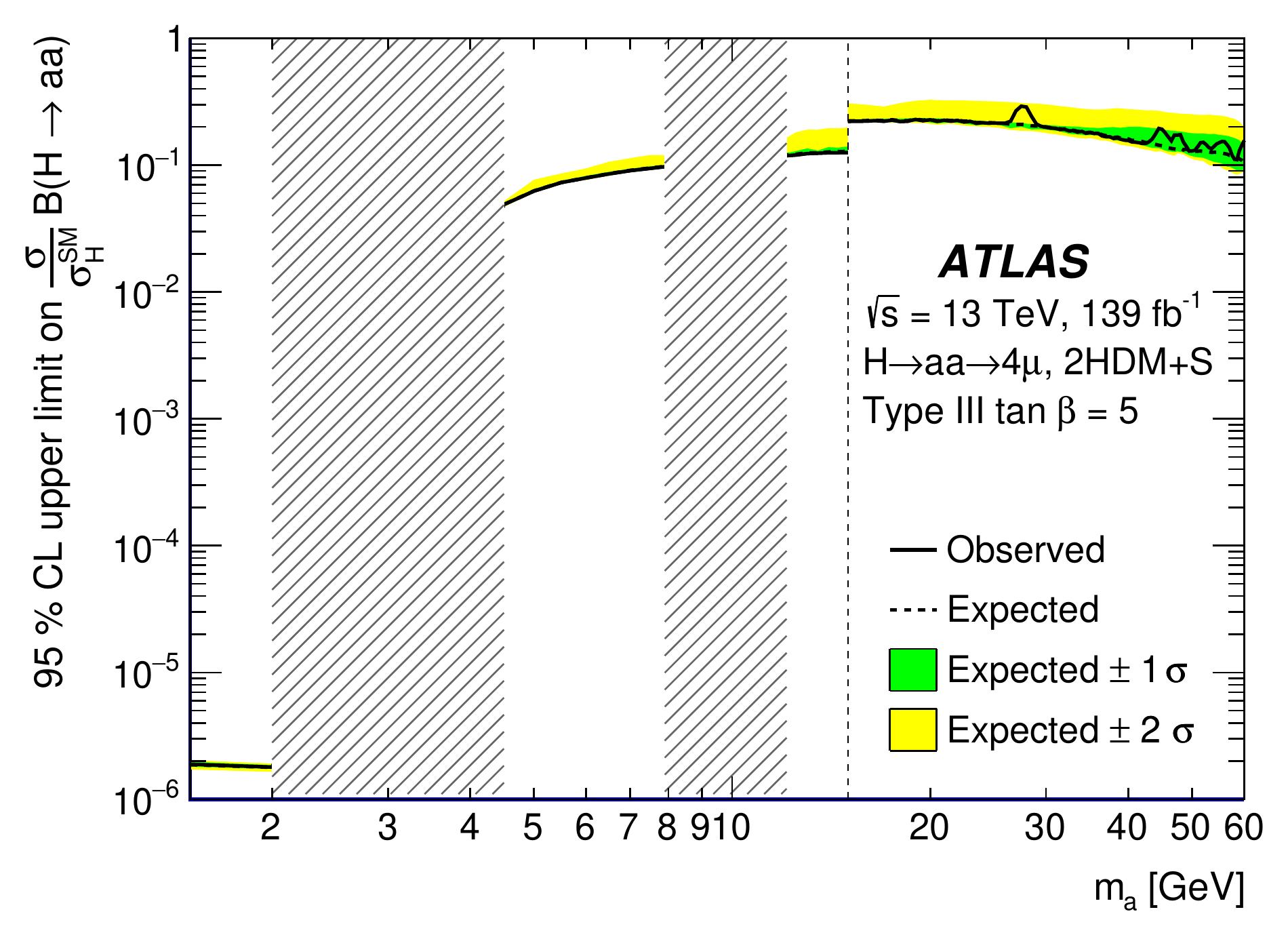}}
\caption{95\popercent CL upper limits on the cross section times
the model-dependent branching ratio divided by the
SM Higgs~boson production cross section
for (a)~the $H\rightarrow \Zd\Zd$ process for the benchmark HAHM
with $\kappa=\epsilon=10^{-4}$ and
(b)~the $H\rightarrow aa$ process for the benchmark 2HDM+S model.
The shaded areas are the quarkonia veto regions.
The step changes at $m_{\Zd}=\SI{15}{\GeV}$
are due to the change in selection from the LM to the HM analysis.}
\label{fig:brLimits}
\end{center}
\end{figure}

A (model-dependent) cross-section limit may be converted
to a branching ratio limit using the relations:
\begin{equation}\nonumber
\mathcal{B}(H\to XX\to 4\ell) = \frac{\sigma_{H\to XX\to 4\ell}}{\sigma_H},
\end{equation}
\begin{equation}\nonumber
\mathcal{B}(H\to XX) = \frac{\mathcal{B}(H\to XX\to 4\ell)}{\sum_{\ell_1=e,\mu}\sum_{\ell_2=\e,\mu} [\mathcal{B}(X\to 2\ell_1)\mathcal{B}(X\to 2\ell_2)]},
\end{equation}
where $\sigma_{H\to XX\to 4\ell}$ is the model-dependent total cross section,
$\sigma_H$ is the SM Higgs~boson production cross section
for the ggF process
($\SI{48.58}{\pb}$ for $m_H=\SI{125}{\GeV}$~\cite{Anastasiou:2016cez}),
and $\mathcal{B}(X\to 2\ell)$ is the model-dependent branching ratio
for each decay to one lepton flavour.
The branching ratios for $\Zd\to \ell\ell$ and $a\to \mu\mu$ are taken
from the benchmark models~\cite{Curtin:2014cca,Curtin:2013fra}, where
for the $\Zd\to\ell\ell$ case, the branching ratios for the two lepton
flavours are taken to be equal.
For the $a\to\mu\mu$ case, the branching ratio varies  considerably
in a model-dependent way over the range of $m_a$ considered here.
The resulting branching ratio limits
are shown in \cref{fig:brLimits}.

\subsection{Limits on Higgs mixing}
 
\begin{figure}[!htbp]
\begin{center}
\includegraphics[width=0.65\textwidth]{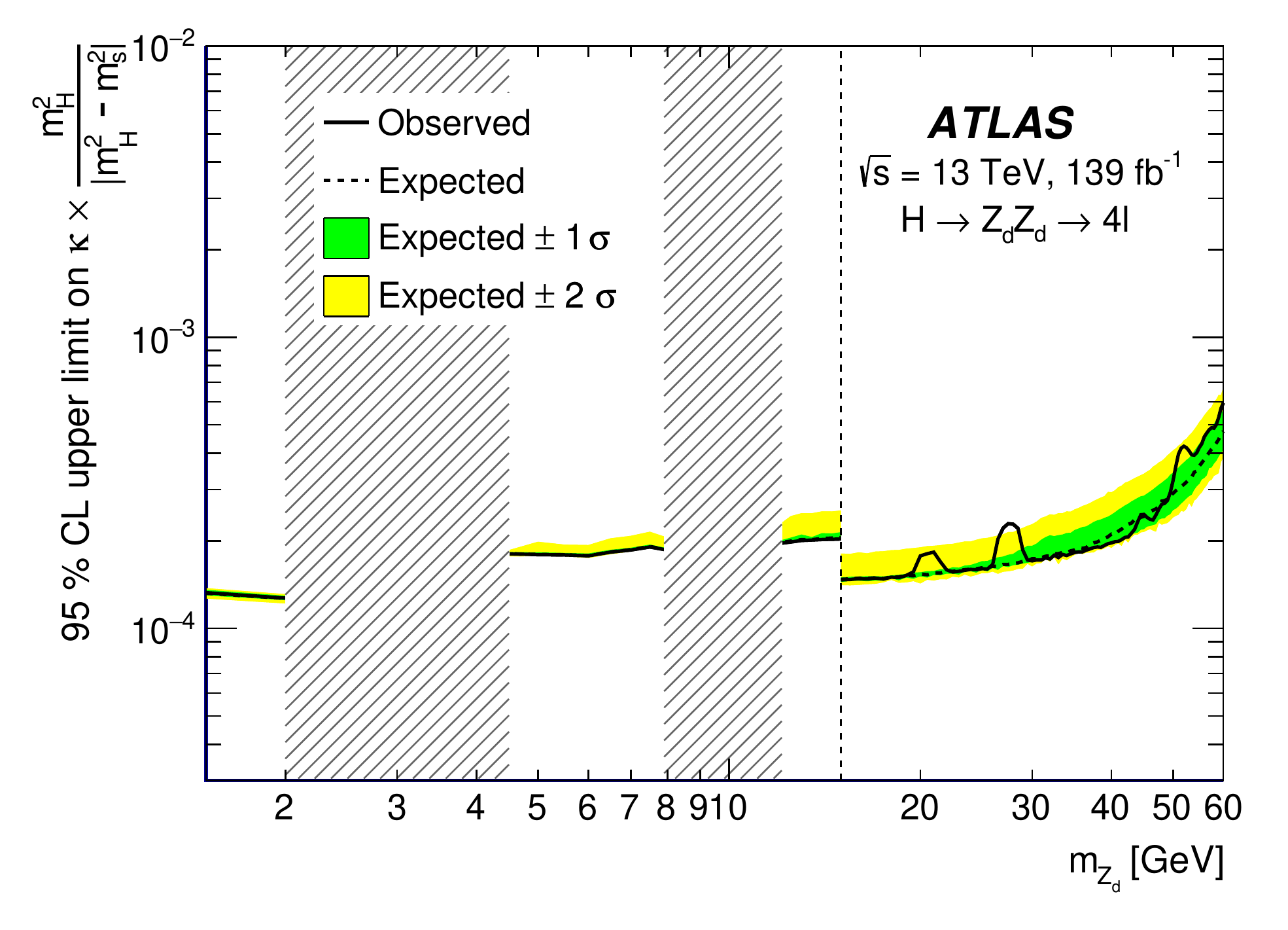}
\caption{Upper limit at
$95\popercent$
CL on
the effective Higgs mixing parameter
$\kappa^{\prime}=\kappa  m_H^2 / |m_H^2-m_S^2|$,
with $\epsilon$ set to $10^{-4}$.
The step change at
$m_{\Zd}=\SI{15}{\GeV}$
is due to the change in
selection from the LM to the HM analysis.
The shaded areas are the quarkonia veto regions.}
\label{fig:kappalimit}
\end{center}
\end{figure}

The branching ratio limit can also be interpreted as a limit on the
effective Higgs mixing parameter $\kappa'$, defined as
\begin{equation}\nonumber
\kappa' = \kappa  \frac{m_H^2}{|m_H^2-m_S^2|},
\end{equation}
where $\kappa$ is the Higgs portal coupling and $m_S$ is the mass
of the dark Higgs~boson.
Using $\kappa'$ rather than $\kappa$ combines the
dependencies on $\kappa$ and $m_S$ into a single parameter.
Then, according to Eq.~(2.33) of Ref.~\cite{Curtin:2014cca}
and assuming $m_S > m_H/2$:
\begin{equation}\nonumber
\kappa'^2 = \frac{\Gamma_{\mathrm{SM}}}{f(m_{\Zd})}\frac{\mathcal{B}(H\rightarrow \Zd\Zd)}{1-\mathcal{B}(H\rightarrow \Zd\Zd)},
\end{equation}
where $\Gamma_{\mathrm{SM}}$ is the SM width of the $\SI{125}{\gev}$ Higgs~boson,
\begin{equation}\nonumber
f(m_{\Zd}) =  \frac{v^2}{32\pi m_H} \sqrt{1-\frac{4m^2_{\Zd}}{m^2_H}}
\frac{\left(m^2_H+2m^2_{\Zd}\right)^2-8\left(m^2_H-m^2_{\Zd}\right)m^2_{\Zd}}{m^4_H},
\end{equation}
and $v\approx \SI{246}{\GeV}$ is the vacuum expectation value of
the Higgs field.
The resulting limit is shown in \cref{fig:kappalimit}.
 
\begin{figure}[!htbp]
\begin{center}
\subfloat[]{
\includegraphics[width=0.48\textwidth]{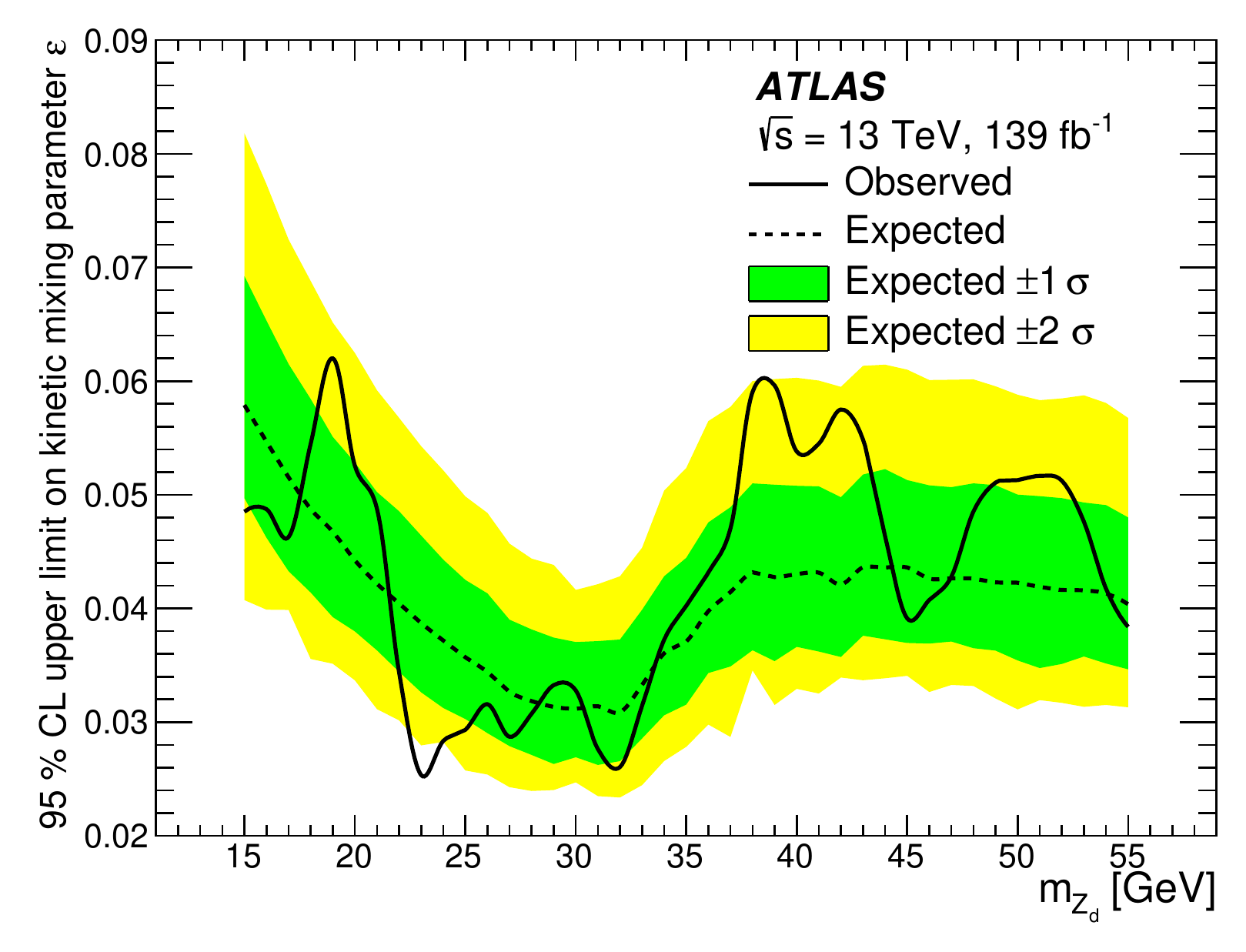}}
\hskip 0.01\textwidth
\subfloat[]{
\includegraphics[width=0.47\textwidth]{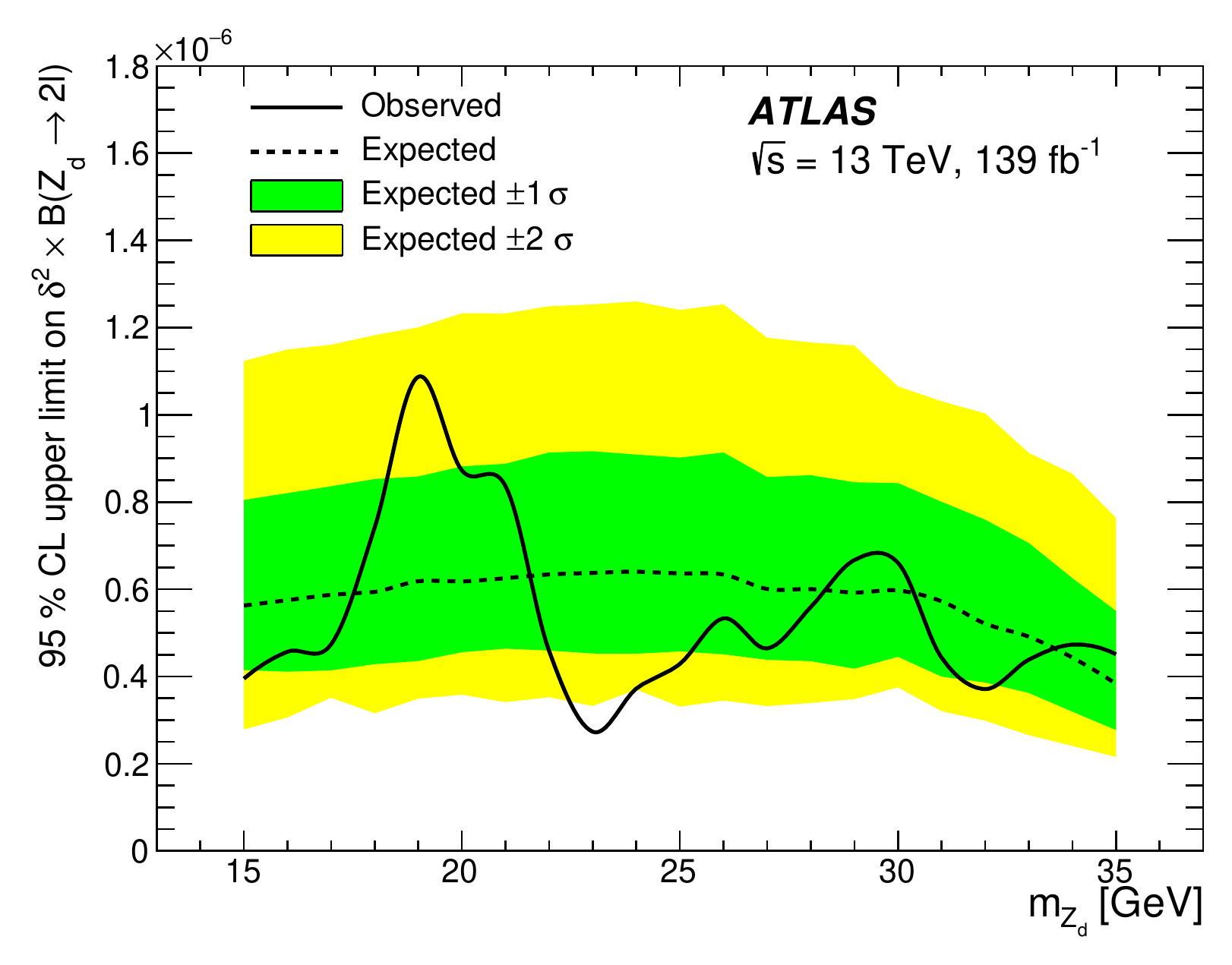}}
\caption{Upper limit at $95\popercent$ CL on (a)~the $\Zd$ mixing
parameter $\epsilon$, with $\kappa$ set to $10^{-10}$, and (b)~the $Z$--$\Zd$ mass mixing parameter $\delta^2 \times \mathcal{B}(\Zd\to \ell\ell)$, assuming the SM Higgs~boson production cross section.}
\label{fig:mixinglimits}
\end{center}
\end{figure}

The $H\to Z \Zd$ analysis can also be used to set limits on the $\Zd$ mixing
parameter $\epsilon$ and on the $Z$--$\Zd$ mass mixing parameter $\delta$,
as described in Refs.~\cite{Curtin:2014cca,EXOT-2013-15}.
These are shown in \cref{fig:mixinglimits}, assuming the SM Higgs~boson
production cross section.

\FloatBarrier
 
\section{Conclusion}
\label{sec:Conclusion}
 
Searches have been conducted for exotic decays of the
Standard Model Higgs~boson into
two new spin-1 particles $H\to \Zd \Zd$, two new spin-0 particles $H\to aa$,
or a $Z$~boson along with a single $\Zd$ or $a$.
The searches used
$\SI{139}{\ifb}$ of proton--proton
collision data
at $\sqrt{s} = \SI{13}{\TeV}$ recorded by the ATLAS experiment at the LHC
during the period 2015--2018.  The first search is for the process
$H\to XX\to 4\ell$, where $X$ is either $\Zd$ or $a$,
with $\SI{15}{\GeV} < m_X < \SI{60}{\GeV}$.
The second search is for the process $H\to XX\to 4\mu$, with
$\SI{1}{\GeV} < m_X < \SI{15}{\GeV}$.
The third search is for the process $H\ra ZX\ra 4\ell$, with
$\SI{15}{\GeV} < m_X < \SI{55}{\GeV}$.
The data are found to be consistent
with the predicted backgrounds in the three aforementioned searches, and
limits on fiducial and total cross sections are set.

Specializing to the benchmark models,
upper limits are set on the branching ratio of the Higgs~boson to
$\Zd\Zd$ and $aa$ as a function of intermediate boson mass,
assuming gluon--gluon fusion Standard Model Higgs
production and prompt decay of the $\Zd$\,/$a$~bosons.
Furthermore, assuming the Hidden Abelian Higgs Model introduced at the Higgs portal level with very weak kinetic mixing, limits are set on the mixing parameters $\kappa'$,
$\epsilon$, and $\delta$.
 
The limits presented in this paper improve on those from
the previous ATLAS search by factors between two and four
due to a larger data sample, improved lepton reconstruction and identification,
and a better optimized event selection.  In addition to the improvements on the
results from the previous search, this paper also presents
limits on total cross sections and on the dark Higgs~boson mixing parameters.


\section*{Acknowledgements}

We thank CERN for the very successful operation of the LHC, as well as the
support staff from our institutions without whom ATLAS could not be
operated efficiently.
 
We acknowledge the support of
ANPCyT, Argentina;
YerPhI, Armenia;
ARC, Australia;
BMWFW and FWF, Austria;
ANAS, Azerbaijan;
SSTC, Belarus;
CNPq and FAPESP, Brazil;
NSERC, NRC and CFI, Canada;
CERN;
ANID, Chile;
CAS, MOST and NSFC, China;
Minciencias, Colombia;
MEYS CR, Czech Republic;
DNRF and DNSRC, Denmark;
IN2P3-CNRS and CEA-DRF/IRFU, France;
SRNSFG, Georgia;
BMBF, HGF and MPG, Germany;
GSRI, Greece;
RGC and Hong Kong SAR, China;
ISF and Benoziyo Center, Israel;
INFN, Italy;
MEXT and JSPS, Japan;
CNRST, Morocco;
NWO, Netherlands;
RCN, Norway;
MEiN, Poland;
FCT, Portugal;
MNE/IFA, Romania;
JINR;
MES of Russia and NRC KI, Russian Federation;
MESTD, Serbia;
MSSR, Slovakia;
ARRS and MIZ\v{S}, Slovenia;
DSI/NRF, South Africa;
MICINN, Spain;
SRC and Wallenberg Foundation, Sweden;
SERI, SNSF and Cantons of Bern and Geneva, Switzerland;
MOST, Taiwan;
TAEK, Turkey;
STFC, United Kingdom;
DOE and NSF, United States of America.
In addition, individual groups and members have received support from
BCKDF, CANARIE, Compute Canada and CRC, Canada;
COST, ERC, ERDF, Horizon 2020 and Marie Sk{\l}odowska-Curie Actions, European Union;
Investissements d'Avenir Labex, Investissements d'Avenir Idex and ANR, France;
DFG and AvH Foundation, Germany;
Herakleitos, Thales and Aristeia programmes co-financed by EU-ESF and the Greek NSRF, Greece;
BSF-NSF and GIF, Israel;
Norwegian Financial Mechanism 2014-2021, Norway;
NCN and NAWA, Poland;
La Caixa Banking Foundation, CERCA Programme Generalitat de Catalunya and PROMETEO and GenT Programmes Generalitat Valenciana, Spain;
G\"{o}ran Gustafssons Stiftelse, Sweden;
The Royal Society and Leverhulme Trust, United Kingdom.
 
The crucial computing support from all WLCG partners is acknowledged gratefully, in particular from CERN, the ATLAS Tier-1 facilities at TRIUMF (Canada), NDGF (Denmark, Norway, Sweden), CC-IN2P3 (France), KIT/GridKA (Germany), INFN-CNAF (Italy), NL-T1 (Netherlands), PIC (Spain), ASGC (Taiwan), RAL (UK) and BNL (USA), the Tier-2 facilities worldwide and large non-WLCG resource providers. Major contributors of computing resources are listed in Ref.~\cite{ATL-SOFT-PUB-2021-003}.
 

\appendix
 
\appendix
\part*{Appendix}
\addcontentsline{toc}{part}{Appendix}
 
The signal region for the HM analysis in Ref.~\cite{EXOT-2016-22} was
defined by $m_{34}/m_{12} > 0.85 $.  This was the result of an optimization
that assumes the $\Zd$ width is narrow, as expected
in the HAHM,
so that the observed width will be dominated by the detector performance.
A $\pm 2\sigma$ change in the energy measurement gives about
a $15\%$ change in the lepton-pair invariant mass, motivating
the coefficient of 0.85 in the selection.
However, the background is low, especially in the lower mass region, so
it is possible to widen the signal region somewhat relative to the
background without significant loss of sensitivity.
This is further motivated by models such as the one discussed
in Chapter~7 of Ref.~\cite{Alimena}.
Accordingly,
the signal region selection was re-optimized to allow for a larger $\Zd$ width.
 
The selection widens up to $3.5\sigma$ at the low end of the
dilepton mass spectrum, where the background is the lowest,
but narrows to the previous value of $2\sigma$ for higher masses.
So the form of the selection is
\begin{equation}\nonumber
m_{34}/m_{12} > 0.85 - 0.1125 f(m_{12}) ,
\end{equation}
where $f(m_{12})$ is a modulating function that is 1 at the lowest
mass and goes to 0 at higher masses.
 
The form of the modulating function is taken from the shape of the
average dilepton mass spectrum in the background.
The function fit to this shape
consists of an exponential tail matched with half of a Gaussian:
\begin{equation}\nonumber
F(x) =  \left\{
\begin{aligned}
B_1+B_2(x-X)+h\,\mathrm{e}^{\frac{-(x-X)^2}{2 \sigma^2}}, \quad x > X-T\\
B_1+B_2(x-X)+h\,\mathrm{e}^{\frac{T(2 x-2 X+T)}{2 \sigma^2}}, \quad x < X-T\\
\end{aligned}
\right. .
\end{equation}
The fit parameters, with $x$ given in \GeV, are $h = 3.73$,
$X = 51.6$,
$\sigma = 16.6$,
$B_1 = -2.62$,
$B_2 = -0.0266$,
and $T = 6.39$,
with the maximum of $F(x)$ occurring at $x_{\max} = \SI{49.64}{\GeV}$.
The modulating function is then defined by
\begin{equation}\nonumber
f(m_{12}) =  \left\{
\begin{aligned}
1 - {\frac{F(m_{12}) - F(10)}{F(x_{\max}) - F(10)}}, \quad m_{12} < x_{\max}\\
0, \quad m_{12} > x_{\max}\\
\end{aligned}
\right. .
\end{equation}

The final shape of the re-optimized signal region is shown in \cref{fig:resultsDataScatterHighMass}.


\clearpage
 
\printbibliography

\clearpage 
 
\begin{flushleft}
\hypersetup{urlcolor=black}
{\Large The ATLAS Collaboration}

\bigskip

\AtlasOrcid[0000-0002-6665-4934]{G.~Aad}$^\textrm{\scriptsize 98}$,    
\AtlasOrcid[0000-0002-5888-2734]{B.~Abbott}$^\textrm{\scriptsize 124}$,    
\AtlasOrcid[0000-0002-7248-3203]{D.C.~Abbott}$^\textrm{\scriptsize 99}$,    
\AtlasOrcid[0000-0002-2788-3822]{A.~Abed~Abud}$^\textrm{\scriptsize 34}$,    
\AtlasOrcid[0000-0002-1002-1652]{K.~Abeling}$^\textrm{\scriptsize 51}$,    
\AtlasOrcid[0000-0002-2987-4006]{D.K.~Abhayasinghe}$^\textrm{\scriptsize 91}$,    
\AtlasOrcid[0000-0002-8496-9294]{S.H.~Abidi}$^\textrm{\scriptsize 27}$,    
\AtlasOrcid[0000-0002-9987-2292]{A.~Aboulhorma}$^\textrm{\scriptsize 33e}$,    
\AtlasOrcid[0000-0001-5329-6640]{H.~Abramowicz}$^\textrm{\scriptsize 157}$,    
\AtlasOrcid[0000-0002-1599-2896]{H.~Abreu}$^\textrm{\scriptsize 156}$
% ...
\bigskip
\\

% ...
$^{27}$Physics Department, Brookhaven National Laboratory, Upton NY; United States of America.\\
% ...
$^{33}$$^{(a)}$Facult\'e des Sciences Ain Chock, R\'eseau Universitaire de Physique des Hautes Energies - Universit\'e Hassan II, Casablanca;$^{(b)}$Facult\'{e} des Sciences, Universit\'{e} Ibn-Tofail, K\'{e}nitra;$^{(c)}$Facult\'e des Sciences Semlalia, Universit\'e Cadi Ayyad, LPHEA-Marrakech;$^{(d)}$LPMR, Facult\'e des Sciences, Universit\'e Mohamed Premier, Oujda;$^{(e)}$Facult\'e des sciences, Universit\'e Mohammed V, Rabat;$^{(f)}$Mohammed VI Polytechnic University, Ben Guerir; Morocco.\\
$^{34}$CERN, Geneva; Switzerland.\\
% ...
$^{51}$II. Physikalisches Institut, Georg-August-Universit\"{a}t G\"ottingen, G\"ottingen; Germany.\\
% ...
$^{91}$Department of Physics, Royal Holloway University of London, Egham; United Kingdom.\\
% ...
$^{98}$CPPM, Aix-Marseille Universit\'e, CNRS/IN2P3, Marseille; France.\\
$^{99}$Department of Physics, University of Massachusetts, Amherst MA; United States of America.\\
% ...
$^{124}$Homer L. Dodge Department of Physics and Astronomy, University of Oklahoma, Norman OK; United States of America.\\
% ...
$^{156}$Department of Physics, Technion, Israel Institute of Technology, Haifa; Israel.\\
$^{157}$Raymond and Beverly Sackler School of Physics and Astronomy, Tel Aviv University, Tel Aviv; Israel.\\
% ...
$^{a}$ Also at Borough of Manhattan Community College, City University of New York, New York NY; United States of America.\\
$^{b}$ Also at Bruno Kessler Foundation, Trento; Italy.\\
$^{c}$ Also at Center for High Energy Physics, Peking University; China.\\
$^{d}$ Also at Centro Studi e Ricerche Enrico Fermi; Italy.\\
$^{e}$ Also at CERN, Geneva; Switzerland.\\
$^{f}$ Also at D\'epartement de Physique Nucl\'eaire et Corpusculaire, Universit\'e de Gen\`eve, Gen\`eve; Switzerland.\\
% ...

\end{flushleft}


\end{document}